\documentclass[12pt]{article}
\usepackage{amsfonts, amsmath, amssymb, cite, graphicx, }

\usepackage{color}
\usepackage[all, knot]{xy}
\usepackage{tikz}
\usepackage{subfigure}
\usepackage{braket}

\usepackage[utf8]{inputenc}
\usepackage{ulem}

\usepackage{bm}
\usepackage{epstopdf}
\usepackage[footnotesize]{caption}
\usepackage{amsthm}
\usepackage{amsthm,amsmath,amssymb}
\usepackage{enumitem}
\usepackage{mathrsfs}
\usepackage{makecell}
\usepackage{diagbox}

\usepackage[margin=3cm]{geometry}

\begin{document}

\begin{titlepage}
\vspace{0.5cm}
\begin{center}
{\Large \bf Quantum cosmology of the flat universe via closed real-time path integral}

\lineskip .75em
\vskip 2.5cm
{\large Hong Wang$^{a}$, Jin Wang$^{b,}$\footnote{jin.wang.1@stonybrook.edu} }
\vskip 2.5em
 {\normalsize\it $^{a}$State Key Laboratory of Electroanalytical Chemistry, Changchun Institute of Applied Chemistry, Chinese Academy of Sciences, Changchun 130022, China\\
 $^{b}$Department of Chemistry and Department of Physics and Astronomy, State University of New York at Stony Brook, NY 11794, USA}
\vskip 3.0em
\end{center}
\begin{abstract}
Quantum cosmology is crucial to understand the evolution of the early universe. Despite significant progress, challenges still remain. For example, the role of time in quantum cosmology is unclear. Furthermore, the influence of the environment on the evolution of the quantum universe is challenging. In this work, we studied the evolution of the quantum universe non-perturbatively using the closed real-time path integral. The environments coupled to the quantum universe being considered are the radiation, the non-relativistic matter, or the dark matter. We evaluated the influence functional of the massless scalar field coupled with the flat FRW universe. We studied the evolution of the quantum universe by setting the initial state of spacetime as a Gaussian wave packet. In different scenarios, we show that the classical trajectory of the universe is consistent with the quantum evolution of the wave packet. The coherence, the absolute quantum fluctuation and the Gibbs entropy all monotonically increase with time, yet the relative quantum fluctuation decreases with time. We show that for a given size of the radiation dominated universe, the lower temperature corresponds to a more quantum universe. We find that the minimal coupling of the free massless scalar field with the flat FRW spacetime generally gives rise to the memory characterized via non-Markovian correlations. Finally, we show that under higher radiation temperatures, a small universe has a higher chance of a transition to a bigger universe.
\end{abstract}
\end{titlepage}

\baselineskip=0.7cm

\tableofcontents
\newpage

\section{Introduction}
\label{sec:1}

There are several issues related with gravity that cannot be solved in the framework of classical general relativity. This includes the singularity point of black holes, the very early times of the universe, and the decoherence of spacetime. Developing a well-defined quantum gravity model is one of the most important challenges in theoretical physics. The specific study of quantum gravity is known as quantum cosmology. Although there are many attempts, a complete theoretical framework for quantum cosmology is still lacking. One of the main obstacles is the problem of time~\cite{CK,CR,AA,KV,CJ,JD}. After quantizing the universe, the super Hamiltonian constraint becomes the Wheeler--DeWitt equation~\cite{CK,CR}
\begin{equation}
\label{eq:1.1}
\hat{H}|\Psi\rangle=0,
\end{equation}
where it appears as if there is no time variable. This leads to difficulty in describing the evolution of any quantum gravitation system, which is the time problem in canonical quantum gravity~\cite{CK,KV,CJ}. While these models try to solve this problem, to the best of our knowledge, none of them have fully resolved this issue.

In 1995, Brown and Kucha$\check{\mathrm{r}}$ introduced a dust field as the time variable~\cite{JD}. By imposing the canonical time gauge fixing condition, one can transform the Wheeler--DeWitt equation to the time-dependent Schr\"{o}dinger equation~\cite{VT}
\begin{equation}
\label{eq:1.2}
i\frac{\partial |\Psi\rangle }{\partial t}=\hat{H}|\Psi\rangle,
\end{equation}
where $t$ represents the coordinate time variable, which is also the dust field~\cite{VT}. Throughout this study, we work in natural units $\hbar=c=G=k_{B}=1$ with the signature (+,-,-,-). Using Eq.~\eqref{eq:1.2} allows studying the unitary evolution of the quantum universe.

In 2009, Amemiya and Koike used Eq.~\eqref{eq:1.2} (in the proper time coordinate) to show that the initial singularity of a flat FRW universe can be avoided by the quantum effect~\cite{FT}. Other works (see~\cite{CK2,TD} and related references therein) have shown that the singularity can be avoided in canonical quantum cosmology. In 2015, Maeda used Eq.~\eqref{eq:1.2} to study the evolution of the quantum universe as driven by a cosmological constant~\cite{HM}. By setting the initial state of the universe as a Gaussian wave packet, the classical trajectory of the universe (driven by the dark energy) was shown to be consistent with the evolution of the wave packet~\cite{HM}.

When the total system degree of freedom is large, it is difficult to study the evolution of Eq.~\eqref{eq:1.2}. In general, one can divide the total system into a subsystem plus the environment~\cite{HC,HF}. The information of the subsystem can be determined from the reduced density matrix, which has an evolution that is dictated by the Liouville--von Neumann equation~\cite{HC,HF}
\begin{equation}
\label{eq:1.3}
\frac{d\rho}{dt}=-i\mathrm{Tr}_{B}[\hat{H},\ \rho_{tot}],
\end{equation}
where $\rho$ and $\rho_{tot}$ represent the density matrices of the subsystem and total system, respectively. The trace is taken over the environment. Strictly solving Eq.~\eqref{eq:1.3} is nearly impossible for usual situations. If coupling between the system and the environment is sufficiently weak, and the memory time of the environment is short enough, Eq.~\eqref{eq:1.3} can be approximated as the Born--Markov quantum master equation~\cite{HC,HF}, which is widely used to study open quantum systems~\cite{HC,HF}. If the interaction Hamiltonian is not small compared to the Hamiltonian of the system and the environment, we cannot use the perturbation approximation and the quantum master equation is not suitable. In this case, the non-perturbative approach is needed.

One of the powerful non-perturbative methods is Feynman's path integral, which also works well in the perturbative regime~\cite{LS,VN,LC}. In the path integral, the time variable can be a real or imaginary number, which corresponds to the Lorentz or Euclidean path integrals, respectively. There is no oscillation problem for the Euclidean path integral, but this is not suitable for nonequilibrium cases~\cite{AM}. For the Euclidean path integral in quantum gravity, there is also the divergence problem as the Euclidean--Einstein--Hilbert action has no lower bound~\cite{GW}. Therefore, the Lorentz path integral has attracted increasing attention~\cite{EW1,EW2,TY,YT1,YT2,HDM,HSY,AGP,ABG1,ABG2,ZG1,ZG2,WYA,JF1,JF2,JF3,JF4,GN,GN2,JDJ,DE,NM1,NM2,NM3,NM4,NM5,NM6,SK,DS1,DS2,DS3}. For nonequilibrium quantum dynamics, the ``trajectory" of the time variable forms a closed contour~\cite{JR,XC,LM,AK}. Thus, it is usually called the closed real-time path integral (or Schwinger--Keldysh path integral).

How the quantum universe evolves when the matter field is heat radiation or non-relativistic particles remains a challenge. Although the universe can sometimes be simplified as homogeneous and isotropic, the degrees of the freedom of the environment are often infinite (the vibrational modes of the particles are infinite)~\cite{LP,WW}. One can treat spacetime and the matter field (radiation or massive particles) as the open system and the environment, respectively~\cite{WW}. The evolution of the reduced density matrix corresponding to quantum spacetime can be considered based on Eq.~\eqref{eq:1.3}.

In our universe, there are various of matter fields, such as the scalar field, the vector field, the Dirac field and so on. For simplicity, one often studies the evolution of the universe driven by the scalar field. In this simple case, the whole universe as an isolated system is composed of the spacetime (gravitational field) and the scalar field. In this work, we are interested in the dynamics of the spacetime. Thus we can trace out the degrees of freedom of the scalar field. Or in another word, we choose the scalar field as the   environment. If the scalar field is massless and in the thermal state, all scalar field quanta compose a heat bath. If the scalar particle is massive, we assume that all massive particles are in the ground state for simplicity, and the particles compose a Bose--Einstein condensate. We study the quantum evolution of homogeneous and isotropic spacetime as driven by heat radiation and the Bose--Einstein condensate. The interaction Hamiltonian is not small, and the memory time of the radiation is long (details to come). Thus, the quantum master equation is not suitable, and we use the closed real-time path integral to study the model.

In this work, we study the evolution of the quantum spacetime by setting the initial state as a Gaussian wave packet. We show that in different coordinates, the trajectory of the wave packet is consistent with the classical evolution of the universe. We find that whatever the form of the matter field is, this conclusion is always valid. In different scenarios, the coherence, the absolute quantum fluctuation and the Gibbs entropy all monotonically increase with time. Trends in variations of these quantities are similar to each other. We show that the relative quantum fluctuation decreases with time. We find that the higher coherence corresponds to the bigger absolute quantum fluctuation. For the radiation dominated universe, the higher initial temperature corresponds to more rapid variations of these quantities. We show that for a given size of the radiation dominated universe, the lower temperature corresponds to a more quantum universe. We also studied the quantum transition of a flat FRW universe. If the initial spacetime state is a Gaussian wave packet, the higher radiation temperature promotes the transition from a small quantum universe into a bigger one. We find that the minimal coupling of the free massless scalar field with the flat FRW spacetime gives rise to non-Markovian dynamics.

\section{Problem of time in quantum gravity and quantum cosmology}
\label{sec:2}
In general, Einstein's theory of gravity is defined by the action~\cite{HM,JJJ4}
\begin{equation}
\label{eq:2.1}
S=\frac{1}{16\pi}\int dx^{4}\sqrt{-g}R+S_{m}+S_{\partial \mathcal{M}}.
\end{equation}
The first term in Eq.~\eqref{eq:2.1} represents the Einstein--Hilbert action, which we denote as $S_{g}$. The second term represents the action of the matter field, and the third term is the Gibbons--Hawking--York surface contribution. Variations in this action to the metric provide the Einstein equation. The surface term does not influence the equation of motion, so neglecting this term is common practice in quantum cosmology~\cite{JF1,JF2,JF3,JF4,GN,JDJ,JJJ1,JJJ2,JJJ3,JJJ4}.

Carrying out the 3+1 decomposition and quantizing spacetime and all matter fields provides~\cite{CK}
\begin{equation}
\label{eq:2.2}
\hat{H}| \Psi \rangle=0,
\end{equation}
\begin{equation}
\label{eq:2.3}
\hat{H}_{i}| \Psi \rangle=0,
\end{equation}
where $i=1, 2, 3$. Equation~\eqref{eq:2.2} is the Wheeler--DeWitt equation, which is usually taken as the foundation of quantum gravity, and Eq.~\eqref{eq:2.3} is the diffeomorphism constraint (or momentum constraint). For the case of a homogeneous and isotropic universe, the diffeomorphism constraint is trivial~\cite{AA,DC,DCC}.

The right-hand side of the Wheeler--DeWitt equation is zero, and the eigenvalue of the Hamiltonian operator is zero. This leads to difficulty understanding the evolution of the quantum gravitational system, which is the so-called problem of time in quantum gravity~\cite{CK,CJ,KV}.
One way to resolve this problem is to divide the total Hamiltonian operator into two parts~\cite{CK,CJ,KV}:
\begin{equation}\begin{split}
\label{eq:2.4}
\hat{H}| \Psi \rangle=\hat{P}_{t}| \Psi \rangle+\hat{h}_{t}| \Psi \rangle=-i\frac{\partial | \Psi \rangle}{\partial t}+\hat{h}(t)| \Psi \rangle=0,
\end{split}
\end{equation}
where $t$ is the global time variable and $\hat{P}_{t}$ represents the conjugate momentum operator of $t$.
Then, the evolution of the quantum gravitation system is based on the Schr\"{o}dinger-like expression of Eq.~\eqref{eq:2.4}. However, there are other more complex issues related to this equation, such as the multiple choice problem, functional evolution problem, and Hilbert space problem~\cite{CJ,KV}.

There have been many attempts to resolve the problem of time~\cite{CJ,KV,VT,RG,HW}. To date, no one has been able to generally overcome these difficulties completely~\cite{KV}. Nevertheless, these difficulties can be bypassed for some simple cases. For a homogeneous and isotropic universe, the Hamiltonian operator does not depend on the spacetime coordinate. Therefore, there is no functional evolution problem. If the influence of cosmological particle production is relatively small, one can approximately define a unique vacuum, which can simplify the definition of the inner product. For the multiple-choice problem, it is unnatural to suppose that there is a privileged time coordinate. Thus, there is a possibility that several of these time coordinates are reasonable. However, strictly specifying this viewpoint is difficult and beyond the scope of this work. Usually, one chooses the time coordinate that can simplify the calculations.

Brown and Kucha$\check{\mathrm{r}}$ showed that the dust field can play the role of the global time variable $t$ in Eq.~\eqref{eq:2.4}~\cite{JD,VT,FT,HM}. Introducing a dust field gives the total action as~\cite{VT,HM}
\begin{equation}
\label{eq:2.7}
S=S_{g}+S_{m}+S_{\partial \mathcal{M}}-\frac{1}{2}\int dx^{4}\sqrt{-g}\rho (g^{\mu\nu}\partial_{\mu} \bm{T}\partial_{\nu} \bm{T}+1).
\end{equation}
The last term in Eq.~\eqref{eq:2.7} is the action of the dust field. The $\rho$ and $\bm{T}$ represent the rest-mass density and dust field, respectively~\cite{VT,HM}.
Using the canonical time gauge fixing condition $\bm{T}=t$~\cite{VT,KC} provides the Hamiltonian constraint~\cite{VT}:
\begin{equation}
\label{eq:2.10}
P_{\bm{T}}+H_{g}+H_{m}=0,
\end{equation}
where $P_{\bm{T}}$ represents both the Hamiltonian and the momentum of the dust field, $H_{g}$ represents the Hamiltonian of spacetime corresponding to the Einstein--Hilbert action, and $H_{m}$ represents the Hamiltonian of the usual matter field. The total Hamiltonian is equal to zero, which is the requirement of the generalized covariance principle~\cite{CK}.

Quantizing $P_{\bm{T}}\rightarrow \hat{P}_{\bm{T}}=-i\partial_{t}$ transforms Eq.~\eqref{eq:2.10} into the Wheeler--DeWitt equation~\cite{FT,HM}
\begin{equation}
\label{eq:2.11}
(\hat{P}_{\bm{T}}+\hat{H}_{g}+\hat{H}_{m})| \Psi \rangle=0,
\end{equation}
or the Liouville--von Neumann equation~\cite{HC,HF}
\begin{equation}
\label{eq:2.13}
\frac{d \rho_{tot}}{dt}=-i[\hat{H}_{g}+\hat{H}_{m},\ \rho_{tot}].
\end{equation}
These two equations are equivalent. If knowledge about the information of the subsystem is solely important, one can trace out the environment and obtain Eq.~\eqref{eq:1.3}.

The philosophy hidden behind Eqs.~\eqref{eq:2.11} and \eqref{eq:2.13} is that the global wave function of the universe (spacetime+matter field+clock/dust field) appears as static. However, if one only cares about the information of the subsystem, the evolution of the state related to the subsystem can be observed. This perspective is commonly adopted by the open quantum system community~\cite{HC,HF}. The total system is in the eigenstate of the total Hamiltonian operator and is static. However, as there are interactions between the subsystem and environment, energy flows in or out of the subsystem, and the subsystem state usually changes with time.

Generally, the clock in any physical theory has no influence on the system evolution. The clock is a tool to record changes in time. However, this is not true for Eqs.~\eqref{eq:2.11} and \eqref{eq:2.13}. These clocks correspond to a non-zero Hamiltonian $P_{\bm{T}}$. Thus, they influence the evolution of the universe. One can eliminate this influence in Eqs.~\eqref{eq:2.11} and \eqref{eq:2.13} by constraining that $|P_{\bm{T}}|\ll|H_{m}|$ (combining with Eq.~\eqref{eq:2.10}, this constraint is equivalent to $|P_{\bm{T}}|\ll|H_{g}|$). This constraint implies that
\begin{equation}
\label{eq:2.14}
\mathrm{Tr}(\rho\hat{H})\rightarrow0,
\end{equation}
where, $\hat{H}=\hat{H}_{g}+\hat{H}_{m}$. This ensures that the clock does not influence the evolution of the universe (spacetime+matter field). The universe can be seen as an isolated system. For any isolated system, the Hamiltonian is conserved. Therefore, if the condition of Eq.~\eqref{eq:2.14} is satisfied at the initial time it will always be satisfied at any time. Noted that when we say ``universe,'' from now on, we do not include the clock as the influence of the clock can be eliminated by the condition of Eq.~\eqref{eq:2.14}.

For the flat FRW (k=0) universe, the metric of spacetime is~\cite{SW,SD}
\begin{equation}
\label{eq:2.15}
ds^{2}=\bm{N}^{2}dt^{2}-a^{2}(t)(dx^{2}+dy^{2}+dz^{2}),
\end{equation}
where $a(t)$ is the scale factor, $\bm{N}=1$ corresponds to the proper time coordinate, and $\bm{N}=a(t)$ corresponds to the conformal time coordinate. It is easily shown that $\sqrt{-g}=\bm{N}a^{3}$ and the Ricci scalar is~\cite{GN}
\begin{equation}
\label{eq:2.16}
R=6(\bm{N}^{-2}a^{-2}\dot{a}^{2}+\bm{N}^{-2}a^{-1}\ddot{a}-\bm{N}^{-3}\dot{\bm{N}}a^{-1}\dot{a}),
\end{equation}
where $\dot{a}=da/dt$ and $\dot{\bm{N}}=d\bm{N}/dt$. Thus the Einstein-Hilbert action becomes
\begin{equation}
\label{eq:2.16a}
S_{g}=-\frac{3}{8\pi}\int dx^{4}\bm{N}^{-1}a\dot{a}^{2}+\frac{3}{8\pi}\int dx^{4}\frac{d}{dt}(\bm{N}^{-1}a^2\dot{a}).
\end{equation}
The second term on the right hand side of Eq.~\eqref{eq:2.16a} has no influence on the dynamics and can be canceled by the Gibbons--Hawking--York surface term. Then the Lagrangian of spacetime is~\cite{WW,DC}
\begin{equation}
\label{eq:2.17}
L_{g}= -\frac{3}{8\pi}V_{0}\bm{N}^{-1}a\dot{a}^{2}.
\end{equation}
Here, we introduce the definition $V_{0}\equiv\int dx^{3}$. The physical meaning of $V_{0}$ is the coordinate volume of the space slice. The $V_{0}$ is divergent for the flat FRW universe. This will initially lead to a difficulty for our model, but we can prove (details to come) that $V_{0}$ as a global factor is not important and can be seen as a global conformal factor. That is, any value of $V_{0}$ corresponds to the same results.

The conjugate momentum of the canonical variable $a$ is given as
\begin{equation}
\label{eq:2.18}
\pi_{a}=-\frac{3}{4\pi}V_{0}\bm{N}^{-1}a\dot{a}.
\end{equation}
The Poisson bracket between $\pi_{a}$ and $a$ is $\{a, \pi_{a}\}=1$. Then, the Hamiltonian of the spacetime is given as~\cite{WW}
\begin{equation}
\label{eq:2.19}
H_{g}=-\frac{2\pi}{3V_{0}}\bm{N}a^{-1}\pi_{a}^{2}.
\end{equation}
For the quantization, the conjugate momentum $\pi_{a}$ becomes $-i\partial_{a}$. In the conformal time coordinate ($\bm{N}=a$), the Hamiltonian operator of the spacetime is equivalent to a free particle with one degree of freedom.

Equation~\eqref{eq:2.11} is a first order differential equation about the time variable. Therefore the definition of the inner product related to the canonical variable $a$ can be simply taken as~\cite{JD,HM}:
\begin{equation}
\label{eq:2.20}
\langle f\mid g\rangle=\int da f^{*}(a)g(a).
\end{equation}
If there exist certain matters, the integral in Eq.~\eqref{eq:2.20} should also be taken over all the degrees of freedom related to these matters.  And if the matters are certain fields with infinite degrees of freedom, usually the integral measure related to these degrees of freedom is complicated~\cite{AJH}. In this work, the matter is a scalar field which the degrees of freedom are infinite. It is convenient to work in the second quantization representation. The inner product corresponding to the scalar field is given in Eq.~\eqref{eq:3.13}. 

It is reasonable to study the evolution of the quantum universe based on any one of Eqs.~\eqref{eq:2.11}, \eqref{eq:2.13}, and \eqref{eq:1.3}.
Husain and Pawlowski~\cite{VT} noted that the theory of quantum gravity based on Eq.~\eqref{eq:2.11} is complete. If we are not interested about the information of the environment, Eq.~\eqref{eq:2.11} reduces to Eq.~\eqref{eq:1.3}. Here, Eq.~\eqref{eq:1.3} is the fundamental equation in this work.

\section{Hamiltonian of the scalar field in flat ($k=0$) FRW spacetime}
\label{sec:3}
As a simple case, a real scalar field plays the role of the environment. The second quantization form of the Hamiltonian operator of the scalar field in Minkowski spacetime is well-known. However, in curved spacetime, it is difficult to find a general form of the Hamiltonian in the second quantization representation (there could be no general form). The main obstacle is the equation of motion of the scalar field in curved spacetime being usually very difficult to solve. For de-Sitter spacetime, the Hamiltonian operator in the particle number representation was derived in~\cite{ETA,SSF}. However, this is not sufficient to study quantum cosmology as the quantum FRW universe may not be in the de-Sitter spacetime. We use a method similar to that in~\cite{SSF} to generalize the form of the Hamiltonian operator obtained in~\cite{ETA,SSF}.

For convenience, we temporarily mark $C(t)=a^{2}(t)$ and fix $\bm{N}=a$. Then, the FRW metric of Eq.~\eqref{eq:2.15} becomes
\begin{equation}
\label{eq:3.1}
ds^{2}=C(t)(dt^{2}-dx^{2}-dy^{2}-dz^{2}).
\end{equation}
We consider a real massless scalar field in curved spacetime. The action is~\cite{LP}
\begin{equation}
\label{eq:3.2}
S_{\phi}=\frac{1}{2}\int dx^{4} \sqrt{-g}g^{\mu\nu}\phi_{,\mu}\phi_{,\nu},
\end{equation}
and the Hamiltonian density is
\begin{equation}
\label{eq:3.6}
\mathscr{H}_{\phi}=\frac{1}{2}C(t)\big(\dot{\phi}^{2}-(\partial_{x}\phi)^2-(\partial_{y}\phi)^2-(\partial_{z}\phi)^2\big).
\end{equation}

The equation of motion of the scalar field is~\cite{LP,SSF}
\begin{equation}
\label{eq:3.7}
\Box\phi=C^{-2}(t)\partial_{t}\big(C(t)\partial_{t}\phi\big)-C^{-1}(t)\nabla^2\phi=0.
\end{equation}
We formally denote the solution of the equation of motion as~\cite{SSF}
\begin{equation}
\label{eq:3.8}
\phi(x,t)=\sum_{\vec{k}}\big(a_{\vec{k}}\Phi_{\vec{k}}(x,t)+a^{\dag}_{\vec{k}}\Phi^{*}_{\vec{k}}(x,t)\big),
\end{equation}
where $a_{\vec{k}}$ is a complex parameter that does not depend on the spacetime coordinate. As the space slice for FRW ($k=0$) spacetime is flat, $\Phi_{\vec{k}}(x,t)$ can be written as~\cite{SSF}
\begin{equation}
\label{eq:3.9}
\Phi_{\vec{k}}(x,t)=(2\pi)^{-\frac{3}{2}}C^{-\frac{1}{2}}(t)f_{\vec{k}}(t)e^{i\vec{k}\cdot\vec{x}}.
\end{equation}
Bringing Eqs.~\eqref{eq:3.8} and \eqref{eq:3.9} into Eq.~\eqref{eq:3.7} gives an expression in which $f_{\vec{k}}(t)$ must be satisfied as
\begin{equation}
\label{eq:3.10}
\ddot{f}_{\vec{k}}(t)+\big[\vec{k}^{2}+\frac{1}{4}(\frac{\dot{C}}{C})^{2}-\frac{1}{2}\frac{\ddot{C}}{C}\big]f_{\vec{k}}(t)=0.
\end{equation}

We first consider this case $C(t)\propto t^{\omega}$, where $\omega$ is a real number. Different values of $\omega$ represent various spacetimes. The $\omega=-2$ represents the de-Sitter spacetime, while $\omega=2$ and $\omega=4$ represent the radiation and non-relativistic matter dominated FRW spacetimes, respectively. For these special cases, $\omega$ is an integer. In most cases, $\omega$ may be a non-integer. Bringing $C(t)\propto t^{\omega}$ into Eq.~\eqref{eq:3.10} and defining $\mathrm{J}_{k}(t)\equiv t^{-1/2}f_{\vec{k}}(t)$ and $z\equiv|\vec{k}|t$ provides the Bessel equation~\cite{SSF}
\begin{equation}
\label{eq:3.11}
\mathrm{J}^{''}_{k}(z)+\frac{1}{z}\mathrm{J}^{'}_{k}(z)+(1-\frac{\nu^{2}}{z^{2}})\mathrm{J}_{k}(z)=0,
\end{equation}
where $\nu^{2}=\frac{1}{4}(\omega-1)^{2}$ and $\mathrm{J}^{'}_{k}(z)=d\mathrm{J}_{k}(z)/d z$, with $\nu$ not being an integer. The two kinds of Hankel functions ($\mathrm{H}^{(1)}_{\nu}(z)$ and $\mathrm{H}^{(2)}_{\nu}(z)$) are two linearly independent solutions of the Bessel expression in Eq.~\eqref{eq:3.11}. Thus, we have
\begin{eqnarray}\begin{split}
\label{eq:3.12}
\Phi_{\vec{k}}(x,t)&=\frac{\sqrt{\pi}}{2}(2\pi)^{-\frac{3}{2}}C^{-\frac{1}{2}}(t)t^{\frac{1}{2}}\mathrm{H}^{(2)}_{\nu}(kt)e^{i\vec{k}\cdot\vec{x}}\\
&\equiv g_{k}(t)e^{i\vec{k}\cdot\vec{x}}.
\end{split}
\end{eqnarray}
Here, we define the function $g_{k}(t)$.

In Eq.~\eqref{eq:3.12}, we can replace $\mathrm{H}^{(2)}_{\nu}(z)$ by $\mathrm{H}^{(1)}_{\nu}(z)$, while $\mathrm{H}^{(1)}_{\nu}(z)$ is related to the Bunch--Davies vacuum~\cite{LP}. Although different kinds of Hankel functions correspond to various vacuums, choosing any one of them can lead to similar forms of the Hamiltonian operator (second quantization representation) for the scalar field. Thus, we can arbitrarily choose $\mathrm{H}^{(2)}_{\nu}(z)$, as was done in~\cite{SSF}. In fact, in the next section, we only consider the simple case where the cosmological particle production is very small and can be neglected. Thus, differences in the vacuum are not important.

We have not fixed the value of $\omega$. In~\cite{SSF}, Feng studied the special case of $\omega=-2$ for a massive scalar field. In this case, Eq.~\eqref{eq:3.10} has an additional mass term, which increases the difficulty of solving the equation for $\omega\neq-2$. This is why we consider only the massless scalar field here.
Carrying out the quantization for the scalar field and introducing the inner product gives~\cite{SSF}
\begin{equation}
\label{eq:3.13}
\langle f_{1}|f_{2}\rangle\equiv i \int dx^{4} C(t) \delta(t-t_{0}) \big[f_{1}^{*}\nabla_{t}f_{2}-(\nabla_{t}f_{1}^{*})f_{2}\big].
\end{equation}
This is an integral over the Klein--Gordon current on the Cauchy surface~\cite{SSF}. One can prove the following property~\cite{SSF}:
\begin{equation}
\label{eq:3.15}
[a_{\vec{k}},a_{\vec{k}^{'}}^{\dag}]=\delta^{3}(\vec{k}-\vec{k}^{'}).
\end{equation}

We introduce the following definitions to represent the Hamiltonian operator in the particle number representation~\cite{SSF}:
\begin{equation}
\label{eq:3.16}
\varepsilon_{k}(t)\equiv(2\pi)^{3}C(t)\big( \dot{g}_{k}(t)\dot{g}_{k}^{*}(t)+k^{2}g_{k}(t)g_{k}^{*}(t) \big),
\end{equation}
\begin{equation}
\label{eq:3.17}
\triangle_{k}(t)\equiv(2\pi)^{3}C(t)\big( \dot{g}_{k}^{2}(t)+k^{2}g_{k}^{2}(t) \big),
\end{equation}
\begin{equation}
\label{eq:3.18}
\omega_{k}(t)\equiv \big(\varepsilon_{k}^{2}(t)-\triangle_{k}^{*}(t)\triangle_{k}(t)\big)^{\frac{1}{2}},
\end{equation}
\begin{equation}
\label{eq:3.19}
u_{k}(t)\equiv \Big(\frac{\varepsilon_{k}(t)+\omega_{k}(t)}{2\omega_{k}(t)}\Big)^{\frac{1}{2}},
\end{equation}
\begin{equation}
\label{eq:3.20}
v_{k}(t)\equiv \frac{\triangle_{k}^{*}(t)}{\varepsilon_{k}(t)+\omega_{k}(t)}u_{k}(t) ,
\end{equation}
\begin{equation}
\label{eq:3.21}
A_{\vec{k}}(t)\equiv u_{k}(t)a_{\vec{k}}+v_{k}(t)a^{\dag}_{-\vec{k}},
\end{equation}
\begin{equation}
\label{eq:3.22}
A_{\vec{k}}^{\dag}(t)\equiv u_{k}^{*}(t)a_{\vec{k}}^{\dag}+v_{k}^{*}(t)a_{-\vec{k}}.
\end{equation}
The physical meaning of $\omega_{k}(t)$ is the frequency (energy) of the scalar particle in the conformal time coordinate. In general, the frequency of the particle on curved spacetime is time-dependent. The $A_{\vec{k}}^{\dag}(t)$ and $A_{\vec{k}}(t)$ represent the creation and annihilation operators of the scalar particles, respectively. Equations \eqref{eq:3.21} and \eqref{eq:3.22} are the Bogolyubov transformation where the Bogolyubov coefficients are $u_{k}(t)$ and $v_{k}(t)$. For some special cases where $\triangle_{k}(t)\rightarrow 0$, the $a^{\dag}_{\vec{k}}$  and $a_{\vec{k}}$ also represent the creation and annihilation operators of the scalar particles, respectively~\cite{ETA}. However for the general cases, $a^{\dag}_{\vec{k}}$  ( $a_{\vec{k}}$) can not be interpreted as the creation (annihilation) operator of the scalar particles~\cite{ETA}. For convenience, we often neglect the operator hat. The reader can easily distinguish what the c-number and q-number are based on the context.

Using Eqs.~\eqref{eq:3.6}, \eqref{eq:3.8}, \eqref{eq:3.9}, and \eqref{eq:3.12}-\eqref{eq:3.22} can transform the Hamiltonian of the scalar field to
\begin{equation}
\label{eq:3.23}
H_{\phi}(t)\equiv \int dx^{3}\mathscr{H}_{\phi}=\sum_{\vec{k}}|\vec{k}|\big(A^{\dag}_{\vec{k}}(t)A_{\vec{k}}(t)+\mathscr{V}_{k}(t)\big),
\end{equation}
where
\begin{equation}
\label{eq:3.24}
\mathscr{V}_{k}(t)\equiv\frac{1}{2}\big(u_{k}(t)u_{k}^{*}(t)-v_{k}(t)v_{k}^{*}(t)\big) .
\end{equation}
The $A^{\dag}_{\vec{k}}(t)A_{\vec{k}}(t)$ is the particle number operator and $\mathscr{V}_{k}(t)$ represents the vacuum energy. Equation\eqref{eq:3.23} shows that the particle number and vacuum energy can change over time, which may lead to cosmological particle production. The calculation process for this Hamiltonian operator is complex, and the main steps are described in Appendix~\ref{sec:B}. For the special case of $\omega=-2$, a similar derivation can be found in~\cite{SSF}.

We note out that the Hamiltonian in Eq.~\eqref{eq:3.23} corresponds to the metric in Eq.~\eqref{eq:3.1}. For a more general form of the FRW metric in Eq.~\eqref{eq:2.15}, the form of the Hamiltonian in Eq.~\eqref{eq:3.23} is incorrect. If we rescale the coordinate time variable, $dt\rightarrow\frac{a}{\bm{N}}dt$, the metric in Eq.~\eqref{eq:3.1} changes into Eq.~\eqref{eq:2.15}. In addition, the eigenvalue of the Hamiltonian operator represents the energy of the matter. In quantum mechanics, the particle energy is proportional to the frequency, and $k^{\mu}x_{\mu}$ is an invariant scalar under reparametrization. Thus, rescaling the coordinate time also requires rescaling the frequency $\omega_{\vec{k}}\rightarrow\frac{\bm{N}}{a}\omega_{\vec{k}}$. Therefore, we must also rescale the Hamiltonian $H\rightarrow\frac{\bm{N}}{a}H$. This gives
\begin{equation}
\label{eq:3.25}
H_{\phi}(t)=\frac{\bm{N}}{a}\sum_{\vec{k}}|\vec{k}|\big(A^{\dag}_{\vec{k}}(t)A_{\vec{k}}(t)+\mathscr{V}_{k}(t)\big).
\end{equation}
This Hamiltonian operator corresponds to the metric in Eq.~\eqref{eq:2.15}. The lapse function $\bm{N}(t)$ can be an arbitrary function of the time variable $t$.

When the quantum evolution of the universe is dominated by heat radiation, the vacuum energy of the scalar field can be neglected. We only consider the simple case where the cosmological particle production is small and can be neglected. For a radiation-dominated universe, the zero point of the effective frequency of the scalar particle is at the singular point of spacetime. The scalar particle cannot go across the Stokes line while the universe is expanding. Thus, even though the spacetime background is not static, scalar particles are not produced from the vacuum state (see~\cite{SHY} for more details). Under these approximations, we can define a unique vacuum, and the particle number operator does not change over time. Then, the Hamiltonian operator in Eq.~\eqref{eq:3.25} becomes
\begin{equation}
\label{eq:3.26}
H_{\phi}=\frac{\bm{N}}{a}\sum_{\vec{k}}|\vec{k}|A^{\dag}_{\vec{k}}A_{\vec{k}}.
\end{equation}

The physical meaning of each part of the Hamiltonian operator of Eq.~\eqref{eq:3.26} is clear. The $A^{\dag}_{\vec{k}}A_{\vec{k}}$ is the particle number operator related to the momentum $\vec{k}$. The $\frac{\bm{N}}{a}$ is the red shift factor, which may sometimes be a blue shift. However, we use the red shift as the unified term for simplicity. Equation \eqref{eq:3.26} shows that the only influence of the FRW ($k=0$) spacetime for the scalar particle is a red shift. In particular, $\bm{N}=a$ can represent both the Minkowski spacetime and FRW spacetime in conformal time coordinates. When $\bm{N}=a$, Eq.~\eqref{eq:3.26} reduces to the Hamiltonian of the scalar field in the flat spacetime when neglecting the vacuum energy.

When deducing the Hamiltonian operator of Eq.~\eqref{eq:3.25}, we assume $C(t)\propto t^{\omega}$. This includes infinitely homogeneous and isotropic spacetimes, but not all. In fact, this form of $C(t)\propto t^{\omega}$ does not include all FRW (k=0) spacetimes. Then, one may ask if the Hamiltonian of Eq.~\eqref{eq:3.26} is correct for any kinds of FRW (k=0) spacetimes (The cosmological particle production can be neglected). Our view is that Eq.~\eqref{eq:3.26} should be reasonable for any kind of flat FRW spacetime (The cosmological particle production can be neglected). Equation~\eqref{eq:2.15} shows that the only possible variation for the FRW spacetime is expansion or contraction, where the scale factor $a(t)$ is a unique variable. Neglecting the cosmological particle production indicates that the only impact for this kind of variation on the scalar particle is from red shifting. This can be quantified by the factor $\frac{\bm{N}}{a}$, which is the only difference between the flat FRW spacetime and Minkowski spacetime. Thus, the Hamiltonian of Eq.~\eqref{eq:3.26} should be reasonable in any flat FRW spacetime. Therefore, the form of Eq.~\eqref{eq:3.26} should be suitable to study quantum cosmology. All our specific examples in this paper support this point.

Taking the continuous limit and replacing the summation in Eq.~\eqref{eq:3.26} with an integral gives
\begin{equation}
\label{eq:3.27}
H_{\phi}=\frac{\bm{N}}{a}\frac{V_{0}}{(2\pi)^{3}}\int d\vec{k}^{3}|\vec{k}|A^{\dag}_{\vec{k}}A_{\vec{k}}.
\end{equation}
If the scalar field plays the role of the environment, the total Hamiltonian of the universe is
\begin{equation}
\label{eq:3.28}
H_{tot}=-\frac{3}{8\pi}V_{0}\bm{N}^{-1}a\dot{a}^{2}+\frac{\bm{N}}{a}\frac{V_{0}}{(2\pi)^{3}}\int d\vec{k}^{3}|\vec{k}|A^{\dag}_{\vec{k}}A_{\vec{k}}.
\end{equation}
Before quantization, $H_{tot}=0$ is equivalent to the Friedmann equation. The $V_{0}$ is a trivial global factor that can be eliminated and does not influence the results. If we rescale the spacetime coordinate by a global parameter, $x_{\mu}\rightarrow x^{'}_{\mu}=\lambda x_{\mu}$, then $V_{0}\rightarrow V_{0}^{'}=\lambda^{3} V_{0}$. Here, $V_{0}^{'}$ represents the coordinate volume related to the coordinate $x^{'}_{\mu}$. The relationship between $\lambda$ and $V_{0}^{'}$ is monotonic with a one-to-one correspondence. Thus, $V_{0}$ can also be seen as a global conformal factor. Although $V_{0}$ does not impact any results, we still retain $V_{0}$ in our formulas and take it as a finite value to show the global scaling symmetry (this treatment is often used by different researchers in quantum cosmology~\cite{FT,HM,DC}).
It is shown in Eq.~\eqref{eq:3.28} that the interaction Hamiltonian is not small, and no term can be viewed as small. Therefore, although gravity is the weakest force, for certain cases, we cannot use the Born-Markov quantum master equation to describe its dynamics.

\section{Heat radiation dominated evolution of the quantum universe}
\label{sec:4}
In the proper time coordinate ($\bm{N}=1$), the metric of the flat FRW spacetime is
\begin{equation}
\label{eq:4.1}
ds^{2}=dt^{2}-a^{2}(t)(dx^{2}+dy^{2}+dz^{2}).
\end{equation}
From Eqs.~\eqref{eq:2.19} and \eqref{eq:3.26}, the Hamiltonian operator of the spacetime and the real scalar field are
\begin{equation}
\label{eq:4.2}
\hat{H}_{g}=\frac{-\pi}{3V_{0}}(\frac{1}{a}\hat{\pi}_{a}^{2}+\hat{\pi}_{a}^{2}\frac{1}{a}),
\end{equation}
\begin{equation}
\label{eq:4.3}
\hat{H}_{\phi}=\frac{1}{a}\sum_{\vec{k}}|\vec{k}|A^{\dag}_{\vec{k}}A_{\vec{k}}.
\end{equation}
In Eq.~\eqref{eq:4.2}, we use symmetrized factor ordering to keep the Hamiltonian operator as Hermitian.

When the scalar field plays the role of the environment, Eq.~\eqref{eq:1.3} can be written in another equivalent form as~\cite{HF}
\begin{equation}
\label{eq:4.4}
\rho(t)=\mathrm{Tr}_{\phi}\big(U(t,0)\rho_{tot}(0)U^{\dag}(t,0)\big).
\end{equation}
Here, we trace out all the degrees of freedom corresponding to the scalar field. The $\rho(t)$ represents the reduced density matrix related to spacetime, $\rho_{tot}(0)$ represents the initial density matrix of the total system (spacetime+scalar field), and $U(t,0)$ is the time evolution operator~\cite{HF},
\begin{equation}
\label{eq:4.5}
U(t,0)=\mathrm{T} \mathrm{exp}\big\{-i\int_{0}^{t}ds\big(\hat{H}_{g}(s)+\hat{H}_{\phi}(s)\big)\big\}.
\end{equation}
The $\mathrm{T}$ denotes the chronological time ordering operator. As $U^{\dag}(t,0)=U(0,t)$, the path integral representation of Eq.~\eqref{eq:4.4} is usually called the closed real-time path integral. In the Schr\"{o}dinger picture, $\hat{H}_{g}$ and $\hat{H}_{\phi}$ are time independent. The infinitesimal time evolution operator $U(\delta t,0)$ can be written as
\begin{eqnarray}\begin{split}
\label{eq:4.6}
U( \delta t,0)&= \mathrm{exp}\big\{-i(\hat{H}_{g}+\hat{H}_{\phi})\delta t\big\}\\
&=\mathrm{exp}(-i\hat{H}_{g}\delta t)\mathrm{exp}(-i\hat{H}_{\phi}\delta t)+o(\delta t^{2}).
\end{split}
\end{eqnarray}
Here, we use the first-order Suzuki--Trotter decomposition~\cite{SJR}.

Bringing Eq.~\eqref{eq:4.6} into Eq.~\eqref{eq:4.4}, discretizing the time variable ($t\rightarrow t_{1}, t_{2},...,t_{N}$) and inserting $\int da_{n}|a_{n}\rangle\langle a_{n}|=\mathbf{1}$ at each time slice gives
\begin{eqnarray}\begin{split}
\label{eq:4.7}
\rho(a^{+}_{N}, a^{-}_{N}) =&\int da_{0}^{\pm}\int da_{1}^{\pm}\cdot\cdot\cdot\int da_{N-1}^{\pm}\mathrm{Tr}_{\phi}\big\{\langle a^{+}_{N}| e^{-i\hat{H}_{g}\delta t}e^{-i\hat{H}_{\phi}\delta t}|a^{+}_{N-1}\rangle\langle a^{+}_{N-1}|e^{-i\hat{H}_{g}\delta t} e^{-i\hat{H}_{\phi}\delta t}|a^{+}_{N-2}\rangle\cdot\cdot\cdot\\
&\times\langle a^{+}_{0}|\rho_{tot}(0)|a^{-}_{0}\rangle\langle a^{-}_{0}|e^{i\hat{H}_{g}\delta t}e^{i\hat{H}_{\phi}\delta t}|a^{-}_{1}\rangle\cdot\cdot\cdot\langle a^{-}_{N-1}|e^{i\hat{H}_{g}\delta t} e^{i\hat{H}_{\phi}\delta t}|a^{-}_{N}\rangle\big\},
\end{split}
\end{eqnarray}
where $\rho(a^{+}_{N}, a^{-}_{N})\equiv\langle a^{+}_{N}|\rho(t_{N})|a^{-}_{N}\rangle$. Note that we use $\bm{N}$ to represent the lapse function and $N$ to represent the number of the discretized time points. In Eq.~\eqref{eq:4.7}, $\int da_{i}^{\pm}$ is the abbreviation of $\int da_{i}^{+} da_{i}^{-}$, and we use $t_{N}$ to represent the final time point. For simplicity, we consider the case where there is no entanglement between spacetime and the scalar field at the initial time. Then, the initial state of the total system can be written as~\cite{DS1,DS2,DS3,WW}
\begin{equation}
\label{eq:4.8}
\rho_{tot}(0)=\rho(0)\otimes\rho_{\phi}(0),
\end{equation}
where $\rho(0)$ and $\rho_{\phi}(0)$ represent the density matrix of spacetime and the scalar field at the initial time, respectively. This condition is used extensively for open quantum systems~\cite{HC,HF}. If the initial state cannot be written in the form of Eq.~\eqref{eq:4.8}, Eq.~\eqref{eq:4.4} is difficult to solve for the usual case.

Using the condition of Eq.~\eqref{eq:4.8}, Eq.~\eqref{eq:4.7} can be written in the more compact form of
\begin{eqnarray}\begin{split}
\label{eq:4.9}
\rho(a^{+}_{N}, a^{-}_{N}) =&\int da_{0}^{\pm}\int da_{1}^{\pm}\cdot\cdot\cdot\int da_{N-1}^{\pm}\langle a^{+}_{N}| e^{-i\hat{H}_{g}\delta t}|a^{+}_{N-1}\rangle\\&\times\langle a^{+}_{N-1}|e^{-i\hat{H}_{g}\delta t} |a^{+}_{N-2}\rangle\cdot\cdot\cdot\langle a^{+}_{0}|\rho(0)|a^{-}_{0}\rangle\\
&\times\langle a^{-}_{0}|e^{i\hat{H}_{g}\delta t} |a^{-}_{1}\rangle\cdot\cdot\cdot\langle a^{-}_{N-1}|e^{i\hat{H}_{g}\delta t}|a^{-}_{N}\rangle\mathbf{I_{rN}},
\end{split}
\end{eqnarray}
where
\begin{eqnarray}\begin{split}
\label{eq:4.10}
\mathbf{I_{rN}}\equiv&\mathrm{ Tr}_{\phi}\big\{e^{-i\hat{H}_{\phi}(a_{N}^{+})\delta t}e^{-i\hat{H}_{\phi}(a_{N-1}^{+})\delta t}\cdot\cdot\cdot e^{-i\hat{H}_{\phi}(a_{1}^{+})\delta t}\\&\times\rho_{\phi}(0)e^{i\hat{H}_{\phi}(a_{1}^{-})\delta t} e^{i\hat{H}_{\phi}(a_{2}^{-})\delta t}\cdot\cdot\cdot e^{i\hat{H}_{\phi}(a_{N}^{-})\delta t}\big\}.
\end{split}
\end{eqnarray}
This is the so-called influence functional~\cite{NM1,NM2,NM3,NM4,NM5,NM6,DS1,DS2,DS3,RP,BL1,BL2,JRA}, which includes all the influences of the environment on the system. If there are no interactions between the system and the environment, the influence functional is trivially equal to one.

Recalling that for a particle (one degree of freedom), $\langle x|p\rangle=\frac{1}{\sqrt{2\pi}}e^{ipx}$. This indicates that the coordinate and the momentum cannot be determined simultaneously. Similarly for spacetime, $\langle a|\pi_{a}\rangle=\frac{1}{\sqrt{2\pi}}e^{i\pi_{a}a}$. Thus, one can easily calculate the following matrix elements:
\begin{eqnarray}\begin{split}
\label{eq:4.11}
\langle a^{+}_{n+1}|e^{-i\hat{H}_{g}\delta t}|a^{+}_{n}\rangle
= &\mathrm{exp}\Big\{\frac{-3iV_{0}a_{n+1}^{+}a_{n}^{+}(a_{n+1}^{+}-a_{n}^{+})^{2}}{4\pi\delta t(a_{n+1}^{+}+a_{n}^{+})}\Big\}\\&\times\frac{1}{2\pi}\big(\frac{3iV_{0}a_{n+1}^{+}a_{n}^{+}}{\delta t (a_{n+1}^{+}+a_{n}^{+})}\big)^{\frac{1}{2}},
\end{split}
\end{eqnarray}
\begin{eqnarray}\begin{split}
\label{eq:4.12}
\langle a^{-}_{n}|e^{i\hat{H}_{g}\delta t}|a^{-}_{n+1}\rangle=&\mathrm{exp}\Big\{\frac{3iV_{0}a_{n+1}^{-}a_{n}^{-}(a_{n+1}^{-}-a_{n}^{-})^{2}}{4\pi\delta t(a_{n+1}^{-}+a_{n}^{-})}\Big\}\\&\times\frac{1}{2\pi}\big(\frac{-3iV_{0}a_{n+1}^{-}a_{n}^{-}}{\delta t (a_{n+1}^{-}+a_{n}^{-})}\big)^{\frac{1}{2}}.
\end{split}
\end{eqnarray}

Noting that for the classical evolution of the FRW universe dominated by heat radiation, $a T$ does not change with time: $a T=a_{0}T_{0}$~\cite{SW,VAR}. The $a_{0}$ and $T_{0}$ represent the scale factor and temperature of the radiation at the initial time, respectively. Assuming at the initial time, the density matrix of the scalar field is given as~\cite{HC}
\begin{equation}
\label{eq:4.13}
\rho_{\phi}(0)=\prod_{j}\mathrm{exp}(\frac{-k_{j}}{\alpha_{0}}A^{\dag}_{k_{j}}A_{k_{j}})\big(1-\mathrm{exp}(\frac{-k_{j}}{\alpha_{0}})\big),
\end{equation}
where $\alpha_{0}$ is an arbitrary parameter. If spacetime is classical, $\alpha_{0}=a_{0}T_{0}$ (in the proper time coordinate) and the density matrix of Eq.~\eqref{eq:4.13} represents an equilibrium state with temperature $T_{0}$. For quantum spacetime, the value of the scale factor can be indefinite. Thus, the temperature of the heat bath may also be indefinite. The relationship between $\alpha_{0}$ and the Hamiltonian of the scalar field is
\begin{equation}
\label{eq:4.14}
\mathrm{Tr}_{\phi}\big(a\hat{H}_{\phi}\rho_{\phi}(0)\big)=\frac{\pi^{2}}{30}V_{0}\alpha_{0}^{4}\equiv\chi_{0}.
\end{equation}
We introduce a parameter $\chi_{0}$ to simplify our formulas later. When $\alpha_{0}$ is equal to zero, the average value of the Hamiltonian operator of the scalar field is zero.

Combining Eqs.~\eqref{eq:4.3}, \eqref{eq:4.10}, and \eqref{eq:4.13} gives an influence functional of
\begin{eqnarray}\begin{split}
\label{eq:4.15}
\mathbf{I_{rN}}=\mathrm{exp}\Big\{\frac{V_{0}\pi^{2}\alpha_{0}^{3}}{90\cdot\big[1+i\alpha_{0}\delta t \sum_{n=1}^{N}(\frac{1}{a_{n}^{+}}-\frac{1}{a_{n}^{-}})\big]^{3}}\Big\}.
\end{split}
\end{eqnarray}
The main steps in the derivations of the influence functional are summarized in Appendix~\ref{sec:C}. This influence functional contains all the influences of the heat scalar field on quantum spacetime. When $\alpha_{0}\rightarrow 0$, the influence functional is $\mathbf{I_{rN}}\rightarrow 1$. This is reasonable as $\alpha_{0}\rightarrow0$ means that there are no radiation particles.

If we introduce the universal dynamical map $\mathscr{J}$ to the reduced density matrix such that $\rho(t_{2})=\mathscr{J}(t_{2},t_{1})\rho(t_{1})$~\cite{BL1,IDV}, then Eqs.~\eqref{eq:4.11}, \eqref{eq:4.12}, and \eqref{eq:4.15} readily show that $\mathscr{J}(t_{3},t_{1})\neq\mathscr{J}(t_{3},t_{2})\mathscr{J}(t_{2},t_{1})$ ($t_{1}\leq t_{2}\leq t_{3}$). That is, the map $\mathscr{J}$ does not satisfy the semigroup property. This indicates that the memory time of the environment is long and non-Markovian dynamics are presented in the evolution of the quantum universe.

It is difficult to handle the integral in Eq.~\eqref{eq:4.9} based on the influence functional of Eq.~\eqref{eq:4.15}. We consider the following simple situation:
\begin{equation}
\label{eq:4.16}
|i\alpha_{0}\delta t \sum_{n=1}^{N}(\frac{1}{a_{n}^{+}}-\frac{1}{a_{n}^{-}})|\ll 1.
\end{equation}
Semiclassically, this condition indicates that the temperature of the radiation is far less than the Planck temperature. In this case, using $(1+x)^{3}=1+3x+o(x^{2})$ simplifies the influence functional of Eq.~\eqref{eq:4.15} as
\begin{eqnarray}\begin{split}
\label{eq:4.17}
\mathbf{I_{rN}}=\mathrm{exp}\Big\{\frac{V_{0}}{90}\pi^{2}\alpha_{0}^{3}\big[1-3i\alpha_{0}\delta t\sum_{n=1}^{N}(\frac{1}{a_{n}^{+}}-\frac{1}{a_{n}^{-}})\big]\Big\}.
\end{split}
\end{eqnarray}

Bringing Eqs.~\eqref{eq:4.11}, \eqref{eq:4.12}, and \eqref{eq:4.17} into Eq.~\eqref{eq:4.9} allows writting the elements of the reduced density matrix as
\begin{eqnarray}\begin{split}
\label{eq:4.18}
\rho(a^{+}_{N}, a^{-}_{N}) =&\mathscr{N}\int da_{0}^{\pm}\int da_{1}^{\pm}\cdot\cdot\cdot\int da_{N-1}^{\pm}\langle a^{+}_{0}|\rho(0)|a^{-}_{0}\rangle\prod_{n=1}^{N}\big(\frac{3iV_{0}a_{n-1}^{+}a_{n}^{+}}{\delta t (a_{n-1}^{+}+a_{n}^{+})}\big)^{\frac{1}{2}}\\
&\times\mathrm{exp}\Big\{\sum_{n=1}^{N}\big[\frac{-3iV_{0}a_{n-1}^{+}a_{n}^{+}(a_{n-1}^{+}-a_{n}^{+})^{2}}{4\pi\delta t(a_{n-1}^{+}+a_{n}^{+})}-i\chi_{0}\delta t \frac{1}{a_{n}^{+}}\big]\Big\}\prod_{n=1}^{N}\big(\frac{-3iV_{0}a_{n-1}^{-}a_{n}^{-}}{\delta t (a_{n-1}^{-}+a_{n}^{-})}\big)^{\frac{1}{2}}\\
&\times\mathrm{exp}\Big\{\sum_{n=1}^{N}\big[\frac{3iV_{0}a_{n-1}^{-}a_{n}^{-}(a_{n-1}^{-}-a_{n}^{-})^{2}}{4\pi\delta t(a_{n-1}^{-}+a_{n}^{-})}+i\chi_{0}\delta t \frac{1}{a_{n}^{-}}\big]\Big\},
\end{split}
\end{eqnarray}
where, $\mathscr{N}$ represents the normalized constant. This may differ for various equations, but we use $\mathscr{N}$ as the unified symbol. In addition, when $\delta t\rightarrow0$,
\begin{equation}
\label{eq:4.19}
\frac{a_{n}^{\pm}a_{n-1}^{\pm}}{a_{n}^{\pm}+a_{n-1}^{\pm}}\rightarrow\frac{a_{n}^{\pm}}{2}.
\end{equation}
Using this equation, Eq.~\eqref{eq:4.18} can be simplified as
\begin{eqnarray}\begin{split}
\label{eq:4.21}
\rho(a^{+}_{N}, a^{-}_{N}) =&\mathscr{N}\int da_{0}^{\pm}\int da_{1}^{\pm}\cdot\cdot\cdot\int da_{N-1}^{\pm}\\&\times\langle a^{+}_{0}|\rho(0)|a^{-}_{0}\rangle\prod_{n=1}^{N}\big(\frac{3iV_{0}a_{n}^{+}}{2\delta t }\big)^{\frac{1}{2}}\\
&\times\prod_{n=1}^{N}\big(\frac{-3iV_{0}a_{n}^{-}}{2\delta t }\big)^{\frac{1}{2}}\cdot \mathrm{exp}\{\mathcal{I}_{r+}+\mathcal{I}_{r-}\},
\end{split}
\end{eqnarray}
where,
\begin{equation}
\label{eq:4.22}
\mathcal{I}_{r\pm}\equiv\sum_{n=1}^{N}\big[\frac{\mp3iV_{0}a_{n}^{\pm}(a_{n-1}^{\pm}-a_{n}^{\pm})^{2}}{8\pi\delta t}\mp i\chi_{0}\delta t \frac{1}{a_{n}^{\pm}}\big].
\end{equation}
The $\mathcal{I}_{r+}$ and $\mathcal{I}_{r-}$ are functions with $N$-1 variables as $\mathcal{I}_{r\pm}=\mathcal{I}_{r\pm}(a_{1}^{\pm},a_{2}^{\pm},...,a_{N-1}^{\pm} )$. For convenience, we introduced the abbreviation $\mathcal{I}_{r\pm}$ to represent $\mathcal{I}_{r+}$ or $\mathcal{I}_{r-}$ (similarly for $a_{n}^{\pm}$).

In conventional path integral quantum mechanics (for isolated systems), any path in the phase space has the same amplitude. Equations \eqref{eq:4.18} and \eqref{eq:4.21} show that in quantum gravity, the amplitudes of different paths in phase space may differ. This is distinct from the usual unitary quantum mechanics. The form of the Hamiltonian operator of Eq.~\eqref{eq:4.2} gives rise to this difference. There is a coupling term between the scale factor and momentum, which leads to the different amplitudes for various paths. Usually (for isolated systems), if there is no coupling term between the canonical variable and the conjugate momentum, all paths in the phase space have the same amplitude. In addition, the forms of the Hamiltonian operators in Eqs.~\eqref{eq:4.2} and \eqref{eq:4.3} give rise to the energy flux between the scalar field and spacetime. An interesting question that goes beyond the scope of this work is whether there is an explicit relationship between the energy flux and the different amplitudes for different paths.

Taking the continuous limit ($N\rightarrow\infty$ or $\delta t=t_{N}/N\rightarrow0 $), the dimension of the integral in Eq. \eqref{eq:4.21} becomes infinity. And $i\mathcal{I}_{r+}$ (or $-i\mathcal{I}_{r-}$) becomes the action of the total system. That is
\begin{equation}
\label{eq:4.m1}
\lim_{N\rightarrow\infty}\mathcal{I}_{r\pm}=\int_{0}^{t_{N}}dt\big[\frac{\mp3iV_{0}a^{\pm}(t)\big(\dot{a}^{\pm}(t)\big)^{2}}{8\pi}\mp \frac{i\chi_{0}}{a^{\pm}(t)}\big].
\end{equation}
Thus, $i\mathcal{I}_{r+}$ or $-i\mathcal{I}_{r-}$ is the discretized version of the action $S_{g}+S_{\phi}$. The classical trajectory corresponding to $\displaystyle\lim_{N\rightarrow\infty}\mathcal{I}_{r\pm}$ is
\begin{equation}
\label{eq:4.m2}
a_{cl}^{\pm}(t)=\Big((a_{0}^{\pm})^{2}+\big((a_{N}^{\pm})^{2}-(a_{0}^{\pm})^{2}\big)\frac{t}{t_{N}}\Big)^{\frac{1}{2}}.
\end{equation}
The steepest descent contour approximation provides~\cite{WX}
\begin{equation}
\label{eq:4.m3}
\int_{a}^{b}g(s)e^{zh(s)}ds\approx i \sqrt{\frac{2\pi}{\sigma z}}g(s_{0})\mathrm{exp}(zh(s_{0})-i\frac{\theta_{0}}{2}),
\end{equation}
where, $s_{0}$ is the saddle point defined by $\partial h(s_{0})/\partial s=0$. The $\sigma$ and $\theta_{0}$ are defined by $\partial^2 h(s_{0})/\partial s^{2}=\sigma e^{i\theta_{0}}$.
Bringing Eq. \eqref{eq:4.m2} into Eq. \eqref{eq:4.m1}, and using the steepest descent contour approximation of Eq. \eqref{eq:4.m3}, then the continuous limit of Eq. \eqref{eq:4.21} ($\delta t\rightarrow0$) becomes
\begin{eqnarray}\begin{split}
\label{eq:4.m4}
\rho(a^{+}_{N}, a^{-}_{N}) \approx&\mathscr{N}(a_{N}^{+}a_{N}^{-})^{\frac{1}{2}}\int da_{0}^{\pm}\langle a^{+}_{0}|\rho(0)|a^{-}_{0}\rangle\mathrm{exp}\Big\{\frac{-2i\chi_{0}t_{N}}{a_{N}^{+}+a_{0}^{+}}\\&-\frac{3iV_{0}}{16\pi t_{N}}(a_{N}^{+}+a_{0}^{+})(a_{N}^{+}-a_{0}^{+})^{2}+\frac{2i\chi_{0}t_{N}}{a_{N}^{-}+a_{0}^{-}}\\&-\frac{3iV_{0}}{16\pi t_{N}}(a_{N}^{-}+a_{0}^{-})(a_{N}^{-}-a_{0}^{-})^{2}\Big\}.
\end{split}
\end{eqnarray}

It is assumed that the initial reduced density matrix of spacetime is given as
\begin{eqnarray}\begin{split}
\label{eq:4.m5}
\rho(a^{+}_{0}, a^{-}_{0})=&\frac{1}{(\pi\alpha^{2})^{\frac{1}{2}}}\mathrm{exp}\big\{-ip_{a}(a_{0}^{+}-a_{0}^{-})\\
&-\frac{1}{2\alpha^{2}}\big((a_{0}^{+})^{2}
+(a_{0}^{-})^{2}\big)\big\},
\end{split}
\end{eqnarray}
where $\rho(a_{0}^{\pm},0)$ is a Gaussian wave packet. The $p_{a}$ and $\alpha$ are two arbitrary parameters where $p_{a}$ represents the initial momentum of spacetime and $\alpha$ measures the initial fluctuations of spacetime. One can show that the initial uncertainty of the scale factor is equal to $\alpha/\sqrt{2}$. Bringing Eq. \eqref{eq:4.m5} into Eq. \eqref{eq:4.m4} and introducing the approximation
\begin{equation}
\label{eq:4.m6}
\frac{1}{a_{N}^{\pm}+a_{0}^{\pm}}\approx \frac{1}{a_{N}^{\pm}}\big[1-\frac{a_{0}^{\pm}}{a_{N}^{\pm}}+o\big((\frac{a_{0}^{\pm}}{a_{N}^{\pm}})^{2}\big)\big],
\end{equation}
one can complete the integrals in Eq. \eqref{eq:4.m4}, which gives
\begin{eqnarray}\begin{split}
\label{eq:4.m7}
\rho(a^{+}_{N}, a^{-}_{N})=&\mathscr{N}\cdot\frac{(a_{N}^{+})^{\frac{1}{2}}}{\big(\frac{1}{2\alpha^{2}}-
i\xi_{1}a_{N}^{+}\big)^{\frac{1}{2}}}\cdot\frac{(a_{N}^{-})^{\frac{1}{2}}}{\big(\frac{1}{2\alpha^{2}}+i\xi_{1}a_{N}^{-}
\big)^{\frac{1}{2}}}\\&\times\mathrm{exp}\Big\{i\xi_{1}\big[(a_{N}^{-})^{3}-(a_{N}^{+})^{3}\big]+2i\chi_{0}t_{N} \\&\times(\frac{1}{a_{N}^{-}}-\frac{1}{a_{N}^{+}})-\frac{(p_{a}-\xi_{1}(a_{N}^{+})^{2}-\frac{2\chi_{0}t_{N}}{(a_{N}^{+})^{2}})^{2}}{\frac{1}{\alpha^{4}}
+4\xi_{1}^{2}(a_{N}^{+})^{2}}\\&\times(\frac{1}{2\alpha^{2}}+i\xi_{1}a_{N}^{+})-\frac{(p_{a}-\xi_{1}(a_{N}^{-})^{2}-\frac{2\chi_{0}t_{N}}{(a_{N}^{-})^{2}})^{2}}{\frac{1}{\alpha^{4}}
+4\xi_{1}^{2}(a_{N}^{-})^{2}}\\&\times(\frac{1}{2\alpha^{2}}-i\xi_{1}a_{N}^{-})\Big\}.
\end{split}
\end{eqnarray}
Here,
\begin{equation}
\label{eq:4.m8}
\xi_{1}\equiv\frac{3V_{0}}{16\pi t_{N}}.
\end{equation}

Equation~\eqref{eq:4.m7} gives the reduced density matrix of spacetime at time $t_{N}$, which can provide all the information about quantum spacetime. The non-diagonal elements of the reduced density matrix of spacetime contain the information of coherence~\cite{TM}, and the diagonal elements represent the probability distribution. Denoting $\rho(a,t_{N})\equiv\langle a|\rho(t_{N})|a\rangle$ as the diagonal element ($a_{N}^{+}=a_{N}^{-}=a$),
\begin{eqnarray}\begin{split}
\label{eq:4.m9}
\rho(a,t_{N})=\frac{\mathscr{N}|a|\cdot\mathrm{exp}\Big\{\frac{-(p_{a}-\frac{2\chi_{0}t_{N}}{a^{2}}-\xi_{1}a^{2})^{2}}{\frac{1}{\alpha^{2}}+4\xi_{1}^{2}\alpha^{2}a^{2}}\Big\}}{\big(\frac{1}{\alpha^{2}}+4\xi_{1}^{2}\alpha^{2}a^{2}
\big)^{\frac{1}{2}}},
\end{split}
\end{eqnarray}
where $\rho(a,t_{N})$ represents the probability distribution of the scale factor at time $t_{N}$. The size of the quantum universe may be uncertain at any given moment.

Combining Eqs.~\eqref{eq:4.2} and \eqref{eq:4.m5} gives
\begin{equation}
\label{eq:4.m10}
\mathrm{Tr}(\hat{\pi}_{a}^{2}\rho(0))=p_{a}^{2}+\frac{1}{2\alpha^{2}}.
\end{equation}
In addition, the condition of Eq.~\eqref{eq:2.14} gives the constraint
\begin{equation}
\label{eq:4.m11}
p_{a}^{2}+\frac{1}{2\alpha^{2}}=\frac{3}{2\pi}V_{0}\chi_{0},
\end{equation}
which can reduce one of the independent parameters.

\begin{figure}[tbp]
\centering
\includegraphics[width=7.5cm]{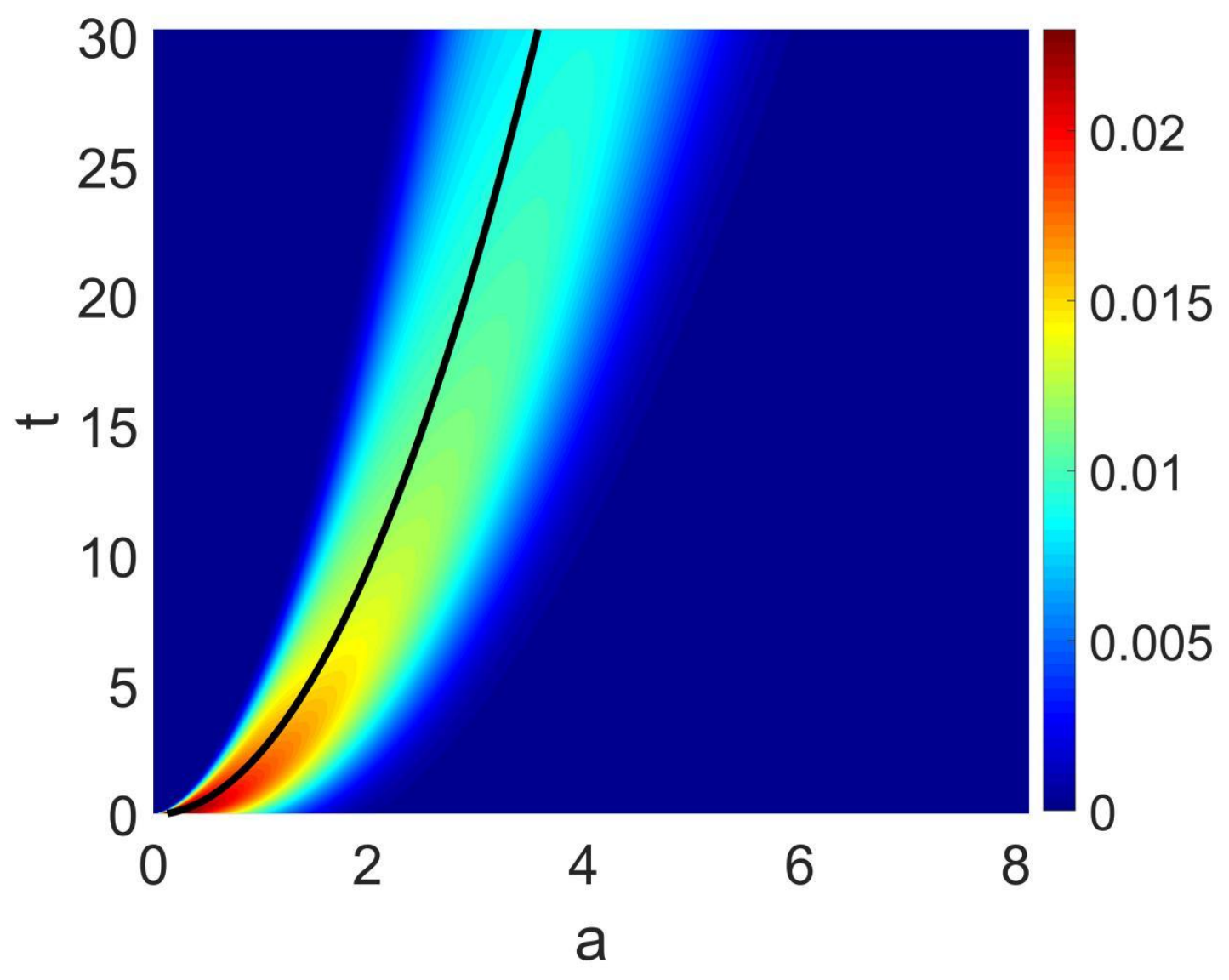}
\caption{\label{fig:1} Variations in the probability distribution of the scale factor. The horizontal and vertical axes represent the scale factor and time variable, respectively. The black curve is the classical trajectory of the universe driven by heat radiation. The parameters are taken as: $p_{a}=10$, $\alpha=0.5$, $V_{0}=200$.}
\end{figure}
\begin{figure}[tbp]
\centering
\includegraphics[width=8cm]{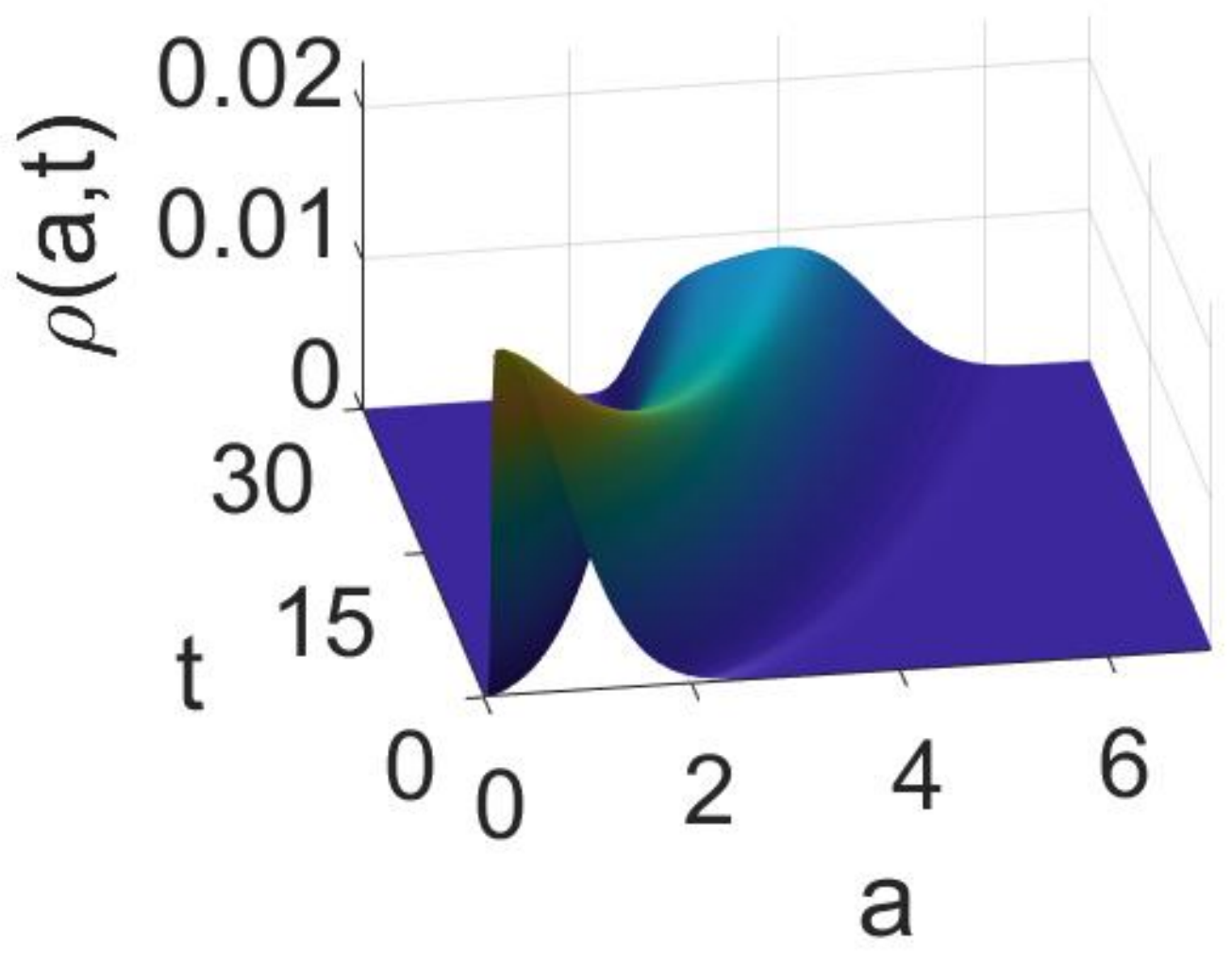}
\caption{\label{fig:2g} 3-dimensional graphic for the evolution of the wave packet.  The parameters are taken as: $p_{a}=10$, $\alpha=0.5$, $V_{0}=200$.}
\end{figure}
\begin{figure}[tbp]
\centering
\includegraphics[width=7cm]{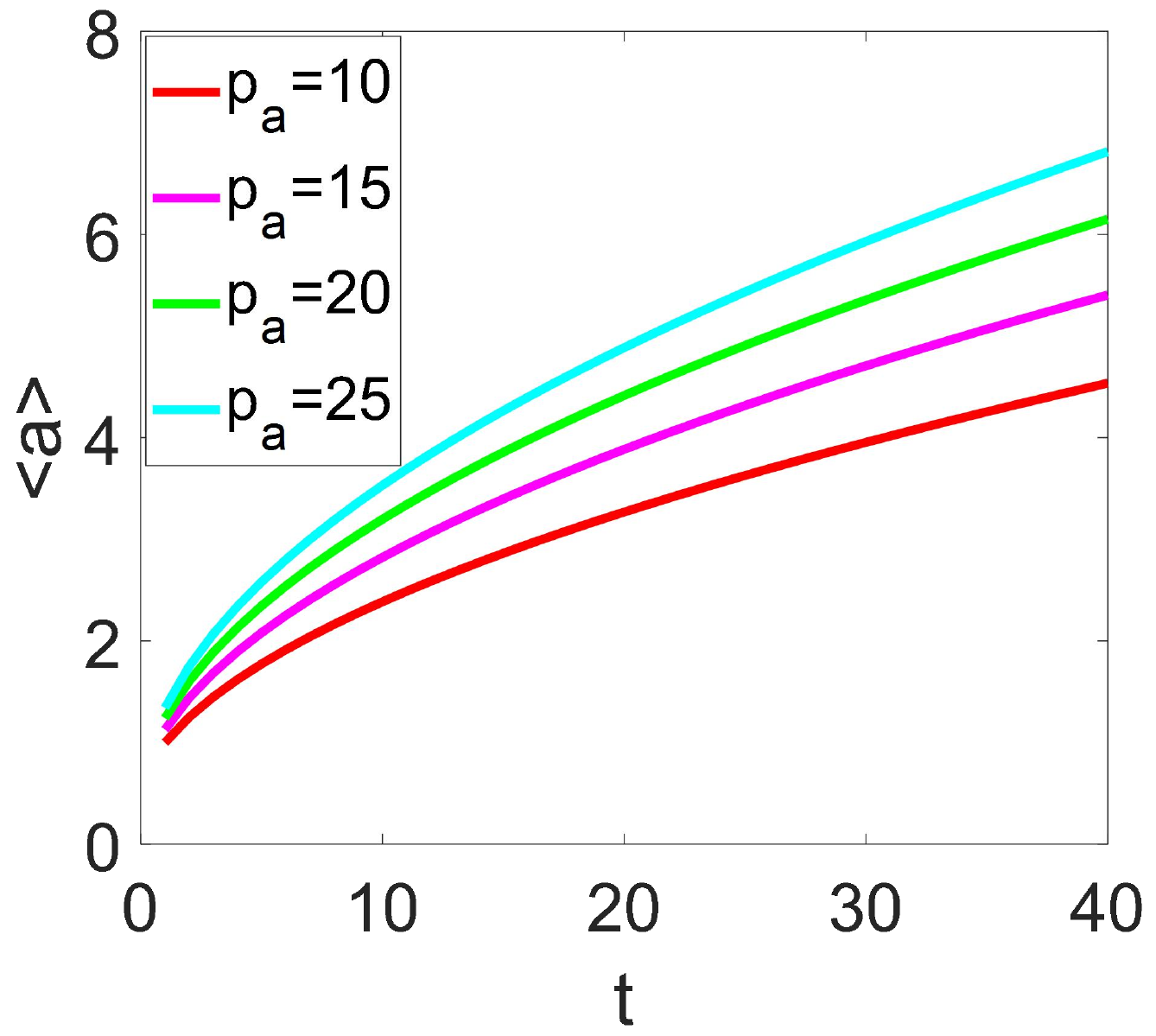}
\caption{\label{fig:2d} Variations in the average value of the scale factor ($\langle a\rangle$). The horizontal and vertical axes represent the time variable and $\langle a\rangle$, respectively. The parameters are taken as: $p_{a}=10$, $\alpha=0.5$, $V_{0}=200$.}
\end{figure}

Variations in the probability distribution are shown in Fig.~\ref{fig:1}. In this figure, the different colors represent various probabilities.
The black curve represents the classical trajectory of the universe as driven by heat radiation (this trajectory can be obtained by solving the classical Friedmann equation where matter is the heat radiation). One can infer that as $t\rightarrow\infty$, the average value of the scale factor will approach infinity.
The evolution of the wave packet is consistent with the classical trajectory, which is similar to the case of a free particle. If the free particle is described by a Gaussian wave packet, the trajectory of the wave packet is consistent with the classical evolution of the free particle. Figure~\ref{fig:2g} is the 3-dimensional representation for the evolution of the wave packet. This figure clearly shows that the wave packet is dispersed with time. Figure~\ref{fig:2d} shows that the average value of the scale factor ($\langle a\rangle= \mathrm{Tr}(a\rho)$) increases with time. Classically, $\alpha_{0}=aT$ does not change with time; thus, as the classical universe grows bigger, the temperature of the radiation decreases. Semi-classically, it is natural to assume that $\alpha_{0}\sim \mathrm{Tr}(a\rho)\cdot T$. Therefore, as the average scale factor increases, it is reasonable to expect the temperature of the radiation to decrease.

\begin{figure}[tbp]
\centering
\includegraphics[width=14cm]{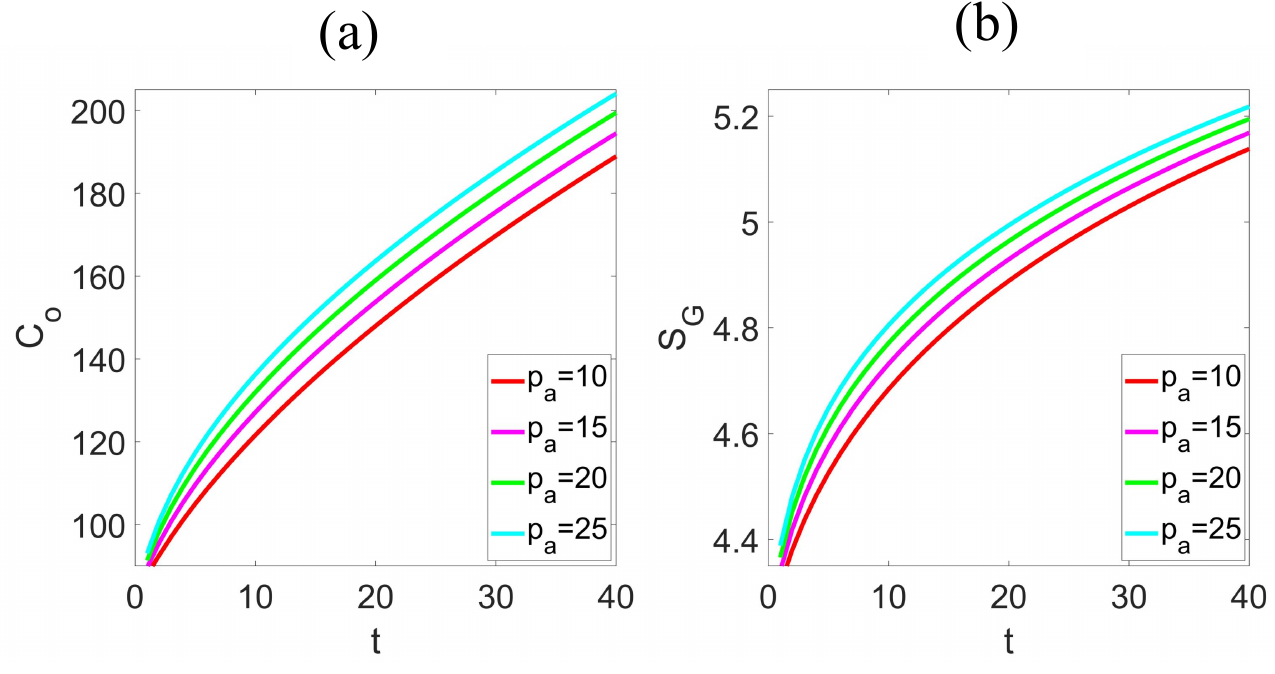}
\caption{\label{fig:2} Variations in the coherence and Gibbs entropy. The horizontal axis for these two figures represents the time variable. The vertical axis represents the (a) coherence and (b) Gibbs entropy. The parameters are taken as: $\alpha=0.5$, $V_{0}=200$.}
\end{figure}

\begin{figure}[tbp]
\centering
\includegraphics[width=7.5cm]{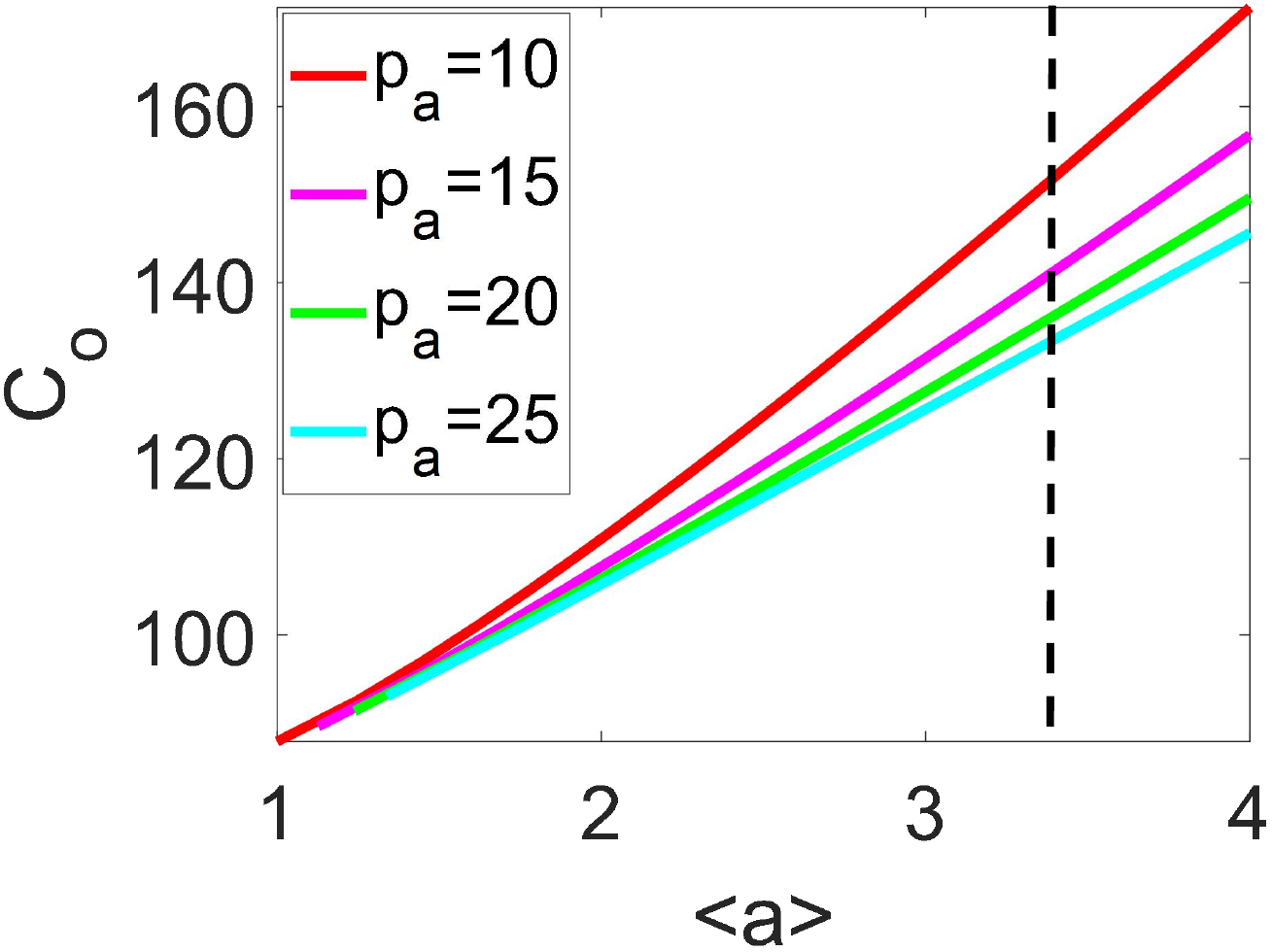}
\caption{\label{fig:2c} Variations in the coherence versus the average value of the scale factor $\langle a\rangle$. The horizontal and vertical axes represent $\langle a\rangle$ and coherence, respectively. The parameters are taken as: $\alpha=0.5$, $V_{0}=200$.}
\end{figure}

\begin{figure}[tbp]
\centering
\includegraphics[width=7cm]{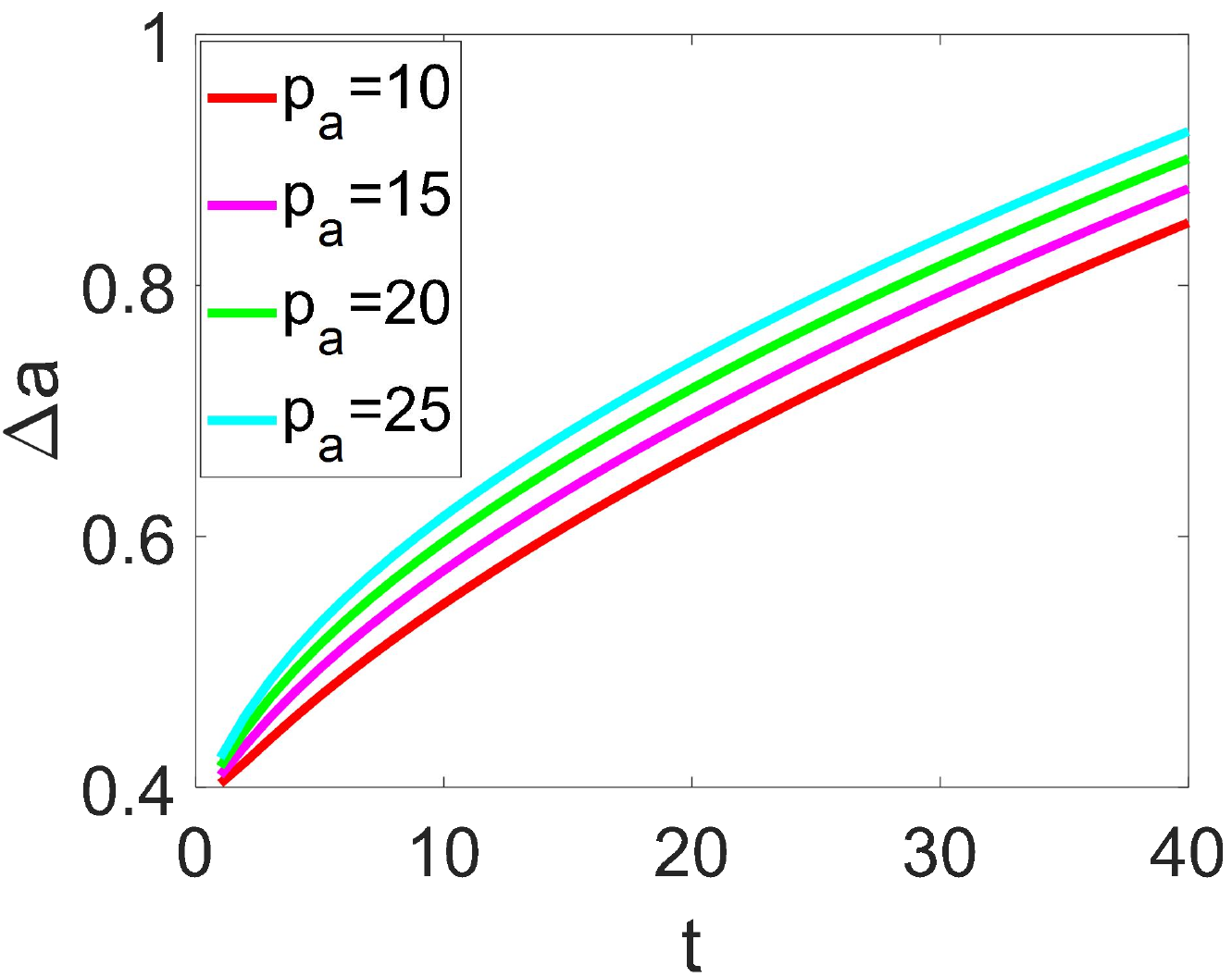}
\caption{\label{fig:2b} Variations in the variance. The horizontal and vertical axes represent the time variable and variance, respectively. The parameters are taken as: $\alpha=0.5$, $V_{0}=200$.}
\end{figure}

\begin{figure}[tbp]
\centering
\includegraphics[width=7cm]{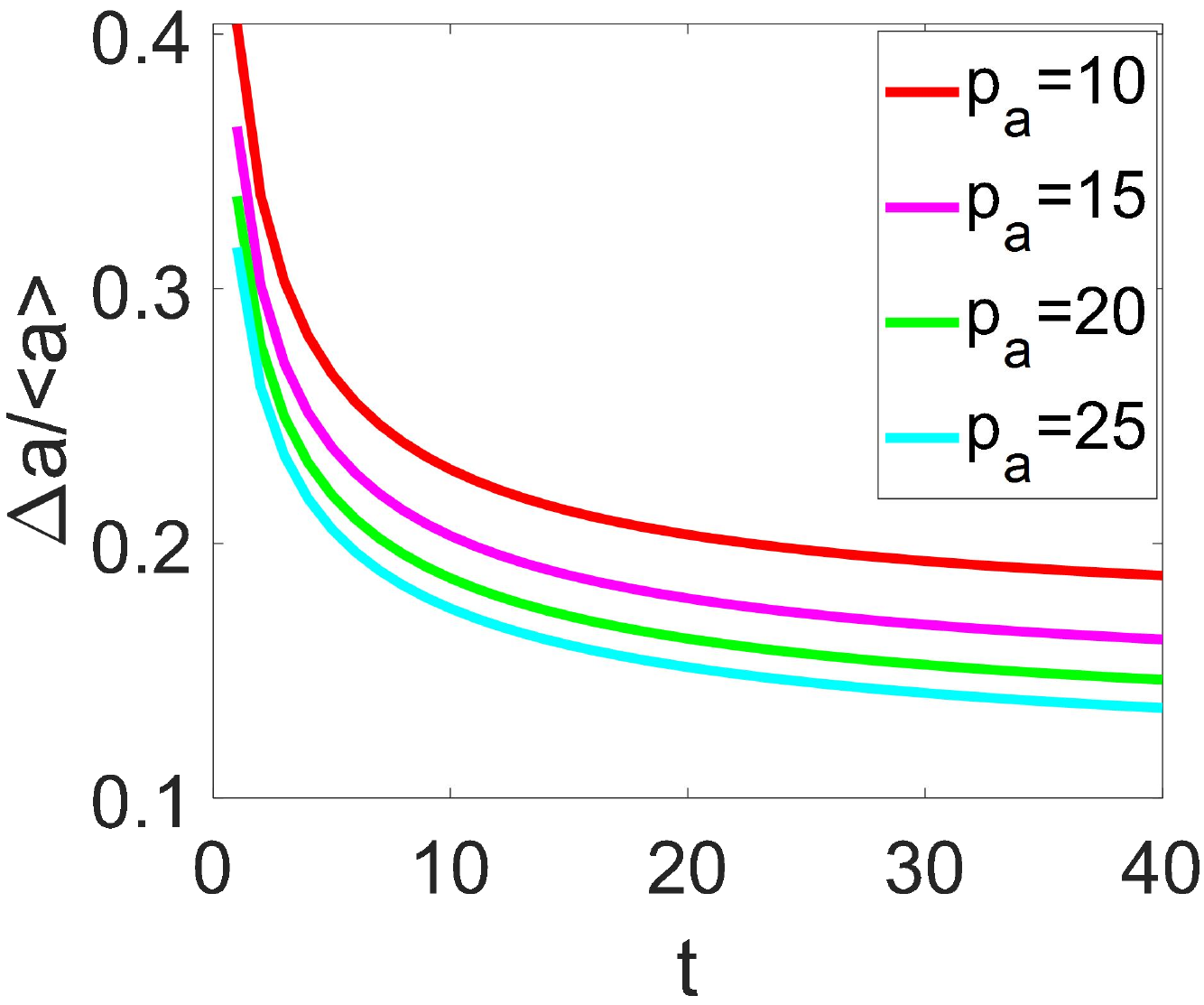}
\caption{\label{fig:2e} Variations in the relative quantum fluctuation ($\triangle a/\langle a\rangle$). The horizontal and vertical axes represent the time variable and $\triangle a/\langle a\rangle$, respectively. The parameters are taken as: $\alpha=0.5$, $V_{0}=200$.}
\end{figure}

\begin{figure}[tbp]
\centering
\includegraphics[width=7cm]{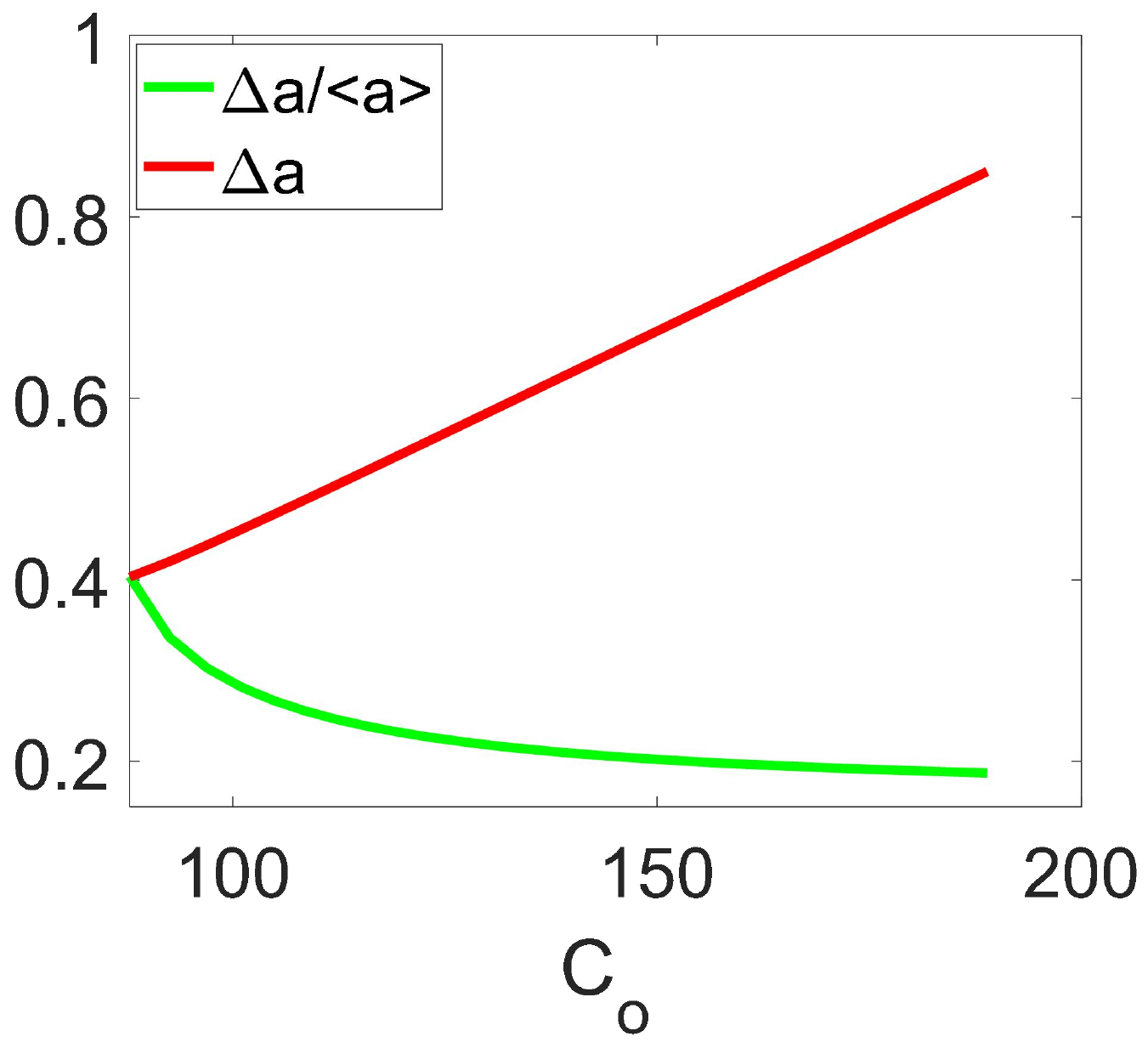}
\caption{\label{fig:2f} Variations in the quantum fluctuation versus the coherence. The horizontal axis represents the coherence.  The parameters are taken as: $p_{a}=10$, $\alpha=0.5$, $V_{0}=200$.}
\end{figure}

Coherence is an important feature of quantum systems. The process of decoherence can be viewed as a decrease in coherence. The coherence is defined by~\cite{TM}
\begin{equation}
\label{eq:4.41}
C_{o}\equiv\sum_{i\neq j}|\rho_{ij}|=\sum_{i, j}|\rho_{ij}|-\sum_{i}|\rho_{ii}|.
\end{equation}
The information of coherence is contained by the non-diagonal elements of the reduced density matrix.
Figure~\ref{fig:2}(a) shows variations in the coherence with time, which monotonic increase over time. Figure~\ref{fig:2c} shows that the bigger the universe is (measured by the average value of the scale factor $\langle a\rangle$), the bigger the coherence is. This indicates that the small quantum universe does not decohere to the classical universe. Figure~\ref{fig:2}(b) and Fig.~\ref{fig:2b} show that variations in the Gibbs entropy (defined by $S_{G}=-\sum_{i}\rho_{ii}\mathrm{ln}\rho_{ii}$) and the variance (defined by $\triangle a=\sqrt{\langle a^{2}\rangle-\langle a\rangle^{2}}$) have similar trends as the coherence. The variance measures the absolute value of the quantum fluctuations. Thus Fig.~\ref{fig:2b} shows that the absolute quantum fluctuations monotonically increase with time. However, the relative quantum fluctuation (defined as $\triangle a/\langle a\rangle$) decreases with time, as shown in Fig.~\ref{fig:2e}. Figure~\ref{fig:2e} shows that the variation rate of the relative quantum fluctuation at first is fast and then gradually becomes slow.
Figure~\ref{fig:2f} shows that the larger absolute quantum fluctuations are associated with the higher coherence. Smaller relative quantum fluctuations show the characteristics of the evolution of the Gaussian-like quantum state. For Gaussian distribution, the relative fluctuations are small since the distribution decays exponentially fast. This leads to the smaller relative fluctuations. Therefore, one can use the average or expectation values and the second order fluctuations to characterize the whole system. On the other hand, if the evolution follows other distributions such as streched exponential or power law, the tail part of the distribution can be fatty and the rare fluctuation events can become important. In such cases, the relative fluctuations are not small but significant. The average or expectation values can no longer be representative. The whole statistical distribution is necessary to characterize the system dynamics.

Figure~\ref{fig:2d}, Fig.~\ref{fig:2}(a), Fig.~\ref{fig:2}(b) and Fig.~\ref{fig:2b} show that when increasing the parameter $p_{a}$, variations in these quantities ($\langle a\rangle$, $C_{o}$, $S_{G}$ and $\triangle a$) are more rapid. Equation~\eqref{eq:4.m11} shows that if we increase the parameter $p_{a}$ (and fixing the parameters $\alpha$ and $V_{0}$), then the parameter $\alpha_{0}$ will be increased. Semiclassically, increasing $\alpha_{0}$ means increasing the temperature of the initial state of the universe ($\chi_{0}\sim\alpha_{0}$ and $\alpha_{0}\sim\langle a\rangle T$). Thus these figures show that the higher initial universe temperature is related to more rapid evolution of these quantities. In addition, Figure~\ref{fig:2c} shows that if we fix the average value of the scale factor and decrease the parameter $p_{a}$, then the coherence increases. Thus for a given size of the universe, the lower temperature corresponds to a more quantum universe.


To sum up, in this section, we studied the evolution of the heat radiation dominated quantum universe in the proper time coordinate. The dynamical information of the quantum spacetime is included in the reduced density matrix. The diagonal elements of the reduced density matrix show that the evolution of the quantum universe described by the wave packet is consistent with the classical evolution trajectory of the universe. And the non-diagonal elements of the reduced density matrix show that the coherence increases with time. The variations in the Gibbs entropy and the absolute quantum fluctuation have similar trends as the coherence. This means that the small quantum universe can grow up to a bigger one, yet the bigger universe still maintains important quantum characteristics such as the coherence in our model. In order to obtain a classical universe from the initial quantum state, more complicated models are needed.

\section{Non-relativistic matter dominated evolution of the quantum universe}
\label{sec:5}
We assume that there are $N_{m}$ particles with mass $m$ that all have the same properties. If these particles are spin-0 and charge-0, they can be viewed as the quanta of a real massive scalar field. All these particles compose a perfect isentropic fluid~\cite{JDB}. Its Hamiltonian operator is~\cite{JDB}
\begin{equation}
\label{eq:5.5}
\hat{H}_{\phi}=V_{0}\bm{N}N_{m}m.
\end{equation}
We introduce the coordinate volume $V_{0}$ from the perspective of quantum field theory [see Eq.~\eqref{eq:5.7}]. The factor $V_{0}$ is needed to keep the entire system with global scaling symmetry. The Hamiltonian operator in Eq.~\eqref{eq:5.5} can also be interpreted as the Hamiltonian of a Bose--Einstein condensate. This condensate is composed of $N_{m}$ real scalar particles with a rest mass $m$. All particles are in the ground state and the kinetic energy can be neglected.

The Hamiltonian operator in Eq.~\eqref{eq:5.5} can also be derived by assuming that the Hamiltonian operator in Eq.~\eqref{eq:3.26} can be generalized for the case of the massive scalar field. The Hamiltonian then becomes
\begin{equation}
\label{eq:5.6}
\hat{H}_{\phi}=\bm{N}\sum_{\vec{k}}\big(\frac{|\vec{k}|^{2}}{a^{2}}+m^{2}\big)^{\frac{1}{2}}A^{\dag}_{\vec{k}}A_{\vec{k}}.
\end{equation}
When $m\rightarrow 0$, this Hamiltonian operator becomes that of Eq.~\eqref{eq:3.26}. For the case $\bm{N}=a$, the Hamiltonian operator of Eq.~\eqref{eq:5.6} is consistent with that in~\cite{ETA,SSF}. In another case $\bm{N}=1$, Eq.~\eqref{eq:5.6} is consistent with the Hamiltonian operator in~\cite{WW}. Thus, the form of the Hamiltonian operator of Eq.~\eqref{eq:5.6} should be rational despite the lack of an exact proof. For a large rest mass, the Hamiltonian operator in Eq.~\eqref{eq:5.6} becomes
\begin{equation}
\label{eq:5.7}
\hat{H}_{\phi}=\bm{N}m\sum_{\vec{k}}A^{\dag}_{\vec{k}}A_{\vec{k}}=\bm{N}m\frac{V_{0}}{(2\pi)^{3}}\int d\vec{k}^{3}A^{\dag}_{\vec{k}}A_{\vec{k}}.
\end{equation}
The particle number for non-relativistic particles is usually conserved. If the particle number is $N_{m}$ ($N_{m}\propto\int d\vec{k}^{3}A^{\dag}_{\vec{k}}A_{\vec{k}}$) and all particles are in the ground state, the Hamiltonian operator of Eq.~\eqref{eq:5.7} reduces to the Hamiltonian operator of Eq.~\eqref{eq:5.5}.

For the case where the universe is dominated by non-relativistic matter, the total Hamiltonian operator becomes
\begin{equation}
\label{eq:5.8}
\hat{H}_{tot}=\hat{H}_{g}+\hat{H}_{\phi}=\frac{-\pi }{3V_{0}}\bm{N}(\frac{1}{a}\hat{\pi}_{a}^{2}+\hat{\pi}_{a}^{2}\frac{1}{a})+V_{0}\bm{N}N_{m}m.
\end{equation}
Solving $H_{tot}=0$ gives the classical evolution trajectory of the universe. In the proper time coordinate ($\bm{N}=1$), $a(t)\propto t^{\frac{2}{3}}$~\cite{SW}. In the conformal time coordinate ($\bm{N}=a$), $a(t)\propto t^{2}$. These results strongly indicate the Hamiltonian operator in Eq.~\eqref{eq:5.5} is reasonable.

In the proper time coordinate, the total Hamiltonian operator is
\begin{equation}
\label{eq:5.9}
\hat{H}_{tot}=\hat{H}_{g}+\hat{H}_{\phi}=\frac{-\pi }{3V_{0}}(\frac{1}{a}\hat{\pi}_{a}^{2}+\hat{\pi}_{a}^{2}\frac{1}{a})+V_{0}N_{m}m.
\end{equation}
In Eq.~\eqref{eq:5.9}, $\hat{H}_{\phi}=V_{0}N_{m}m$. Bringing $\hat{H}_{\phi}$ into the definition of influence functional of Eq.~\eqref{eq:4.10}, we have
\begin{eqnarray}\begin{split}
\label{eq:5.9b}
\mathbf{I_{mN}}=&\mathrm{ Tr}_{\phi}\big\{e^{-iV_{0}mN_{m}\delta t}\cdot\cdot\cdot e^{-iV_{0}mN_{m}\delta t}\\&\times\rho_{\phi}(0)e^{iV_{0}mN_{m}\delta t}\cdot\cdot\cdot e^{iV_{0}mN_{m}\delta t}\big\}.
\end{split}
\end{eqnarray}
Here, we use $\mathbf{I_{mN}}$ to represent the influence functional in the case of the non-relativistic particles dominated universe. Using $\mathrm{Tr(ABC)=Tr(CAB)}$ , one can show that the influence functional is equal to one. Assuming the initial state of the total system can be written as $\rho_{tot}(0)=\rho(0)\otimes\rho_{\phi}(0)$, similar to Eq.~\eqref{eq:4.9}, we have
\begin{eqnarray}\begin{split}
\label{eq:5.10}
\rho(a^{+}_{N}, a^{-}_{N}) =&\int da_{0}^{\pm}\int da_{1}^{\pm}\cdot\cdot\cdot\int da_{N-1}^{\pm}\langle a^{+}_{N}|e^{-i\hat{H}_{g}\delta t}|a^{+}_{N-1}\rangle\\&\times\langle a^{+}_{N-1}|e^{-i\hat{H}_{g}\delta t}|a^{+}_{N-2}\rangle\cdot\cdot\cdot\langle a^{+}_{0}|\rho(0)|a^{-}_{0}\rangle\\&\times\langle a^{-}_{0}|e^{i\hat{H}_{g}\delta t}|a^{-}_{1}\rangle\cdot\cdot\cdot\langle a^{-}_{N-1}|e^{i\hat{H}_{g}\delta t}|a^{-}_{N}\rangle,
\end{split}
\end{eqnarray}
where $\langle a^{+}_{n+1}|e^{-i\hat{H}_{g}\delta t}|a^{+}_{n}\rangle$ and $\langle a^{-}_{n}|e^{i\hat{H}_{g}\delta t}|a^{-}_{n+1}\rangle$ are the same as in Eqs.~\eqref{eq:4.11} and \eqref{eq:4.12}, respectively. Bringing Eqs.~\eqref{eq:4.11} and \eqref{eq:4.12} into Eq.~\eqref{eq:5.10}, and using the approximation in Eq.~\eqref{eq:4.19} gives
\begin{eqnarray}\begin{split}
\label{eq:5.11}
\rho(a^{+}_{N}, a^{-}_{N}) =&\mathscr{N}\int da_{0}^{\pm}\int da_{1}^{\pm}\cdot\cdot\cdot\int da_{N-1}^{\pm}\\&\times\langle a^{+}_{0}|\rho(0)|a^{-}_{0}\rangle\prod_{n=1}^{N}\sqrt{\frac{3iV_{0}a_{n}^{+}}{2\delta t }}\\&\times\prod_{n=1}^{N}\sqrt{\frac{-3iV_{0}a_{n}^{-}}{2\delta t }} \mathrm{exp}\{\mathcal{I}_{m+}+\mathcal{I}_{m-}\},
\end{split}
\end{eqnarray}
where,
\begin{equation}
\label{eq:5.12}
\mathcal{I}_{m\pm}\equiv\sum_{n=1}^{N}\big[\frac{\mp3iV_{0}a_{n}^{\pm}(a_{n-1}^{\pm}-a_{n}^{\pm})^{2}}{8\pi\delta t}\big],
\end{equation}

If $\chi_{0}=0$, Eq.~\eqref{eq:4.22} is reduced to Eq.~\eqref{eq:5.12}. Taking the continuous limit, $i\mathcal{I}_{m+}$ (or $-i\mathcal{I}_{m-}$) in Eq.~\eqref{eq:5.12} provides the action of spacetime defined by the Hamiltonian of Eq.~\eqref{eq:4.2}. That is,
\begin{equation}
\label{eq:5.m1}
\lim_{N\rightarrow\infty}\mathcal{I}_{m\pm}=\int_{0}^{t_{N}}dt\big[\frac{\mp3iV_{0}a^{\pm}(t)\big(\dot{a}^{\pm}(t)\big)^{2}}{8\pi}\big].
\end{equation}
The classical trajectory corresponding to $\displaystyle\lim_{N\rightarrow\infty}\mathcal{I}_{m\pm}$ is
\begin{equation}
\label{eq:5.m2}
a_{cl}^{\pm}(t)=\Big((a_{0}^{\pm})^{\frac{3}{2}}+\big((a_{N}^{\pm})^{\frac{3}{2}}-(a_{0}^{\pm})^{\frac{3}{2}}\big)\frac{t}{t_{N}}\Big)^{\frac{2}{3}}.
\end{equation}

\begin{figure}[tbp]
\centering
\includegraphics[width=7.5cm]{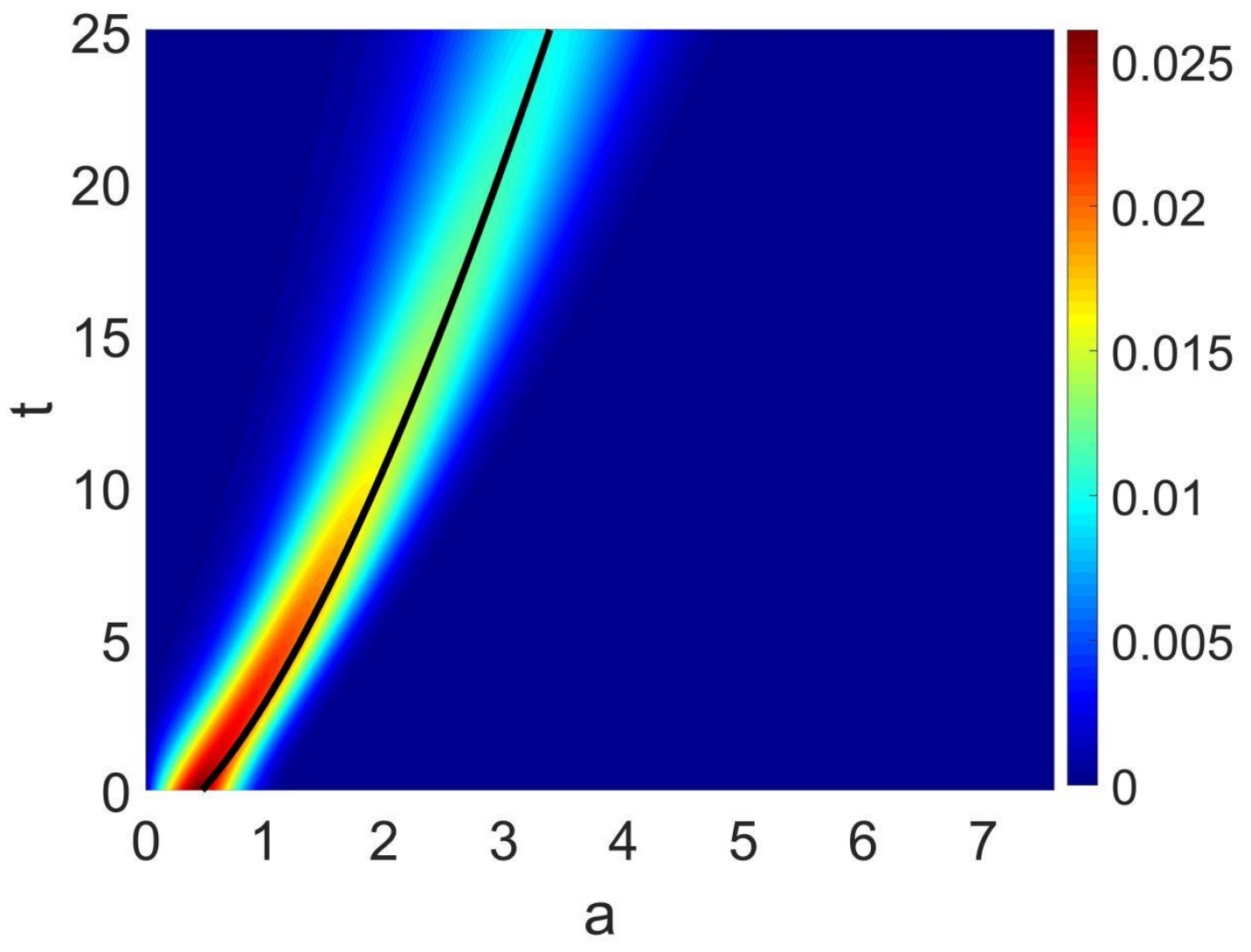}
\caption{\label{fig:3} Evolution of the probability distribution of the scale factor. The horizontal axis represents the scale factor, and the vertical axis represents the time variable. The black curve is the classical trajectory of the universe driven by non-relativistic particles. The parameters are taken as: $p_{a}=5$, $\alpha=0.5$, $V_{0}=200$, $mN_{m}=3\pi p_{a}^{2}/(2V_{0}^{2})$.}
\end{figure}

\begin{figure}[tbp]
\centering
\includegraphics[width=8cm]{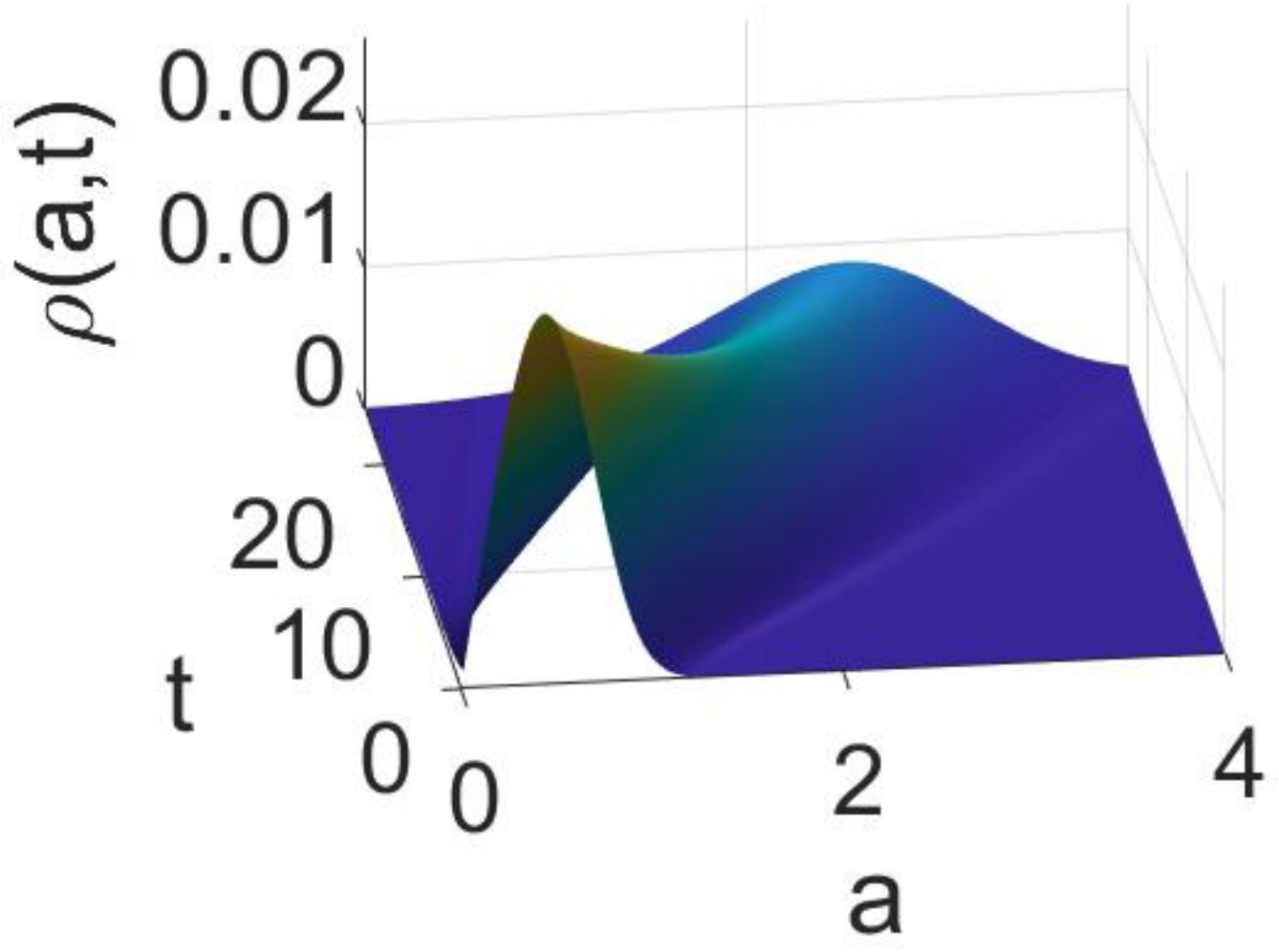}
\caption{\label{fig:3b} 3-dimensional graphic for the evolution of the wave packet.  The parameters are taken as: $p_{a}=5$, $\alpha=0.5$, $V_{0}=200$.}
\end{figure}

\begin{figure}[tbp]
\centering
\includegraphics[width=14cm]{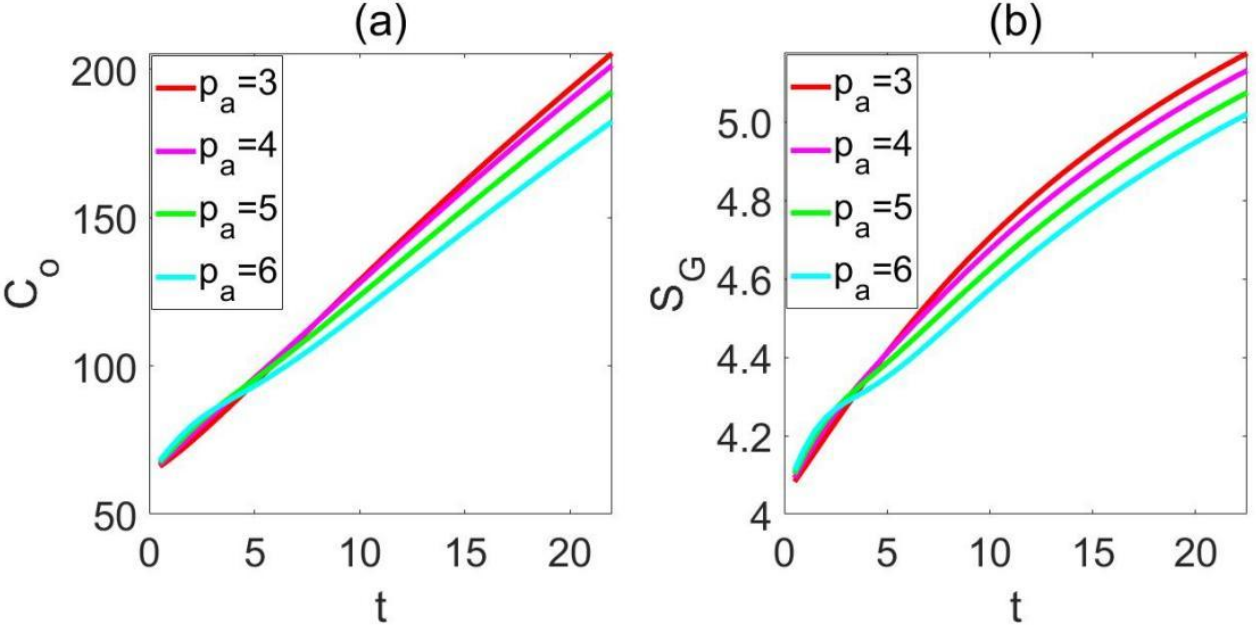}
\caption{\label{fig:4} Variations in the (a) coherence and (b) Gibbs entropy as functions of the time variable. The parameters are taken as: $\alpha=0.5$, $V_{0}=200$.}
\end{figure}

\begin{figure}[tbp]
\centering
\includegraphics[width=7cm]{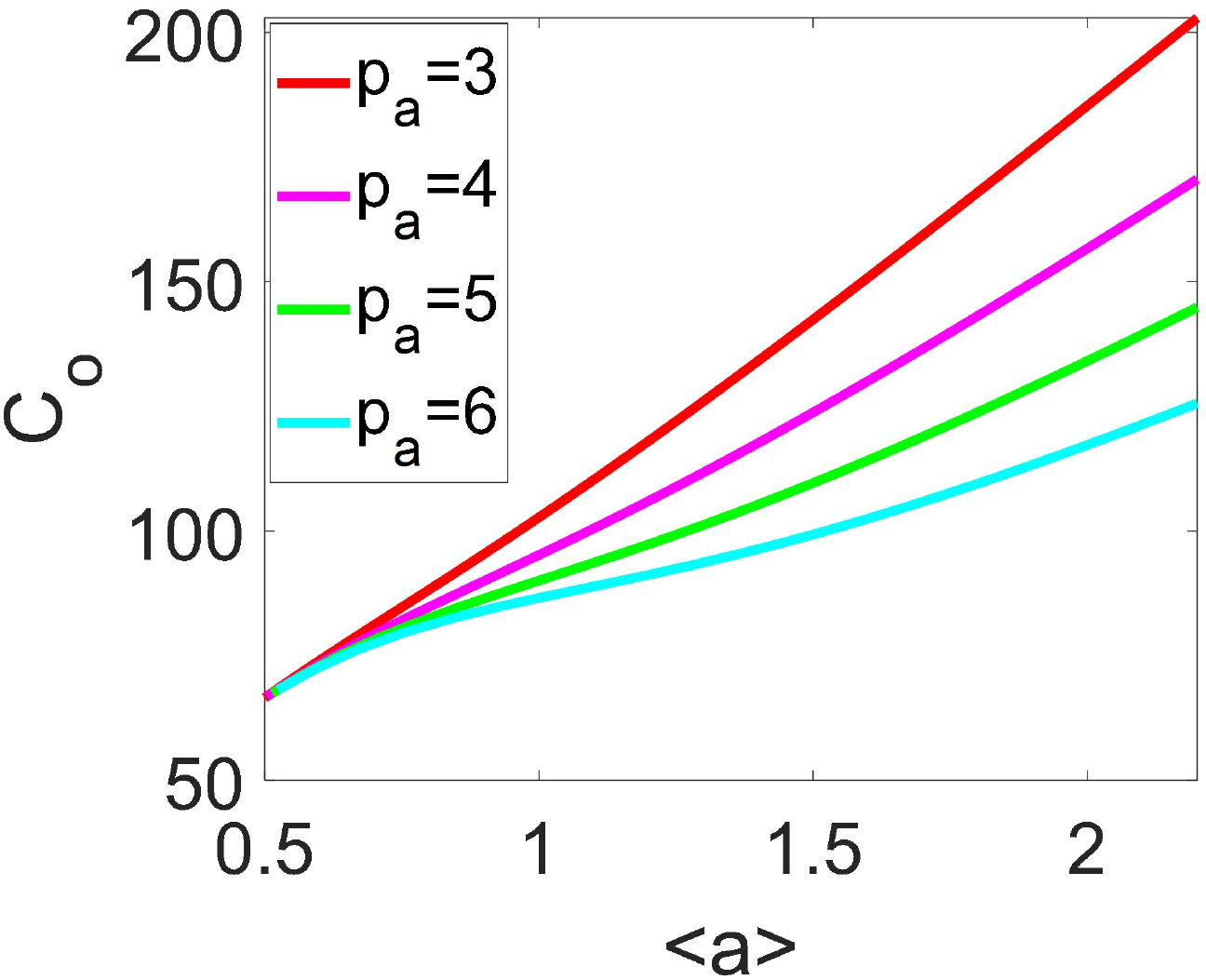}
\caption{\label{fig:4c} Variations in the coherence versus the average value of the scale factor $\langle a\rangle$. The horizontal and vertical axes represent $\langle a\rangle$ and coherence, respectively. The parameters are taken as: $\alpha=0.5$, $V_{0}=200$.}
\end{figure}

\begin{figure}[tbp]
\centering
\includegraphics[width=7cm]{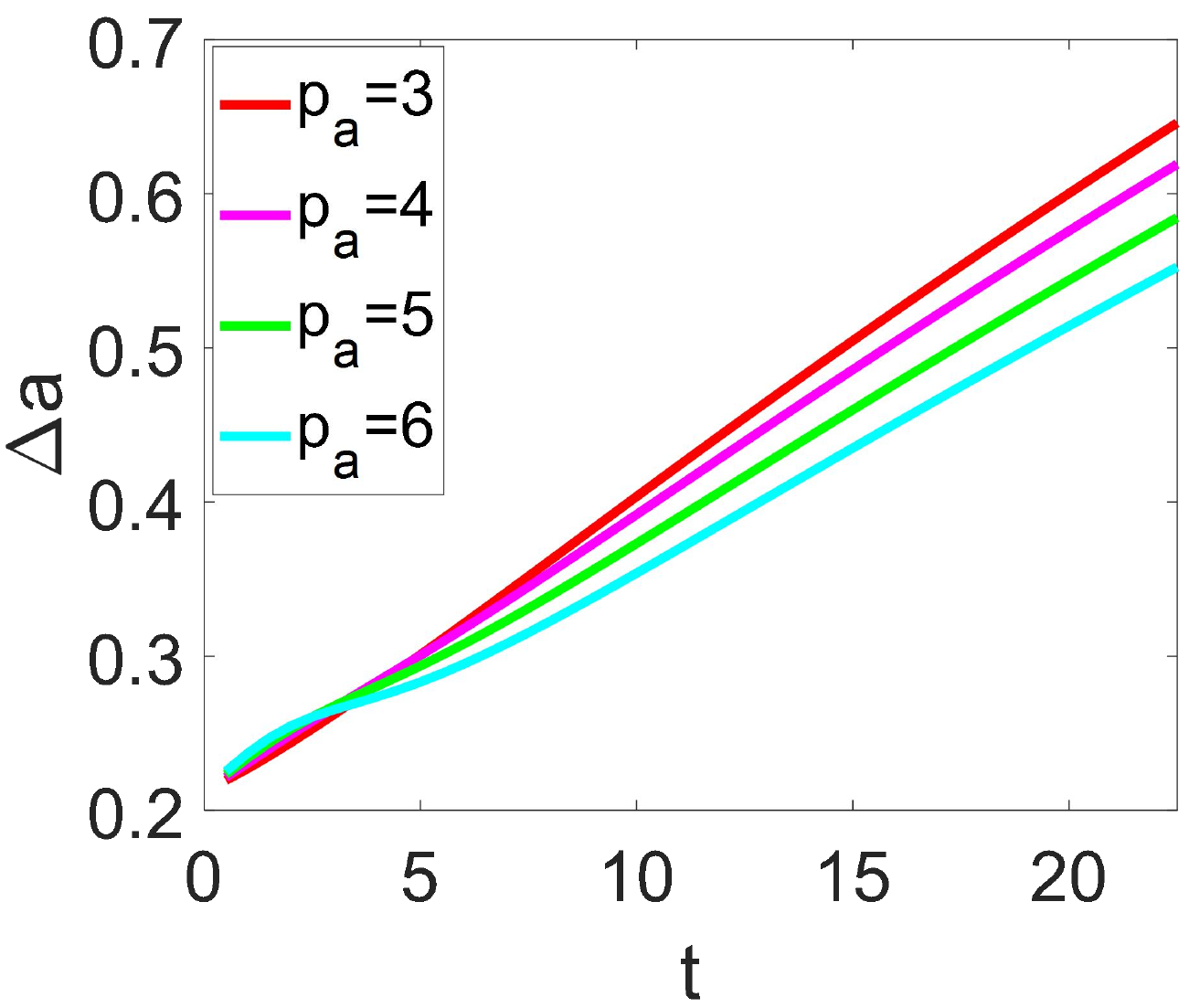}
\caption{\label{fig:4b} Variations in the variance. The horizontal and vertical axes represent the time variable and variance, respectively. The parameters are taken as: $\alpha=0.5$, $V_{0}=200$.}
\end{figure}
\begin{figure}[tbp]
\centering
\includegraphics[width=7cm]{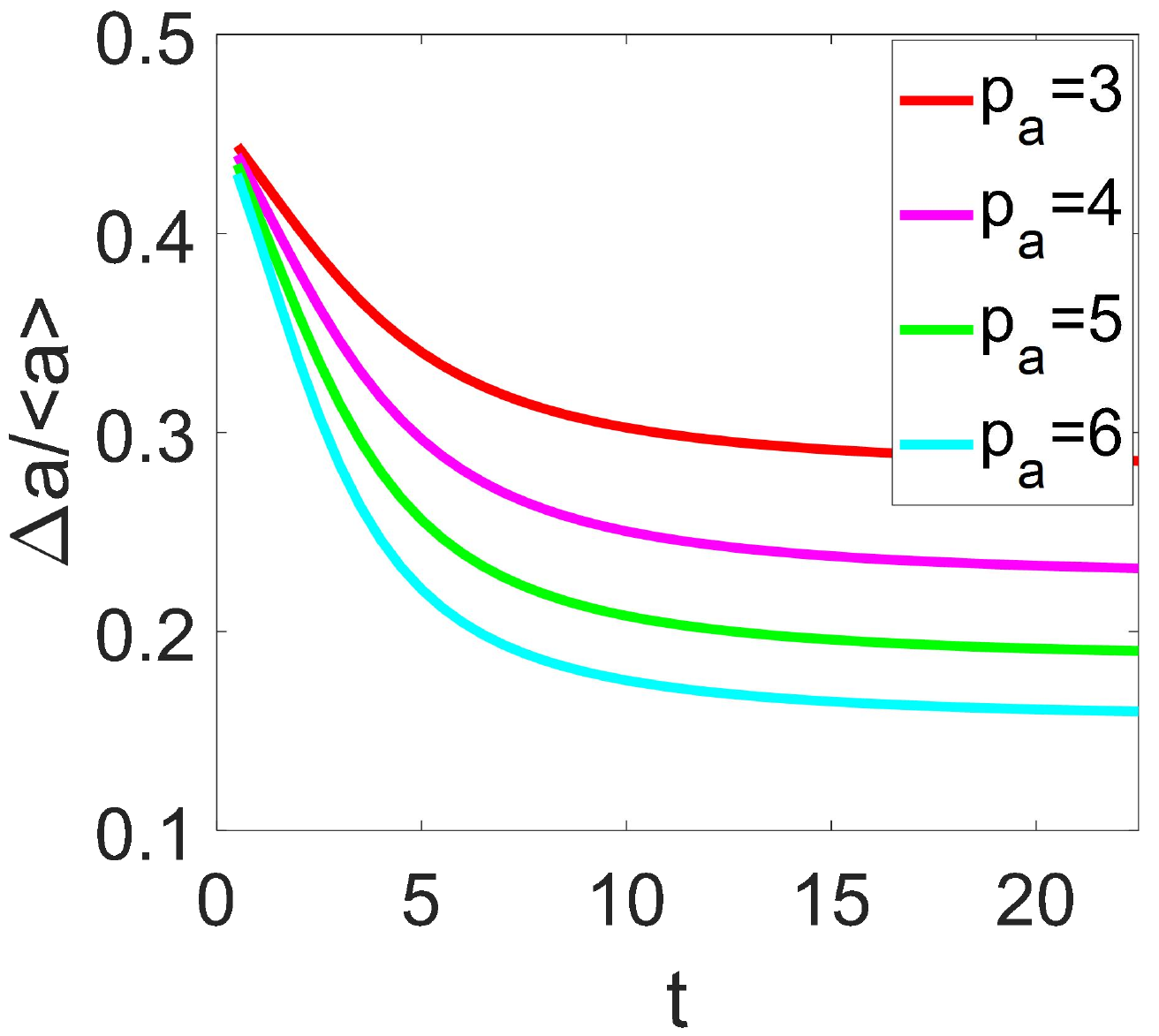}
\caption{\label{fig:4d} Variations in the relative quantum fluctuation. The horizontal and vertical axes represent the time variable and relative fluctuation, respectively. The parameters are taken as: $\alpha=0.5$, $V_{0}=200$.}
\end{figure}
\begin{figure}[tbp]
\centering
\includegraphics[width=7cm]{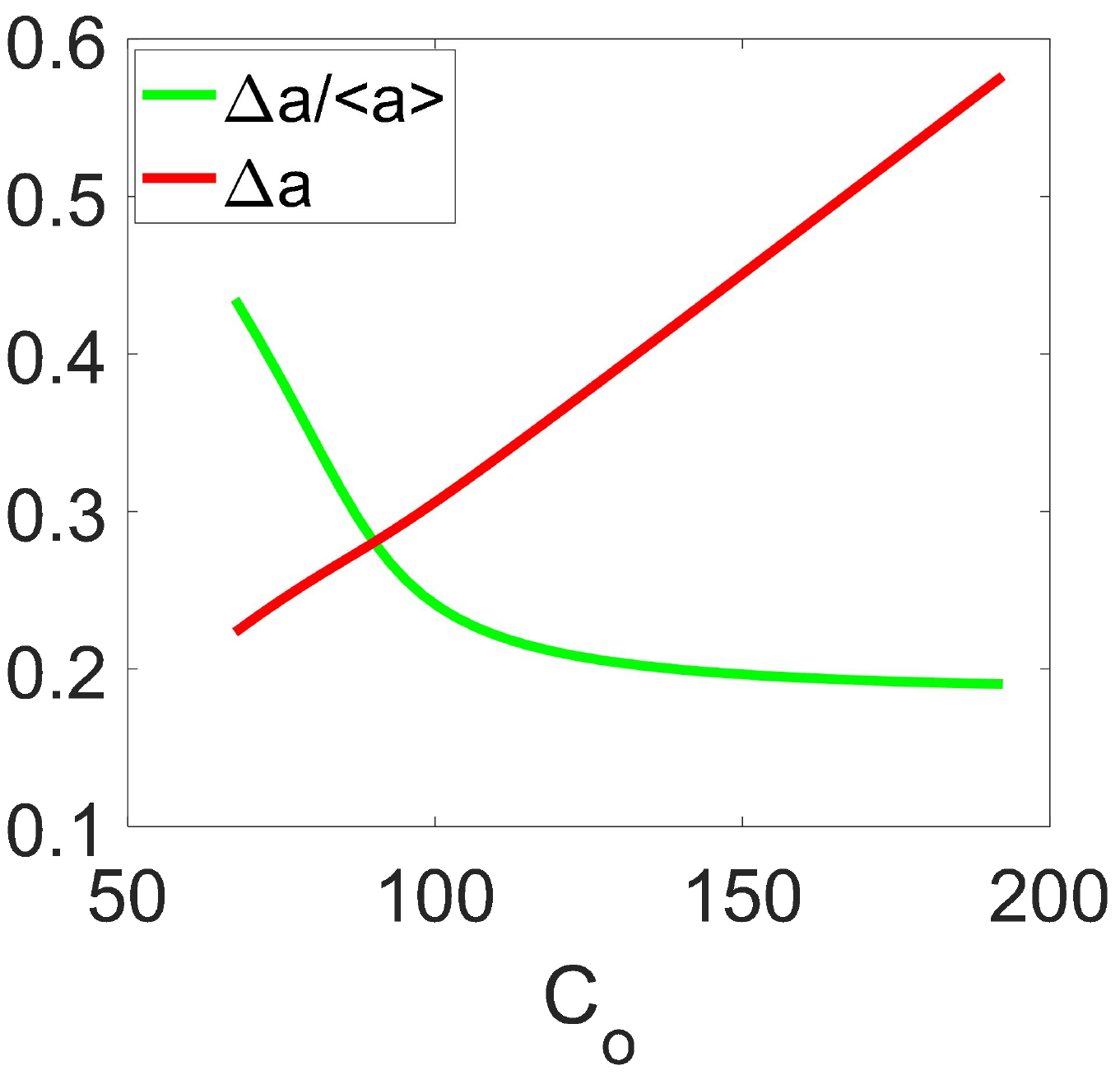}
\caption{\label{fig:4f} Variations in the quantum fluctuation versus the coherence. The horizontal axis represents the coherence.  The parameters are taken as: $p_{a}=5$, $\alpha=0.5$, $V_{0}=200$.}
\end{figure}

Expanding $\displaystyle\lim_{N\rightarrow\infty}\mathcal{I}_{m\pm}$ in Eq.~\eqref{eq:5.m1} around the classical trajectory in Eq.~\eqref{eq:5.m2}, and using the steepest descent contour approximation of Eq.~\eqref{eq:4.m3} allows solving the integrals in Eq.~\eqref{eq:5.11} (taking the continuous limit $N\rightarrow\infty$) over the variables $(a_{1}^{\pm}, a_{1}^{\pm},...,a_{N-1}^{\pm})$ to give
\begin{eqnarray}\begin{split}
\label{eq:5.m3}
\rho(a^{+}_{N}, a^{-}_{N}) =&\mathscr{N}(a_{N}^{+}a_{N}^{-})^{\frac{1}{2}}\int da_{0}^{\pm}\langle a^{+}_{0}|\rho(0)|a^{-}_{0}\rangle\\&\times \mathrm{exp}\Big\{-\frac{iV_{0}}{6\pi t_{N}}
((a_{N}^{+})^{\frac{3}{2}}-(a_{0}^{+})^{\frac{3}{2}})^{2}\\&+\frac{iV_{0}}{6\pi t_{N}}
((a_{N}^{-})^{\frac{3}{2}}-(a_{0}^{-})^{\frac{3}{2}})^{2}\Big\}.
\end{split}
\end{eqnarray}

If there is no entanglement between quantum spacetime and the non-relativistic particles at the initial time, then the initial density matrix of the universe can be written as $\rho_{tot}(0)=\rho(0)\otimes\rho_{\phi}(0)$. For simplicity, The initial reduced density matrix $\rho(0)$ is chosen as
\begin{eqnarray}\begin{split}
\label{eq:5.m4}
\rho(a^{+}_{0}, a^{-}_{0})=&(a_{0}^{+})^{\frac{1}{2}}(a_{0}^{-})^{\frac{1}{2}}\mathrm{exp}\big\{-ip_{a}\big[(a_{0}^{+})^{\frac{3}{2}}\\&-(a_{0}^{-})^{\frac{3}{2}}\big]-
\frac{1}{2\alpha^{2}}\big[(a_{0}^{+})^{3}+(a_{0}^{-})^{3}\big]\big\}.
\end{split}
\end{eqnarray}
Bringing Eq.~\eqref{eq:5.m4} into Eq.~\eqref{eq:5.m3}, one can complete the integrals over $a_{0}^{\pm}$ in Eq.~\eqref{eq:5.m3}. This gives
\begin{eqnarray}\begin{split}
\label{eq:5.m5}
\rho(a^{+}_{N}, a^{-}_{N})=&\mathscr{N}\cdot\frac{(a_{N}^{+})^{\frac{1}{2}}}{\big(\frac{1}{2\alpha^{2}}+i\xi_{2}
\big)^{\frac{1}{2}}}\cdot\frac{(a_{N}^{-})^{\frac{1}{2}}}{\big(\frac{1}{2\alpha^{2}}-i\xi_{2}\big)^{\frac{1}{2}}}\mathrm{exp}\Big\{i\xi_{2}\\&\times\big[(a_{N}^{-})^{3}-(a_{N}^{+})^{3}\big] -\frac{(p_{a}-2\xi_{2}(a_{N}^{+})^{\frac{3}{2}})^{2}}{\frac{1}{\alpha^{4}}
+4\xi_{2}^{2}}\\&\times(\frac{1}{2\alpha^{2}}-i\xi_{2})-\frac{(p_{a}-2\xi_{2}(a_{N}^{-})^{\frac{3}{2}})^{2}}{\frac{1}{\alpha^{4}}
+4\xi_{2}^{2}}\\&\times(\frac{1}{2\alpha^{2}}+i\xi_{2})\Big\},
\end{split}
\end{eqnarray}
where
\begin{equation}
\label{eq:5.m6}
\xi_{2}=\frac{V_{0}}{6\pi t_{N}}.
\end{equation}
Equation~\eqref{eq:5.m5} provides the diagonal elements of the reduced density matrix ($a_{N}^{+}=a_{N}^{-}=a$) as
\begin{equation}
\label{eq:5.m7}
\rho(a,t)=\frac{\mathscr{N}|a|}{\big(\frac{1}{\alpha^{2}}+4\alpha^{2}\xi_{2}^{2}\big)^{\frac{1}{2}}} \mathrm{exp}\Big\{\frac{-(p_{a}-2\xi_{2}a^{\frac{3}{2}})^{2}}{\frac{1}{\alpha^{2}}+
4\alpha^{2}\xi_{2}^{2}}\Big\}.
\end{equation}

In addition, one noted that the classical trajectory of the non-relativistic particles dominated universe (in the proper time coordinate) can also be written as
\begin{equation}
\label{eq:5.m8}
a(t)=\eta_{1}\big(t+\frac{(a(0))^{\frac{3}{2}}}{\eta_{1}^{\frac{3}{2}}}\big)^{\frac{2}{3}}.
\end{equation}
The parameter $\eta_{1}$ is
\begin{equation}
\label{eq:5.m9}
\eta_{1}=(6\pi m N_{m})^{\frac{1}{3}}.
\end{equation}

The evolution of the diagonal element of the reduced density matrix (representing the probability distribution of the scale factor) are shown in Fig.~\ref{fig:3}. In this figure, the different colors represent the various diagonal elements $\rho(a,t)$ in Eq.~\eqref{eq:5.m7}. The black curve represents the classical trajectory of the universe driven by non-relativistic matter. Figure~\ref{fig:3} shows that the classical evolution of the universe is consistent with the trajectory of the wave packet. As time progresses, the wave packet evolves to where the scale factor becomes larger, which indicates that the universe grows. Figure~\ref{fig:3b} is the 3-dimensional plot for the evolution of the wave packet. This figure shows that the wave packet is dispersed with time. Noted that the condition \eqref{eq:2.14} can give rise to a constraint among $p_{a}$, $\alpha$ and $V_{0}mN_{m}$.  However, it is not very easy to analytically exactly solve the condition \eqref{eq:2.14} in this case. Our numerical simulations indicate that this constraint can be approximately taken as $mN_{m}=3\pi p_{a}^{2}/(2V_{0}^{2})$. 

Figure~\ref{fig:4}(a) shows the variations in the coherence, which increase with time. Figure~\ref{fig:4c} shows that a bigger universe is associated with a bigger coherence. Therefore, the small quantum universe does not grow up to a classical one. Variations of the Gibbs entropy  and the absolute quantum fluctuation are shown in Fig.~\ref{fig:4}(b) and Fig.~\ref{fig:4b}, respectively. These two figures show that both the Gibbs entropy and the absolute quantum fluctuation increase with time. However, Fig.~\ref{fig:4d} shows that the relative quantum fluctuation decreases with time. And the variation rate of the relative quantum fluctuation gradually becomes slower. Figure~\ref{fig:4f} shows that the larger absolute quantum fluctuations are associated with the higher coherence.

In the proper time coordinate, we have shown that the coherence increases with time both in heat radiation and non-relativistic matter dominated universe. Thus, quantum spacetime can not evolve to a classical one in this model of minimally massless scalar field and non-relativistic  particles in flat FRW universe. We introduced some approximations to obtain this result. One may think that the result will be different without these approximations. However, we will show in the next section that in the conformal time coordinate, without any approximation, one still reaches the same conclusion.

\section{Evolution of the reduced density matrix in the conformal time coordinate}
\label{sec:E}
\subsection{Heat radiation dominated universe}

\begin{figure}[tbp]
\centering
\includegraphics[width=7.5cm]{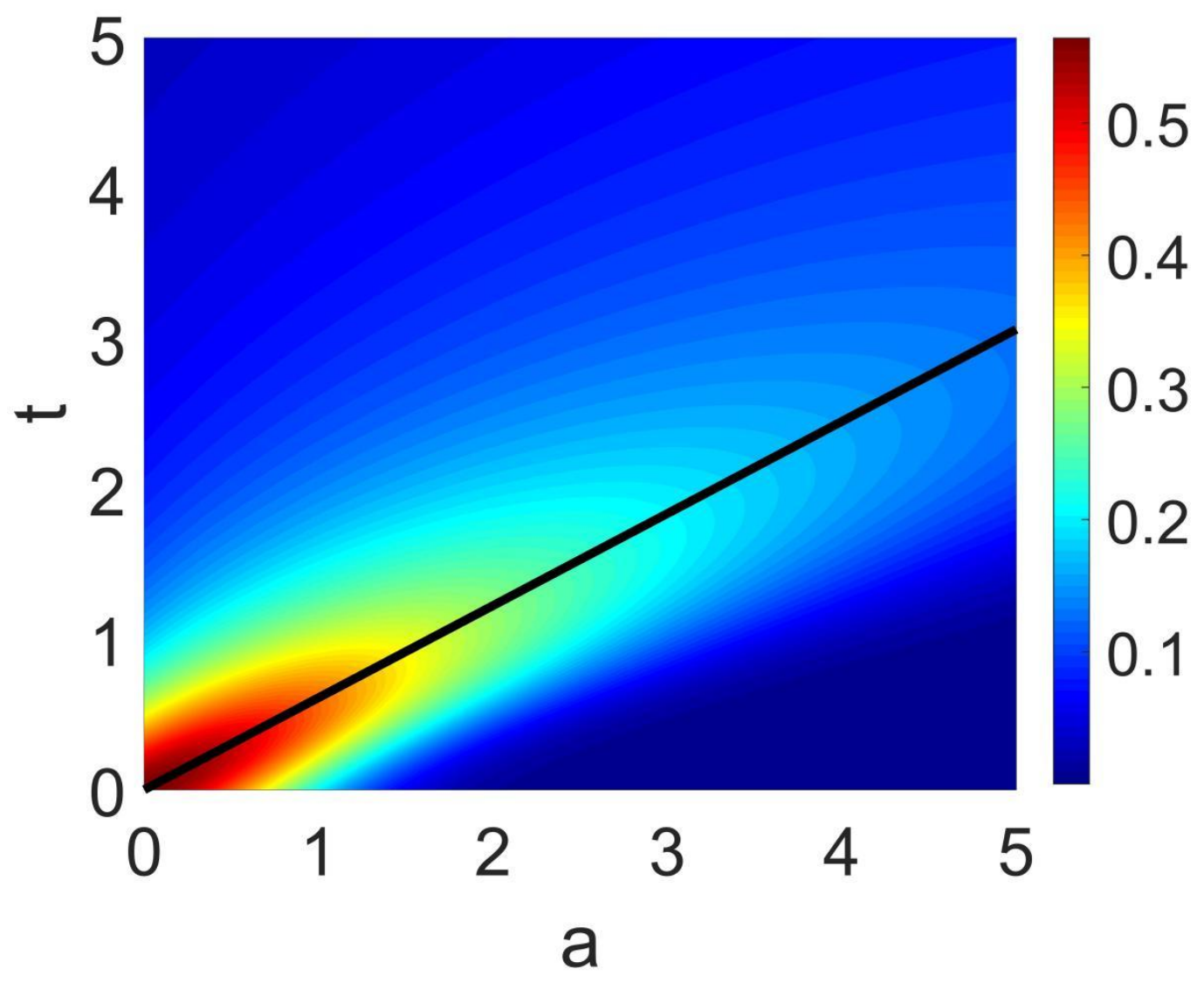}
\caption{\label{fig:5} Evolution of the probability distribution of the scale factor. The horizontal axis represents the scale factor, and the vertical axis represents the time variable. The black curve is the classical trajectory of the universe. The parameters are taken as: $p_{a}=1$, $\alpha=1$, $V_{0}=\pi$.}
\end{figure}
\begin{figure}[tbp]
\centering
\includegraphics[width=14cm]{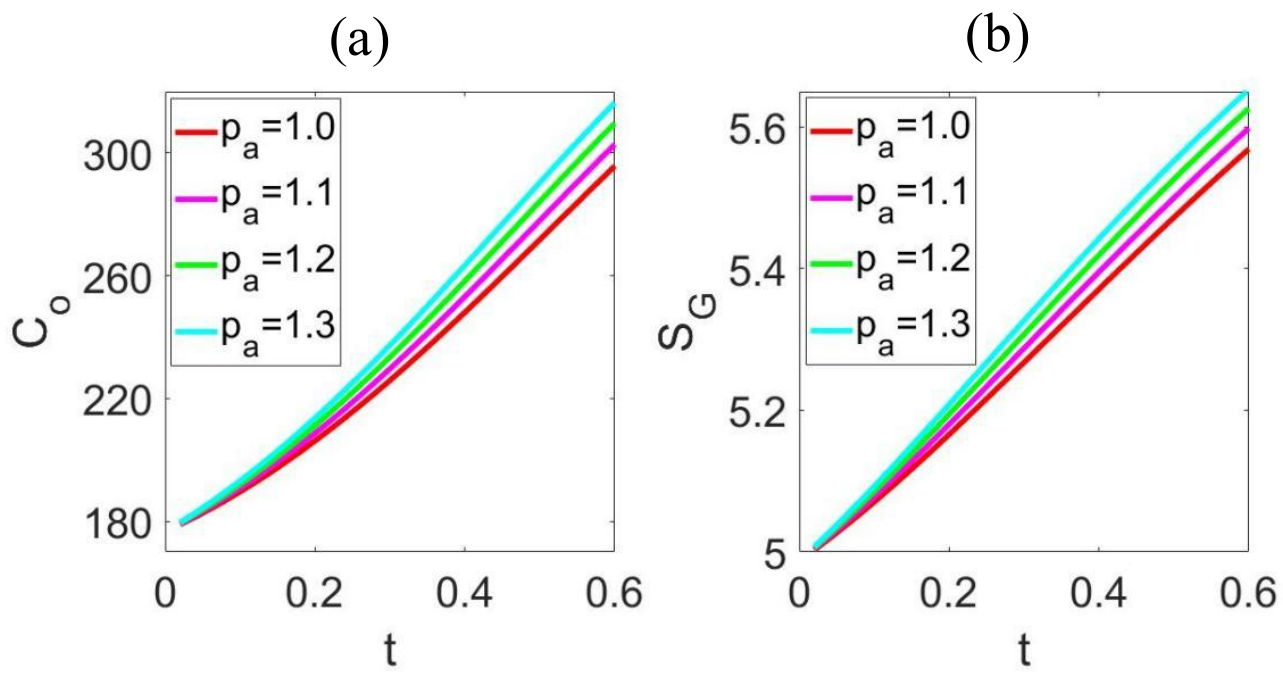}
\caption{\label{fig:6} Variations in the (a) coherence and (b) Gibbs entropy as functions of the time variable. The parameters are taken as: $\alpha=1$, $V_{0}=\pi$.}
\end{figure}

\begin{figure}[tbp]
\centering
\includegraphics[width=7cm]{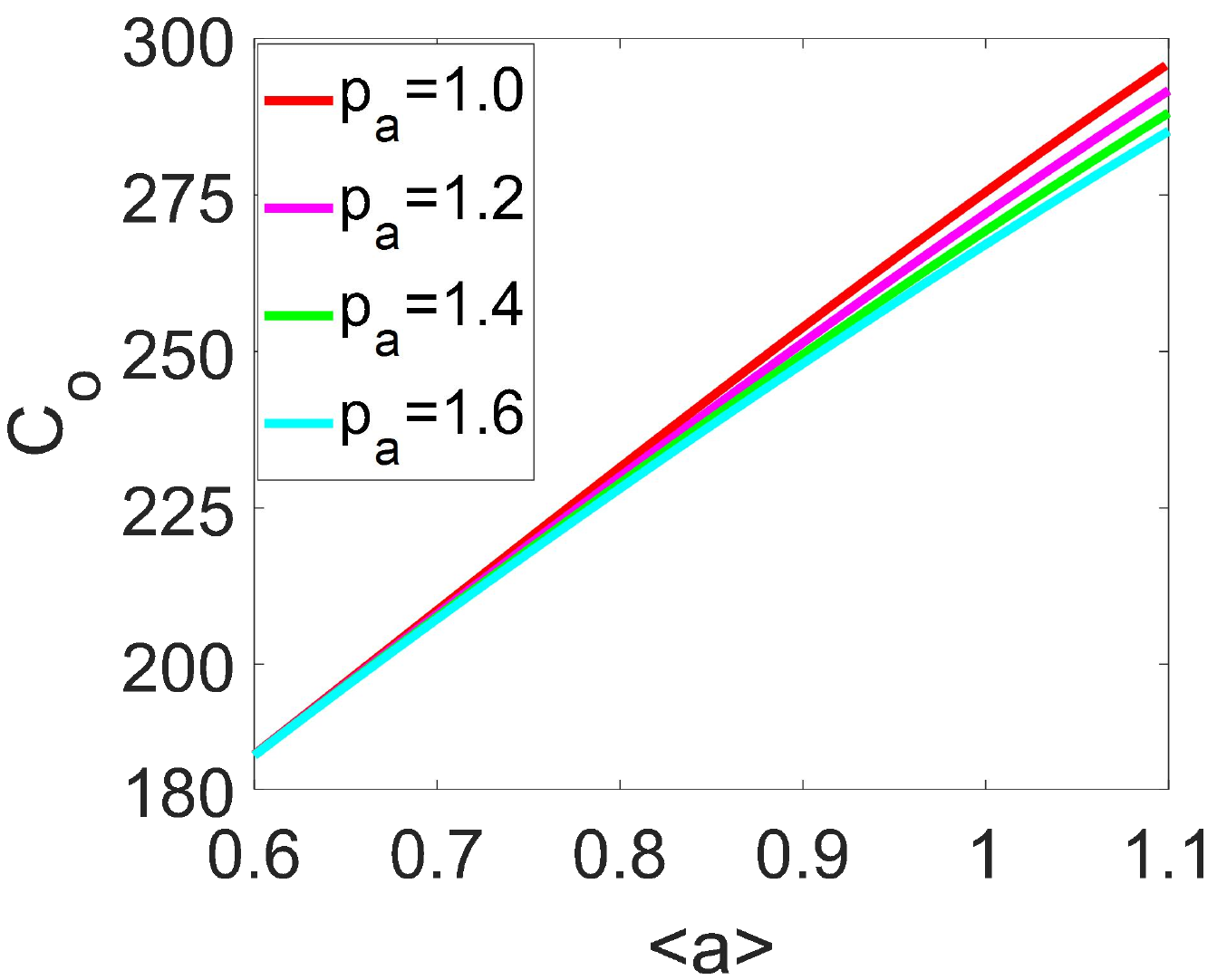}
\caption{\label{fig:6c} Variations in the coherence versus the average value of the scale factor $\langle a\rangle$. The horizontal and vertical axes represent $\langle a\rangle$ and coherence, respectively. The parameters are taken as: $\alpha=1$, $V_{0}=\pi$.}
\end{figure}
\begin{figure}[tbp]
\centering
\includegraphics[width=7cm]{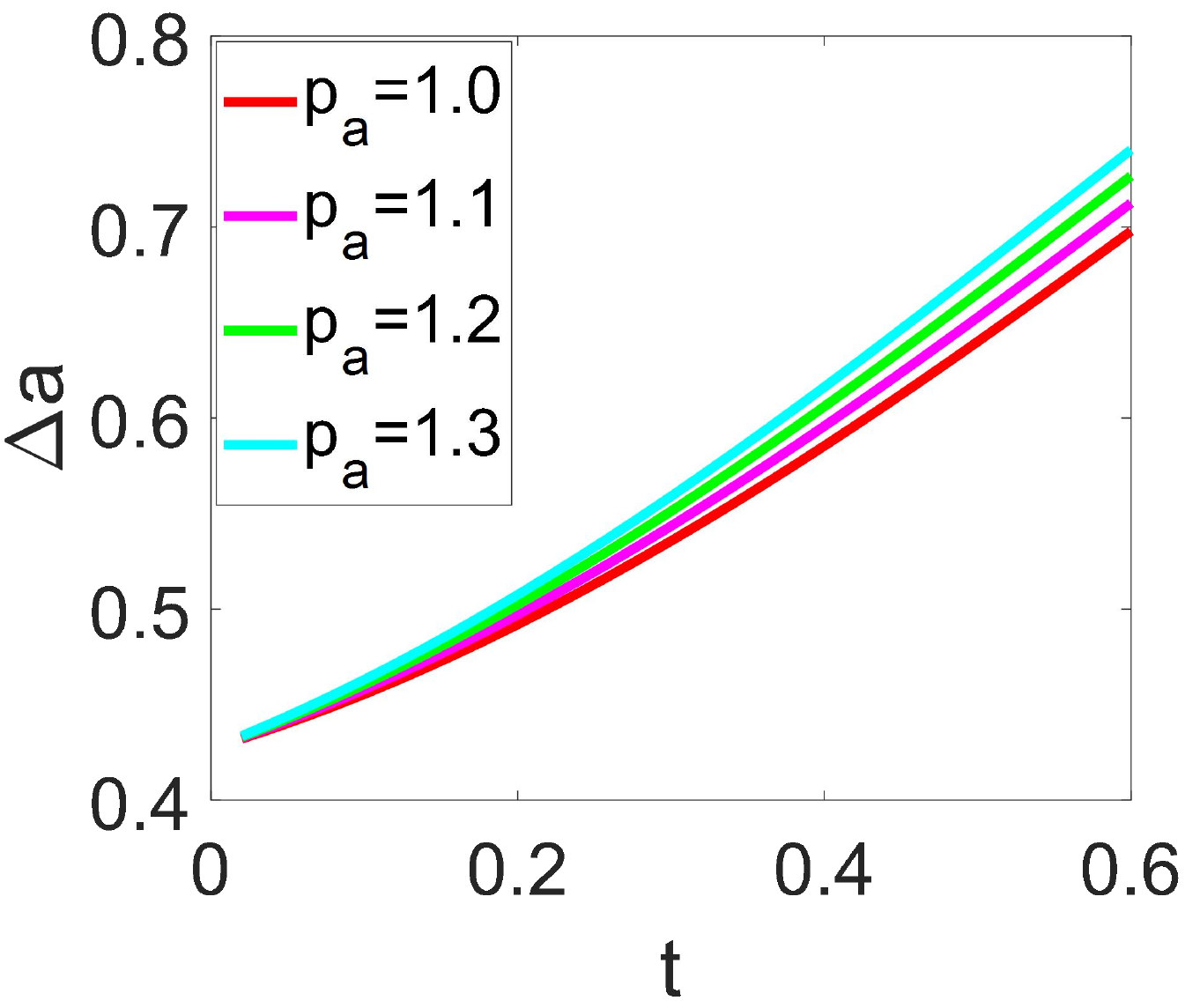}
\caption{\label{fig:6b} Variations in the variance. The horizontal and vertical axes represent the time variable and variance, respectively. The parameters are taken as: $\alpha=1$, $V_{0}=\pi$.}
\end{figure}
\begin{figure}[tbp]
\centering
\includegraphics[width=7cm]{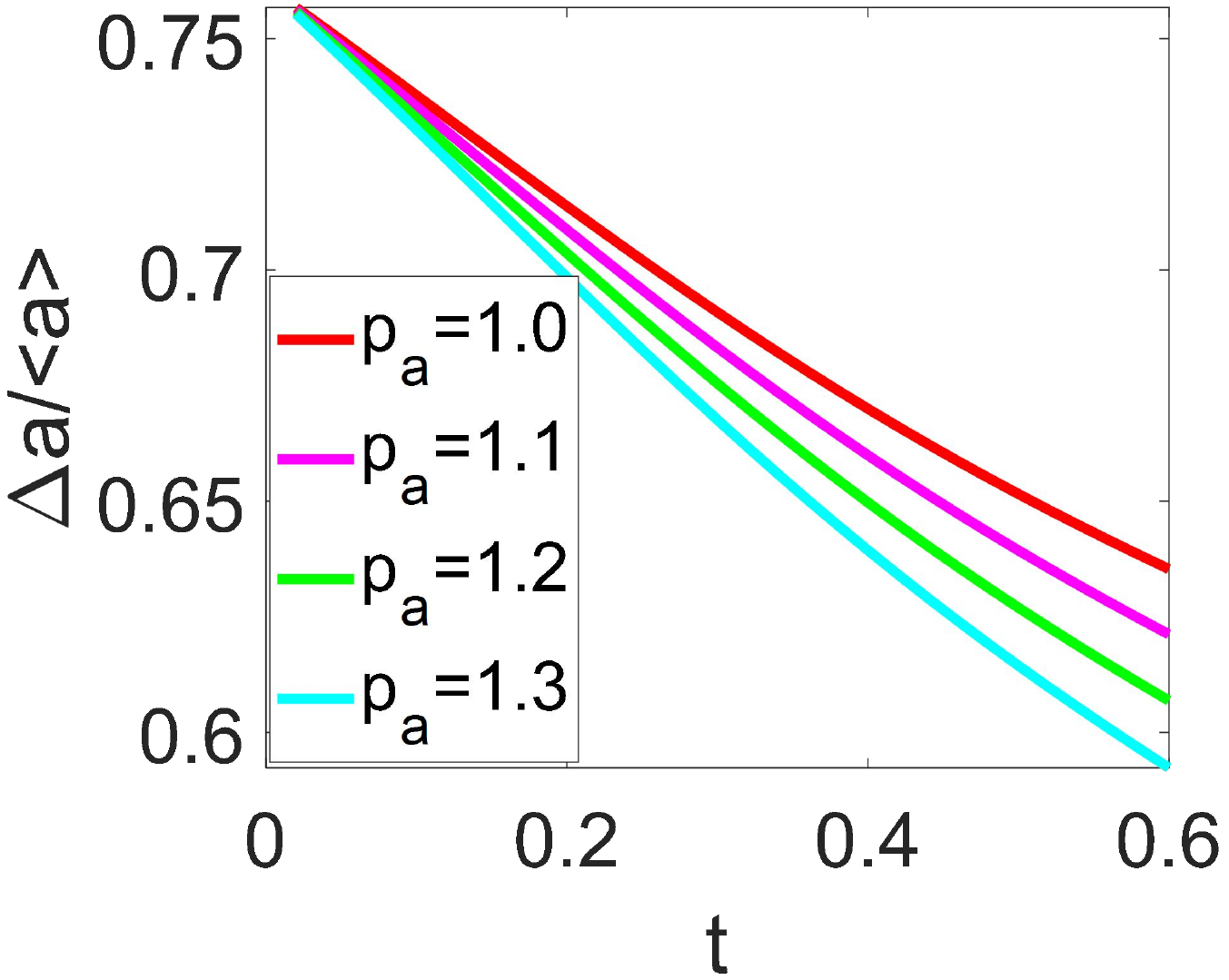}
\caption{\label{fig:6d} Variations in the relative quantum fluctuation. The horizontal and vertical axes represent the time variable and relative fluctuation, respectively. The parameters are taken as: $\alpha=1$, $V_{0}=\pi$.}
\end{figure}
\begin{figure}[tbp]
\centering
\includegraphics[width=7cm]{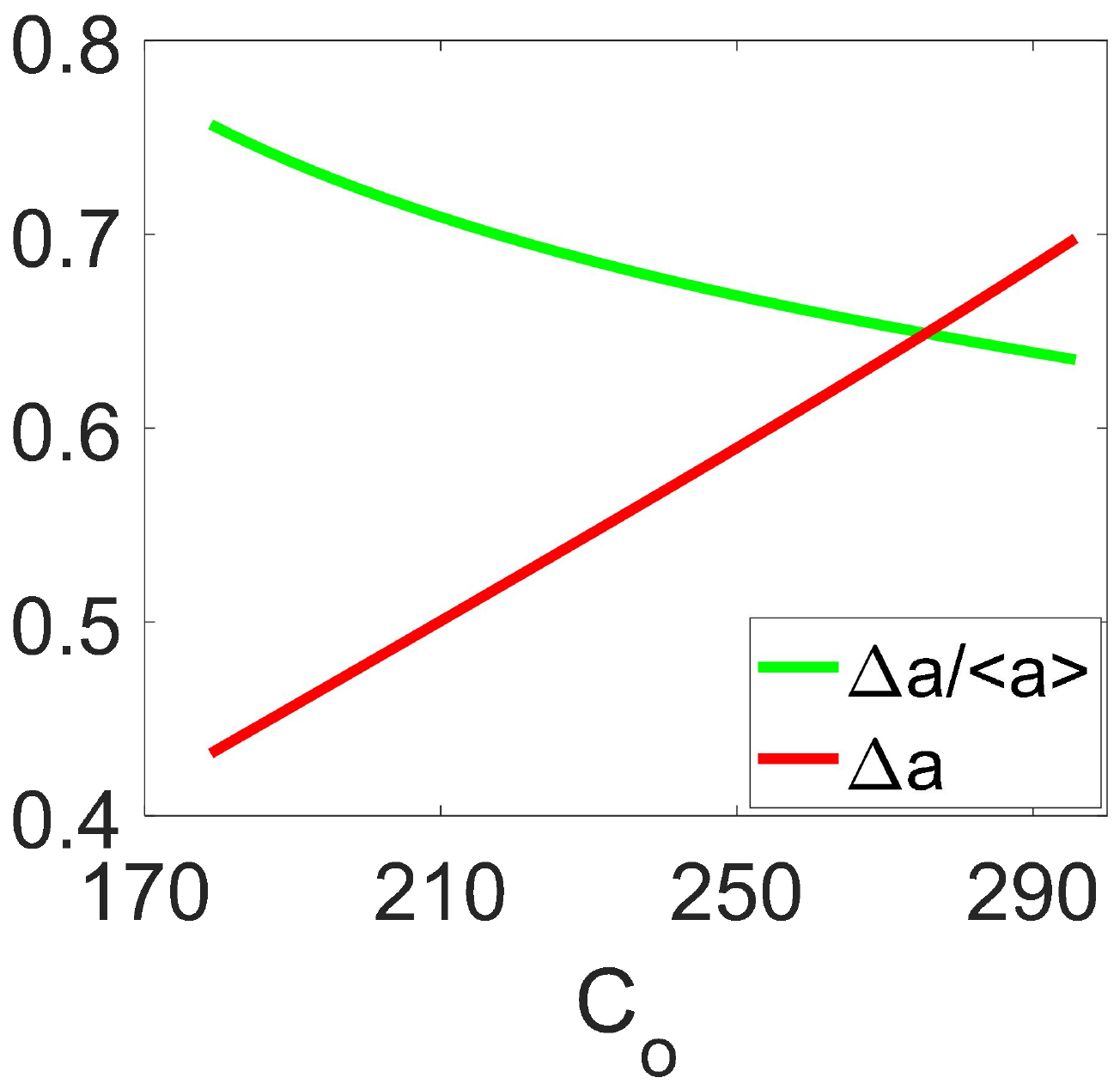}
\caption{\label{fig:6f} Variations in the quantum fluctuation versus the coherence. The horizontal axis represents the coherence.  The parameters are taken as: $p_{a}=1$, $\alpha=1$, $V_{0}=\pi$.}
\end{figure}

According to Eqs.~\eqref{eq:2.19} and \eqref{eq:3.26}, in the conformal time coordinate ($\bm{N}=a$), the Hamiltonian operator of spacetime, and the real massless scalar field are
\begin{equation}
\label{eq:4.43}
\hat{H}_{g}=\frac{-2\pi}{3V_{0}}\hat{\pi}_{a}^{2},
\end{equation}
\begin{equation}
\label{eq:4.44}
\hat{H}_{\phi}=\sum_{\vec{k}}|\vec{k}|A^{\dag}_{\vec{k}}A_{\vec{k}}.
\end{equation}
Assuming that the initial state of the total system can be written as $\rho_{tot}(0)=\rho(0)\otimes\rho_{\phi}(0)$, the similarity to Eq.~\eqref{eq:4.9} provides the reduced density matrix as
\begin{eqnarray}\begin{split}
\label{eq:4.45}
\rho(a^{+}_{N}, a^{-}_{N}) =&\int da_{0}^{\pm}\int da_{1}^{\pm}\cdot\cdot\cdot\int da_{N-1}^{\pm}\langle a^{+}_{N}|e^{-i\hat{H}_{g}\delta t}|a^{+}_{N-1}\rangle\\&\times\langle a^{+}_{N-1}|e^{-i\hat{H}_{g}\delta t}|a^{+}_{N-2}\rangle\cdot\cdot\cdot\langle a^{+}_{0}|\rho(0)|a^{-}_{0}\rangle\\&\times\langle a^{-}_{0}|e^{i\hat{H}_{g}\delta t}|a^{-}_{1}\rangle\cdot\cdot\cdot\langle a^{-}_{N-1}|e^{i\hat{H}_{g}\delta t}|a^{-}_{N}\rangle.
\end{split}
\end{eqnarray}
We assume that the initial state of spacetime is a Gaussian wave packet described by Eq.~\eqref{eq:4.m5}. Then, bringing Eq.~\eqref{eq:4.m5} into Eq.~\eqref{eq:4.45} and integrating over $a_{0}^{\pm},a_{1}^{\pm},...,a_{N-1}^{\pm}$  (taking $\delta t\rightarrow0$) gives
\begin{eqnarray}\begin{split}
\label{eq:4.48}
\rho(a^{+}_{N}, a^{-}_{N}) =&\frac{1}{(\frac{16\pi^{3}t_{N}^{2}}{9V_{0}^{2}\alpha^{2}}+\pi\alpha^{2})^{\frac{1}{2}}} \mathrm{exp}\Big\{\frac{-3iV_{0}}{8\pi t_{N}}\big[(a_{N}^{+})^{2}-(a_{N}^{-})^{2}\big]\Big\}\\&\times\mathrm{exp}\Big\{\frac{-(p_{a}-\frac{3V_{0}a_{N}^{+}}{4\pi t_{N}})^{2}(\frac{1}{2\alpha^{2}}-\frac{3iV_{0}}{8\pi t_{N}})}{\frac{1}{\alpha^{4}}+\frac{9V_{0}^{2}}{16\pi^{2}t_{N}^{2}}}\Big\}\\&\times\mathrm{exp}\Big\{\frac{-(p_{a}-\frac{3V_{0}a_{N}^{-}}{4\pi t_{N}})^{2}(\frac{1}{2\alpha^{2}}+\frac{3iV_{0}}{8\pi t_{N}})}{\frac{1}{\alpha^{4}}+\frac{9V_{0}^{2}}{16\pi^{2}t_{N}^{2}}}\Big\}.
\end{split}
\end{eqnarray}

The diagonal element of the reduced density matrix is then ($a_{N}^{+}=a_{N}^{-}=a$)
\begin{equation}
\label{eq:4.49}
\rho(a,t_{N})=\frac{1}{(\frac{16\pi^{3}t_{N}^{2}}{9V_{0}^{2}\alpha^{2}}+\pi\alpha^{2})^{\frac{1}{2}}}\cdot\mathrm{exp}\Big\{\frac{-(a-\frac{4\pi p_{a}t_{N}}{3V_{0}})^{2}}{\alpha^{2}+\frac{16\pi^{2}t_{N}^{2}}{9V_{0}^{2}\alpha^{2}}}\Big\}.
\end{equation}
It is worthwhile to point out that in the derivation process of the reduced density matrix \eqref{eq:4.48}, we do not introduce any approximation. The reduced density matrix \eqref{eq:4.48} is exact. Replacing the scale factor in \eqref{eq:4.48} by the space coordinate, one can obtain the density matrix corresponding to a free particle with one degree of freedom. Thus all the features of the quantum spacetime in the conformal time coordinate is similar to a free particle.

As the Hamiltonian of the scalar field in the conformal time coordinate is conserved, we denote it by $\langle H_{\phi}\rangle$.  The condition \eqref{eq:2.14} gives
\begin{equation}
\label{eq:4.53}
\mathrm{Tr}(\rho\hat{\pi}_{a}^{2})=p_{a}^{2}+\frac{1}{2\alpha^{2}}=\frac{3V_{0}}{2\pi}\langle H_{\phi}\rangle.
\end{equation}
Thus, the parameters $p_{a}$, $\alpha$, $V_{0}$ and $\langle H_{\phi}\rangle$ are not independent. The evolution of the diagonal element of the reduced density matrix is shown in Fig.~\ref{fig:5}.

In Fig.~\ref{fig:5}, different colors represent different values of the diagonal elements in \eqref{eq:4.49}. The black line represents the classical trajectory of the universe. The evolution of the  wave packet is consistent with the classical trajectory of the universe. Figure~\ref{fig:5} clearly shows that the peak values decreases with time. This shows that the absolute quantum fluctuation becomes more and more important. This figure indicates that the universe grows to become bigger as time goes by. Figure~\ref{fig:6}(a) shows that the coherence monotonically increases with time. Figure~\ref{fig:6c} shows that the bigger the universe is, the bigger the coherence is. Thus in the conformal time coordinate, the heat radiation dominated quantum universe also does not decohere to a classical universe. Figure~\ref{fig:6}(b) and Fig.~\ref{fig:6b} show that the Gibbs entropy and the absolute quantum fluctuation also increase with time. Noted that Eq.~\eqref{eq:4.49} is the Gaussian distribution. Thus the analytic expressions for the average value of the scale factor and the absolute quantum fluctuation are $\langle a\rangle=4\pi p_{a}t_{N}/3V_{0}$ and $\triangle a=(\alpha^{2}/2+8\pi^{2}t^{2}_{N}/9V^{2}_{0}\alpha^{2})^{1/2}$, respectively. Both $\langle a\rangle$ and $\triangle a$ monotonically increase with time. In addition, similar analysis with those in Sec.~\ref{sec:4} show that in the conformal time coordinate, the higher initial temperature also corresponds to more rapid variations in these quantities ($\langle a\rangle$, $C_{o}$, $S_{G}$ and $\triangle a$). And for a given size of the universe, the lower temperature corresponds to a more quantum universe. Figure~\ref{fig:6d} shows that the relative quantum fluctuation monotonically decreases with time. In the conformal time coordinate, the analytic expression of the relative quantum fluctuation is $\triangle a/\langle a\rangle=(1/(2p_{a}^{2}\alpha^{2})+9V_{0}^{2}\alpha^{2}/(32\pi^{2}p_{a}^{2}t_{N}^{2}))^{1/2}$.  Figure~\ref{fig:6f} shows that the larger absolute quantum fluctuations are associated with the higher coherence. Comparing these results with those in Sec.~\ref{sec:4}, we can see that all conclusions in the conformal time coordinate are consistent with those in the proper time coordinate. 

In~\cite{AAA}, Ashtekar and other researchers introduced the non-commutativity to measure the quantum nature of a system. The fading of the non-commutativity characteristics the process of the decoherence. Working in the conformal time coordinate and assuming that the matter is the radiation field, they found that the non-commutativity increases with time. The larger the universe, the larger the non-commutativity is~\cite{AAA,JDE}. This is consistent with our results.

\subsection{Non-relativistic particles dominated universe}
\label{sec:Eb}

\begin{figure}[tbp]
\centering
\includegraphics[width=7.5cm]{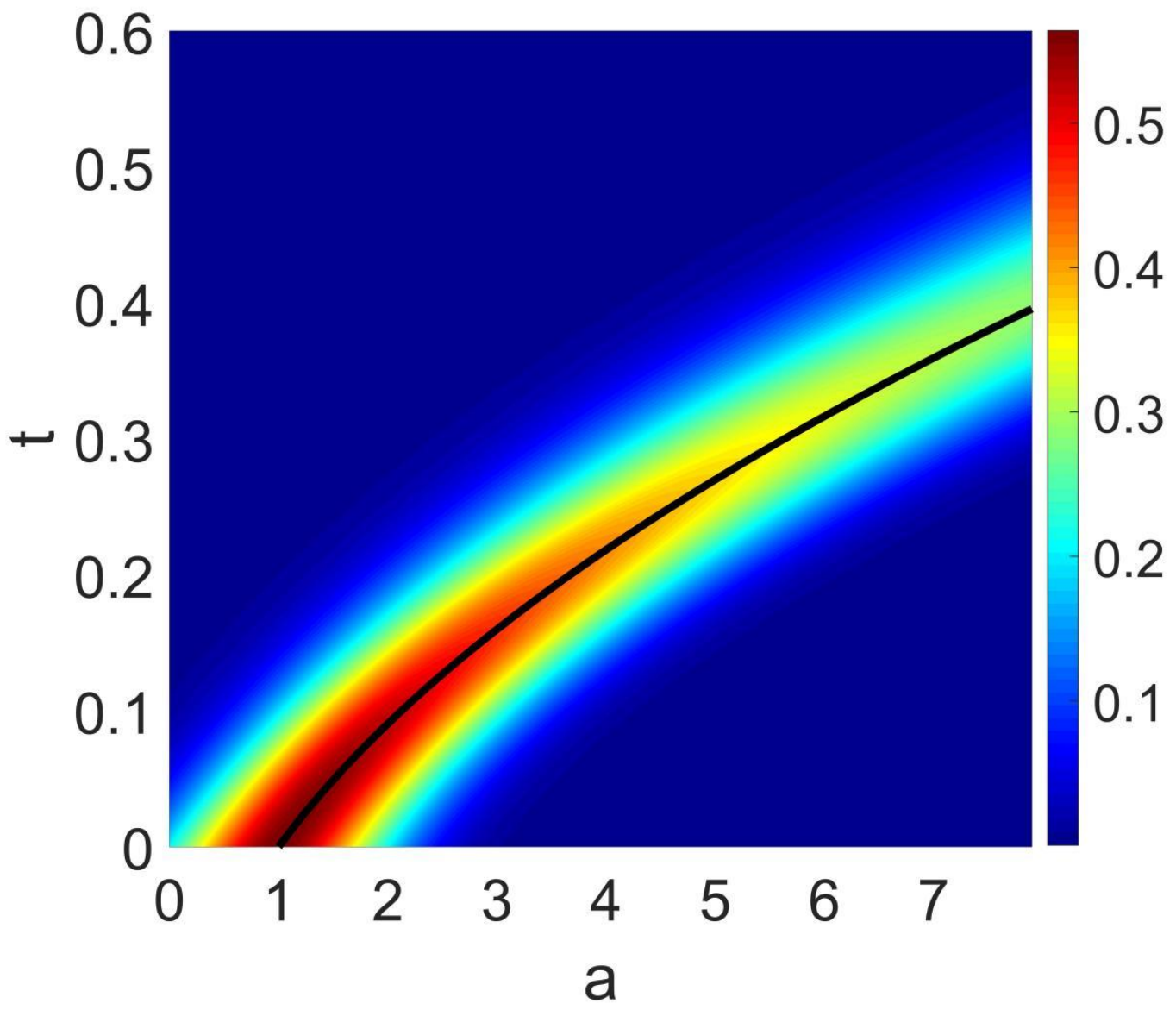}
\caption{\label{fig:7}Variations in the probability distribution of the scale factor. The horizontal and vertical axes represent the scale factor and time variable, respectively. The black curve shows the evolution for the classical trajectory of the universe. The parameters are taken as: $\alpha=1$, $V_{0}=1$, $A_{0}=1$, $mN_{m}=10$.} 
\end{figure}

In the conformal time coordinate, the Hamiltonian operator of spacetime and non-relativistic matter are
\begin{eqnarray}\begin{split}
\label{eq:5.23}
&\hat{H}_{g}=-\frac{2\pi }{3V_{0}}\hat{\pi}_{a}^{2},\\&
\hat{H}_{\phi}=aV_{0}N_{m}m.
\end{split}
\end{eqnarray}
Observing the Hamiltonian operator in Eq. \eqref{eq:5.23}, it is similar to a charged particle
in an uniform electrostatic field. This is also an exactly solvable model~\cite{LS}. Assuming that the initial density matrix of the total system can be written as $\rho_{tot}(0)=\rho(0)\otimes\rho_{\phi}(0)$, similar to Eq.~\eqref{eq:4.9}, the evolution of the reduced density matrix is
\begin{eqnarray}\begin{split}
\label{eq:5.24}
\rho(a^{+}_{N}, a^{-}_{N}) =&\int da_{0}^{\pm}\int da_{1}^{\pm}\cdot\cdot\cdot\int da_{N-1}^{\pm}\langle a^{+}_{N}|e^{-i\hat{H}_{g}\delta t}|a^{+}_{N-1}\rangle\\&\times\langle a^{+}_{N-1}|e^{-i\hat{H}_{g}\delta t}|a^{+}_{N-2}\rangle\cdot\cdot\cdot\\&\times\langle a^{+}_{0}|\rho(0)|a^{-}_{0}\rangle\langle a^{-}_{0}|e^{i\hat{H}_{g}\delta t}|a^{-}_{1}\rangle\cdot\cdot\cdot\\&\times\langle a^{-}_{N-1}|e^{i\hat{H}_{g}\delta t}|a^{-}_{N}\rangle \cdot\mathbf{I_{mN}},
\end{split}
\end{eqnarray}
where
\begin{eqnarray}\begin{split}
\label{eq:5.25}
\mathbf{I_{mN}}\equiv&\mathrm{ Tr}_{\phi}\big\{e^{-i\hat{H}_{\phi}(a_{N}^{+})\delta t}\cdot\cdot\cdot e^{-i\hat{H}_{\phi}(a_{1}^{+})\delta t}\\
&\times\rho_{\phi}(0)e^{i\hat{H}_{\phi}(a_{1}^{-})\delta t}\cdot\cdot\cdot e^{i\hat{H}_{\phi}(a_{N}^{-})\delta t}\big\}
\end{split}
\end{eqnarray}
is the influence functional. All particles are assumed to be in the ground state (kinetic energy is zero). Bringing the Hamiltonian operator $\hat{H}_{\phi}$ of Eq.~\eqref{eq:5.23} into Eq.~\eqref{eq:5.25} gives the influence functional as
\begin{eqnarray}\begin{split}
\label{eq:5.26}
\mathbf{I_{mN}}&=\mathrm{Tr}_{\phi}\Big\{\mathrm{exp}\big[-i\delta t mV_{0}N_{m} \sum_{n=1}^{N}(a_{n}^{+}-a_{n}^{-})\big]\rho_{\phi}(0)\Big\}\\&=\mathrm{exp}\Big\{-i\delta t mV_{0}N_{m}\sum_{n=1}^{N}(a_{n}^{+}-a_{n}^{-})\Big\}.
\end{split}
\end{eqnarray}

We assume that the initial state of spacetime is a Gaussian wave packet as
\begin{equation}\begin{split}
\label{eq:5.28}
\rho(a^{+}_{0}, a^{-}_{0})=&\frac{1}{(\pi\alpha^{2})^{\frac{1}{2}}}\mathrm{exp}\big\{-ip_{a}(a_{0}^{+}-a_{0}^{-})\\&-\frac{1}{2\alpha^{2}}\big((a_{0}^{+}-A_{0})^{2}
+(a_{0}^{-}-A_{0})^{2}\big)\big\},
\end{split}
\end{equation}
Bringing Eqs.~\eqref{eq:5.26} and \eqref{eq:5.28} into Eq.~\eqref{eq:5.24} completes the integrals over these variables $(a_{0}^{\pm}, a_{1}^{\pm},...,a_{N-1}^{\pm})$ (taking $\delta t\rightarrow0$) to finally obtain
\begin{eqnarray}\begin{split}
\label{eq:5.29}
\rho(a^{+}_{N}, a^{-}_{N}) = &\frac{1}{(\frac{16\pi^{3}t_{N}^{2}}{9V_{0}^{2}\alpha^{2}}+\pi\alpha^{2})^{\frac{1}{2}}}\cdot\mathrm{exp}\Big\{\frac{-3iV_{0}}{8\pi t_{N}}\big[(a_{N}^{+})^{2}-(a_{N}^{-})^{2}\big]+(\frac{3iA_{0}}{4\pi t_{N}}-\frac{i}{2}mN_{m}t_{N})V_{0}(a_{N}^{+}-a_{N}^{-})\Big\}
\\&\times\mathrm{exp}\Big\{\frac{-(p_{a}-\frac{3V_{0}a_{N}^{+}}{4\pi t_{N}}+\frac{3V_{0}A_{0}}{4\pi t_{N}}+\frac{1}{2}mV_{0}N_{m}t_{N})^{2}(\frac{1}{2\alpha^{2}}-\frac{3iV_{0}}{8\pi t_{N}})}{\frac{1}{\alpha^{4}}+\frac{9V_{0}^{2}}{16\pi^{2}t_{N}^{2}}}\Big\}\\&
\times\mathrm{exp}\Big\{\frac{-(p_{a}-\frac{3V_{0}a_{N}^{-}}{4\pi t_{N}}+\frac{3V_{0}A_{0}}{4\pi t_{N}}+\frac{1}{2}mV_{0}N_{m}t_{N})^{2}(\frac{1}{2\alpha^{2}}+\frac{3iV_{0}}{8\pi t_{N}})}{\frac{1}{\alpha^{4}}+\frac{9V_{0}^{2}}{16\pi^{2}t_{N}^{2}}}\Big\}.
\end{split}
\end{eqnarray}
In the derivation process of the reduced density matrix \eqref{eq:5.29}, we do not introduce any approximation. Then the diagonal element of the reduced density matrix becomes ($a_{N}^{+}=a_{N}^{-}=a$)
\begin{equation}
\label{eq:5.31}
\rho(a,t_{N})=\frac{\mathrm{exp}\Big\{\frac{-(a-\frac{4\pi p_{a}t_{N}}{3V_{0}}-\frac{2}{3}\pi mN_{m}t_{N}^{2}-A_{0})^{2}}{\alpha^{2}+\frac{16\pi^{2}t_{N}^{2}}{9V_{0}^{2}\alpha^{2}}}\Big\}}{(\frac{16\pi^{3}t_{N}^{2}}{9V_{0}^{2}\alpha^{2}}+\pi\alpha^{2})^{\frac{1}{2}}}.
\end{equation}

\begin{figure}[tbp]
\centering
\includegraphics[width=14cm]{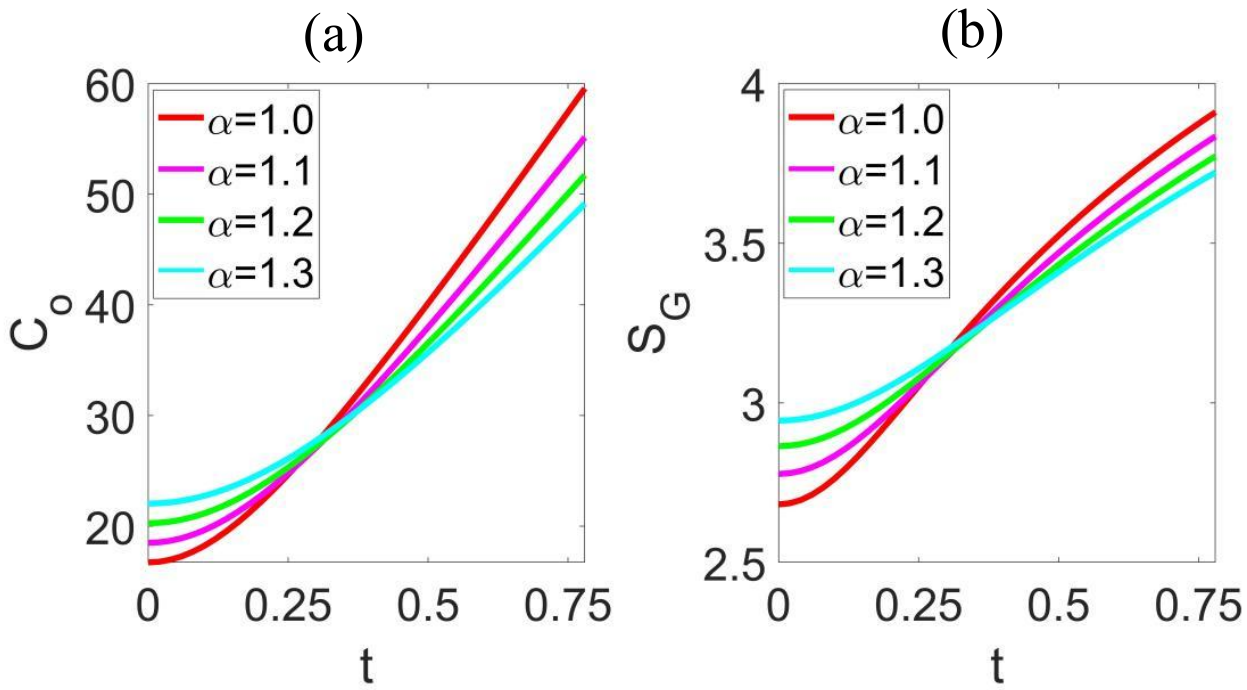}
\caption{\label{fig:8} Variations in the (a) coherence and (b) Gibbs entropy as functions of the time variable. The parameters are taken as: $V_{0}=1$, $A_{0}=1$, $mN_{m}=10$.}
\end{figure}

\begin{figure}[tbp]
\centering
\includegraphics[width=7cm]{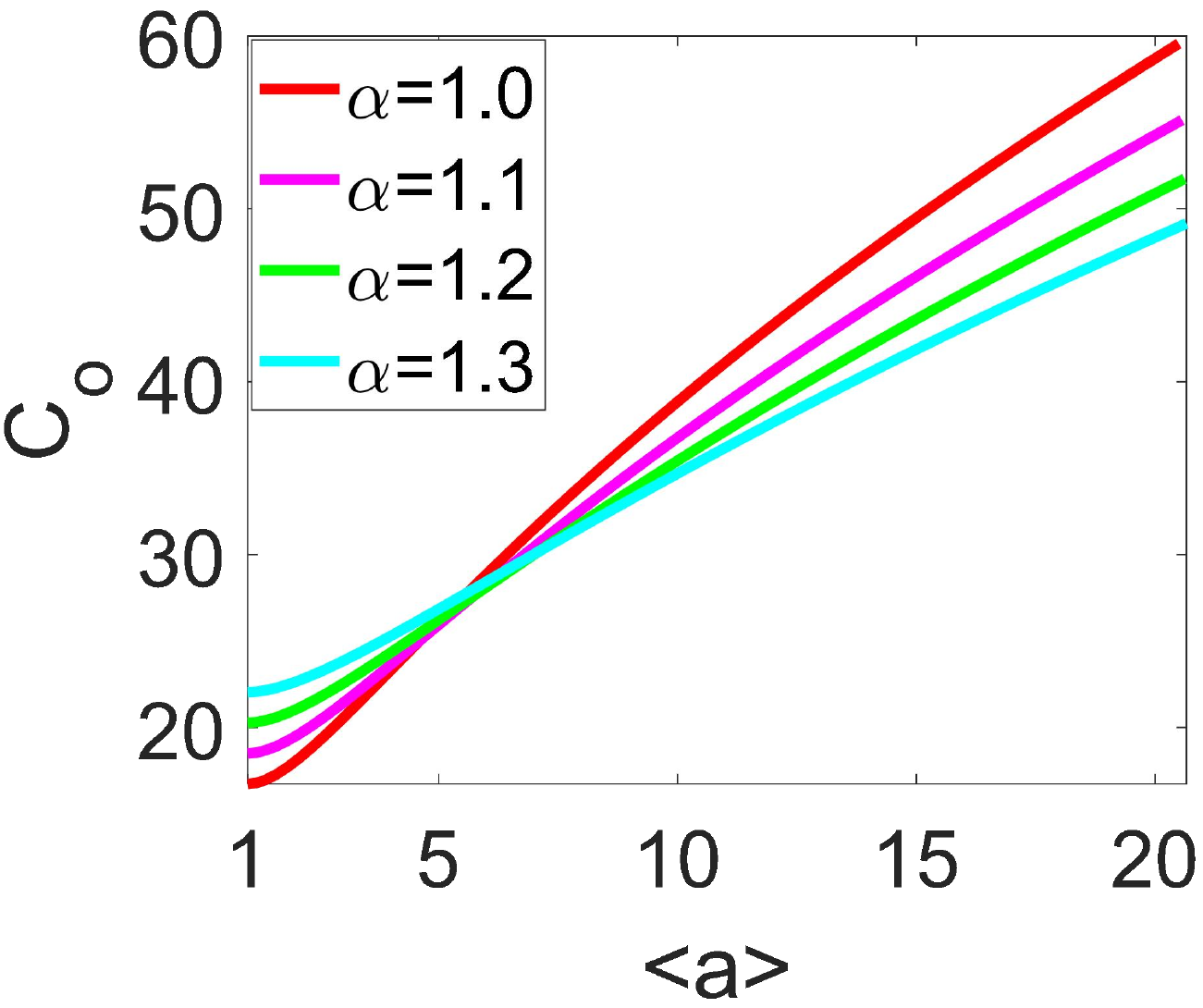}
\caption{\label{fig:8a} Variations in the coherence versus the average value of the scale factor $\langle a\rangle$. The parameters are taken as: $V_{0}=1$, $A_{0}=1$, $mN_{m}=10$.}
\end{figure}

\begin{figure}[tbp]
\centering
\includegraphics[width=7cm]{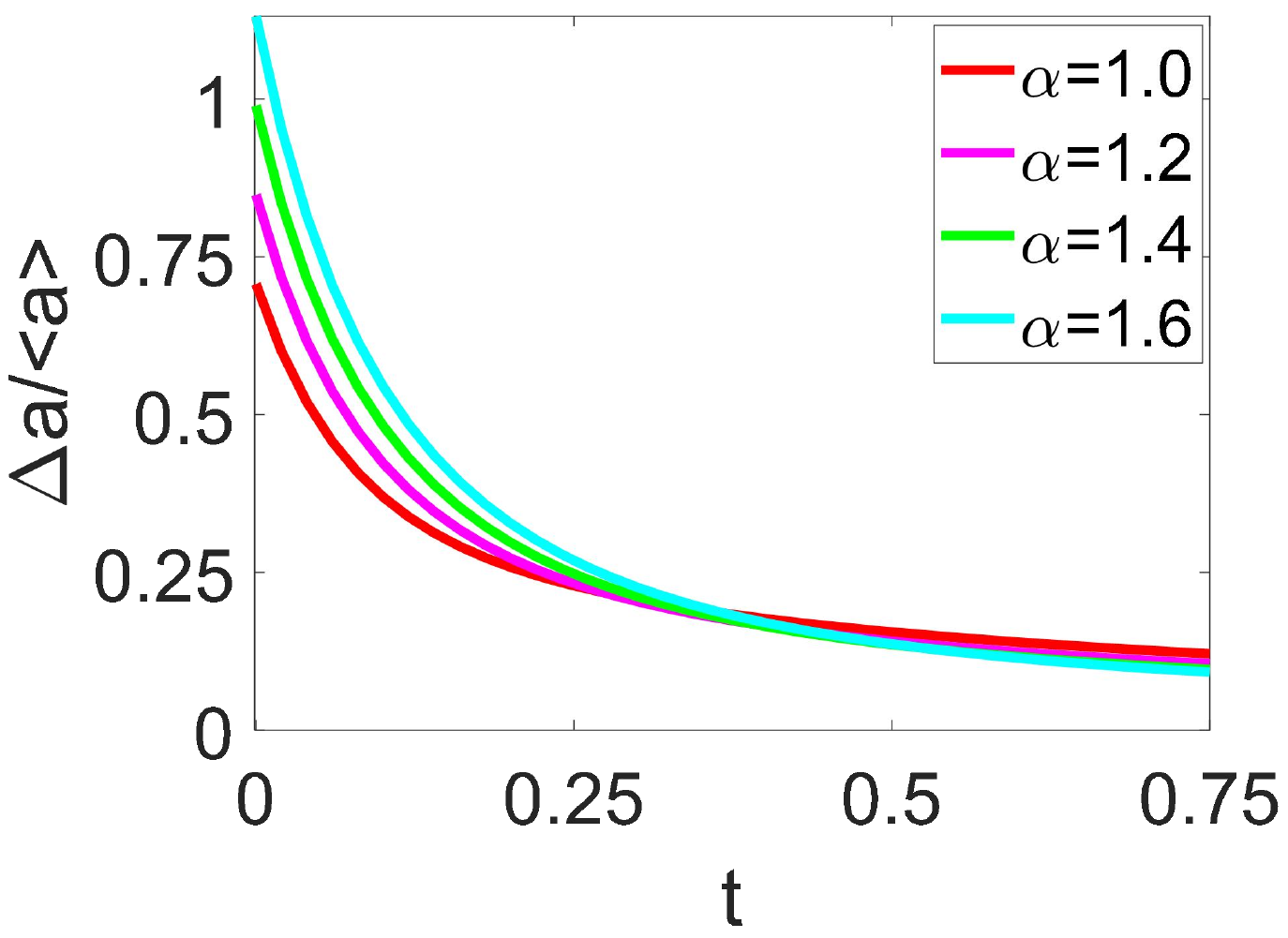}
\caption{\label{fig:8b} Variations in the relative quantum fluctuation. The horizontal and vertical axes represent the time variable and relative fluctuation, respectively. The parameters are taken as: $V_{0}=1$, $A_{0}=1$, $mN_{m}=10$.}
\end{figure}

\begin{figure}[tbp]
\centering
\includegraphics[width=7cm]{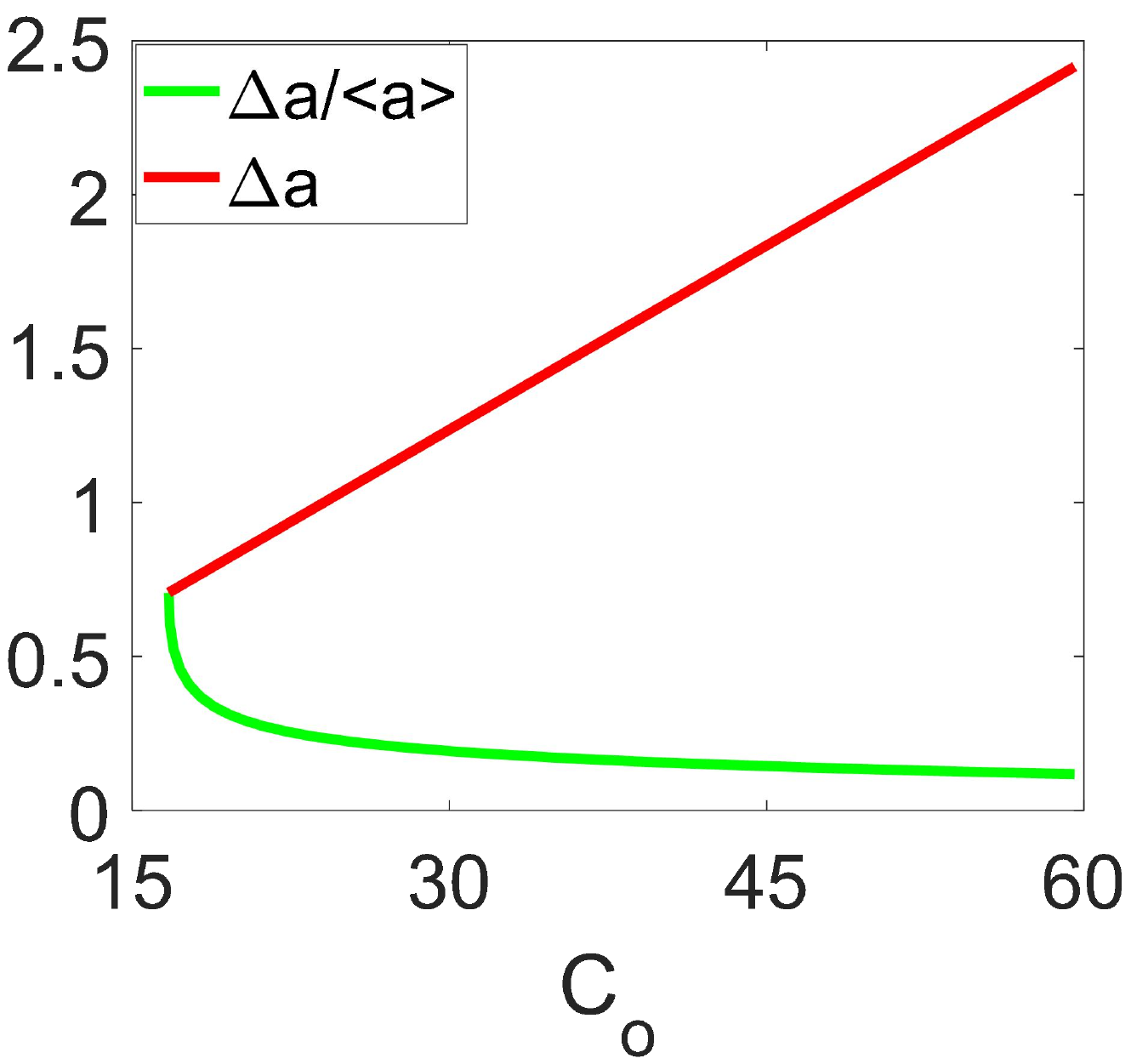}
\caption{\label{fig:8c} Variations in the quantum fluctuation versus the coherence. The horizontal axis represents the coherence.  The parameters are taken as: $\alpha=1$, $V_{0}=1$, $A_{0}=1$, $mN_{m}=10$.}
\end{figure}

The condition of Eq.~\eqref{eq:2.14} gives rise to the following constraint
\begin{equation}
\label{eq:5.36}
p_{a}^{2}+\frac{1}{2\alpha^{2}}=\frac{3m}{2\pi}A_{0}N_{m}V_{0}^{2}.
\end{equation}
Variations in the diagonal element are shown in Fig.~\ref{fig:7}. From this figure, we can see that the classical trajectory is consistent with the evolution of the wave packet. Figure~\ref{fig:8} shows that the coherence and the Gibbs entropy monotonically increase with time. Figure~\ref{fig:8a} shows that the bigger the universe is, the bigger the coherence is. Therefore in the conformal time coordinate, the non-relativistic particles dominated quantum universe also can not decohere to a classical universe. From Eq.~\eqref{eq:5.31}, one can obtain the absolute quantum fluctuation as $\triangle a=(\alpha^{2}/2+8\pi^{2}t^{2}_{N}/9V^{2}_{0}\alpha^{2})^{1/2}$. Thus the trend of the variations in the absolute quantum fluctuation are the same with that in Fig~\ref{fig:6b}. However, Fig.~\ref{fig:8b} shows that the relative quantum fluctuation monotonically decreases with time. And the variation rate of the relative quantum fluctuation gradually becomes slower. Figure~\ref{fig:8c} shows that the larger absolute quantum fluctuations are associated with the higher coherence.

To sum up, we have studied the evolution of the heat radiation or the non-relativistic particle dominated universe both in the proper time coordinate and in the conformal time coordinate. In different cases, the classical trajectory of the universe is consistent with the evolution of the wave packet. The coherence increases with time, thus the quantum universe can not decohere to a classical one. Comparing Fig.~\ref{fig:2}, Fig.~\ref{fig:2b}, Fig.~\ref{fig:4}, Fig.~\ref{fig:4b}, Fig.~\ref{fig:6}, Fig.~\ref{fig:6b} and Fig.~\ref{fig:8}, we find that in different cases, the variations of the absolute quantum fluctuation and the Gibbs entropy  have the similar trend with the variations of the coherence. This is the characteristic of the evolution of the Gaussian-like quantum state. All these quantities are monotonically increase with time in different scenarios. In the conformal time coordinate, we have not introduced any approximation. Thus the results in conformal time coordinate are exact.


\section{Dark energy dominated evolution of the quantum universe}

We have not yet quite understood the nature of the dark energy. One way to describe the dark energy is the cosmological constant. The action of the dark energy can be characterized by~\cite{HM}
\begin{equation}
\label{eq:6.1}
S_{\Lambda}=-\frac{1}{8\pi}\int dx^{4}\sqrt{-g}\Lambda,
\end{equation}
where $\Lambda$ represents the cosmological constant. In this work, we constrain that $\Lambda>0$. If the metric of the flat FRW spacetime is defined by \eqref{eq:2.15}, then the Hamiltonian of the dark energy is given by
\begin{equation}
\label{eq:6.2}
H_{\Lambda}=\frac{\Lambda}{8\pi}V_{0}\bm{N}a^{3}.
\end{equation}
Again, the total Hamiltonian $H_{tot}=H_{g}+H_{\Lambda}$ has the global scaling symmetry. The coordinate volume $V_{0}$ can be viewed as a global scaling factor. Thus the actual value of  $V_{0}$ is not important.

The dark energy is a specific type of (or a class of) matter, one can represent it by the cosmological constant. The only possible observable degree of freedom of the total system considered here is the scale factor.

In the proper time coordinate or the conformal time coordinate, it is not convenient to study the evolution of the dark energy dominated quantum universe by the way of the path integral. One often uses the gauge condition $\bm{N}=1/a$~\cite{JF1,GN}, then the metric \eqref{eq:2.15} becomes
\begin{equation}
\label{eq:6.3}
ds^{2}=\frac{1}{a^{2}}dt^{2}-a^{2}(dx^{2}+dy^{2}+dz^{2}).
\end{equation}
Introducing $q=a^{2}$, then the metric \eqref{eq:6.3} becomes~\cite{JF1,GN}
\begin{equation}
\label{eq:6.4}
ds^{2}=\frac{1}{q}dt^{2}-q(dx^{2}+dy^{2}+dz^{2}).
\end{equation}
The Hamiltonian of the spacetime and the dark energy are then given as
\begin{equation}
\label{eq:6.5}
H_{g}=-\frac{3V_{0}}{32\pi}\dot{q}^{2}=-\frac{8\pi}{3V_{0}}\pi_{q}^{2},
\end{equation}
\begin{equation}
\label{eq:6.6}
H_{\Lambda}=\frac{V_{0}\Lambda}{8\pi}q.
\end{equation}
In Eq.~\eqref{eq:6.5}, $\pi_{q}$ represents the conjugate momentum of the canonical variable $q$,
\begin{equation}
\label{eq:6.7}
\pi_{q}=-\frac{3V_{0}}{16\pi}\dot{q}.
\end{equation}

Observing the Hamiltonian operator \eqref{eq:6.5} and \eqref{eq:6.6}, the whole system is similar to the case \eqref{eq:5.23} discussed  in the Sec.~\ref{sec:Eb}. We can use the same method as the Sec.~\ref{sec:Eb} to solve it. Assuming that the initial state of the spacetime is given by
\begin{equation}\begin{split}
\label{eq:6.8}
\rho(q_{0}^{+},q_{0}^{-})=&\frac{1}{(\pi\alpha^{2})^{\frac{1}{2}}}\mathrm{exp}\big\{-ip_{q}(q_{0}^{+}-q_{0}^{-})\\&-\frac{1}{2\alpha^{2}}\big((q_{0}^{+}-Q_{0})^{2}
+(q_{0}^{-}-Q_{0})^{2}\big)\big\},
\end{split}
\end{equation}
where $Q_{0}=\mathrm{Tr}(q\rho(0))$ represents the average value of the scale factor $q$ at the initial time. According to \eqref{eq:6.6}, $Q_{0}$  being equal to zero means that the Hamiltonian of the dark energy is zero. In order to ensure the Hamiltonian of the dark energy to be non-zero, we set $Q_{0}$ as a finite positive value. The condition \eqref{eq:2.14} gives rise to
\begin{equation}
\label{eq:6.9}
 \frac{8\pi}{3V_{0}}(p_{q}^{2}+\frac{1}{2\alpha^{2}})=\frac{\Lambda}{8\pi}V_{0}Q_{0}.
\end{equation}

Noted that $q\geq 0$, or equivalently to say, when $q\leq 0$, all elements of the density matrix are equal to zero at any time. We can choose the parameters to ensure that $\langle q^{+}_{N}|\rho(t_{N})|q^{-}_{N}\rangle =0$ while $q \leq 0$. In this case, the condition $q\geq 0$ is not important thus can be neglected. Then $q$ can take values from $-\infty$ to $+\infty$. This can simplify our analysis. Using the same method as in the Sec.~\ref{sec:Eb}, one can obtain
\begin{eqnarray}\begin{split}
\label{eq:6.10}
\rho(q^{+}_{N},q^{-}_{N}) = &\mathscr{N}\cdot\mathrm{exp}\Big\{\frac{3iV_{0}}{32\pi t_{N}}\big[(q_{N}^{-})^{2}-(q_{N}^{+})^{2}\big]+(\frac{3iQ_{0}}{16\pi t_{N}}-\frac{i\Lambda t_{N}}{16\pi})V_{0}(q_{N}^{+}-q_{N}^{-})\Big\}
\\&\cdot\mathrm{exp}\Big\{\frac{-(p_{q}-\frac{3V_{0}q_{N}^{+}}{16\pi t_{N}}+\frac{3V_{0}Q_{0}}{16\pi t_{N}}+\frac{V_{0}\Lambda t_{N}}{16\pi})^{2}(\frac{1}{2\alpha^{2}}-\frac{3iV_{0}}{32\pi t_{N}})}{\frac{1}{\alpha^{4}}+\frac{9V_{0}^{2}}{16^{2}\pi^{2}t_{N}^{2}}}\Big\}\\&
\cdot\mathrm{exp}\Big\{\frac{-(p_{q}-\frac{3V_{0}q_{N}^{-}}{16\pi t_{N}}+\frac{3V_{0}Q_{0}}{16\pi t_{N}}+\frac{V_{0}\Lambda t_{N}}{16\pi})^{2}(\frac{1}{2\alpha^{2}}+\frac{3iV_{0}}{32\pi t_{N}})}{\frac{1}{\alpha^{4}}+\frac{9V_{0}^{2}}{16^{2}\pi^{2}t_{N}^{2}}}\Big\}.
\end{split}
\end{eqnarray}
The diagonal element is ($q_{N}^{+}=q_{N}^{-}=q$)
\begin{equation}
\label{eq:6.11}
\rho(q,t_{N})=\frac{\mathrm{exp}\Big\{\frac{-(q-\frac{16\pi p_{q}t_{N}}{3V_{0}}-\frac{\Lambda}{3} t_{N}^{2}-Q_{0})^{2}}{\alpha^{2}+\frac{16^{2}\pi^{2}t_{N}^{2}}{9V_{0}^{2}\alpha^{2}}}\Big\}}{(\frac{16^{2}\pi^{3}t_{N}^{2}}{9V_{0}^{2}\alpha^{2}}+\pi\alpha^{2})^{\frac{1}{2}}}.
\end{equation}

\begin{figure}[tbp]
\centering
\includegraphics[width=7.5cm]{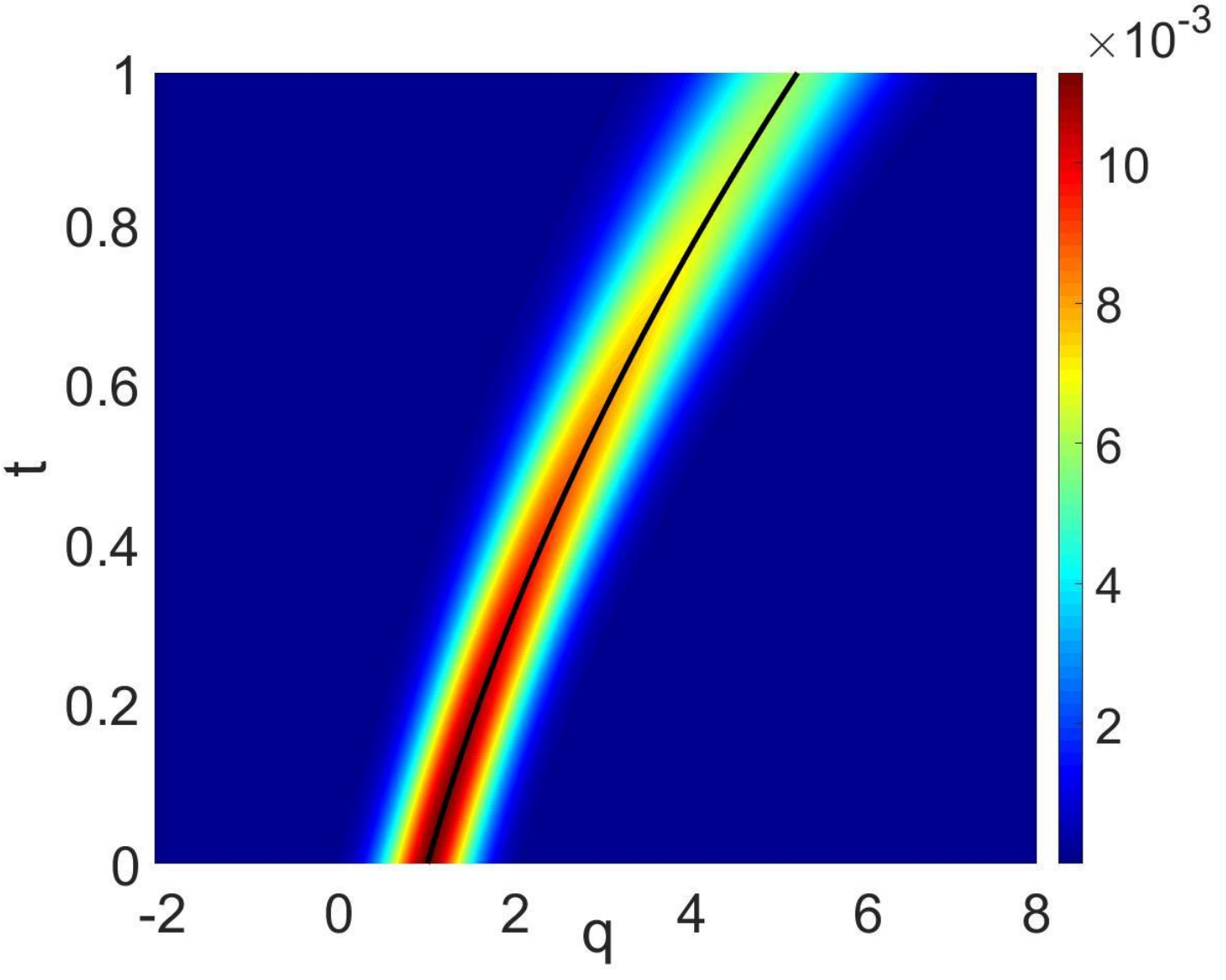}
\caption{\label{fig:10a}Variations in the probability distribution of the scale factor. The horizontal and vertical axes represent the scale factor ($q$) and time variable, respectively. The black curve shows the evolution for the classical trajectory of the universe. The parameters are taken as: $\alpha=0.5$, $V_{0}=40$, $Q_{0}=1$, $\Lambda=5$.} 
\end{figure}

\begin{figure}[tbp]
\centering
\includegraphics[width=13cm]{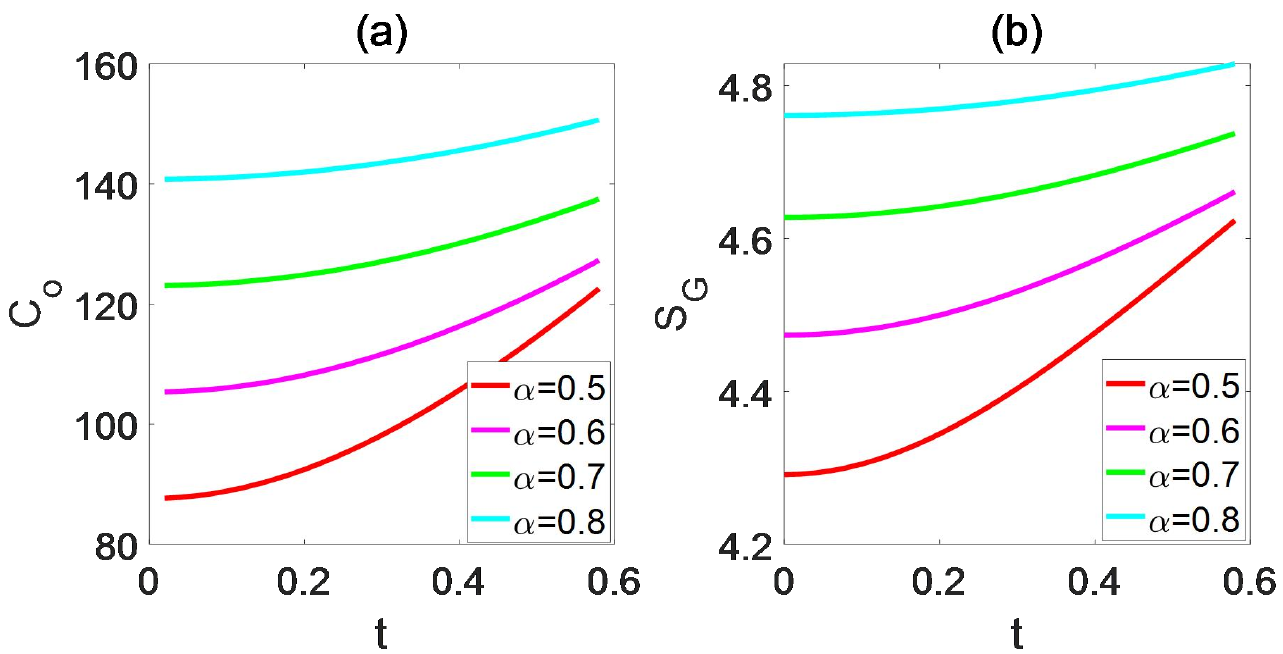}
\caption{\label{fig:11} Variations in the (a) coherence and (b) Gibbs entropy as functions of the time variable. The parameters are taken as: $V_{0}=40$, $Q_{0}=1$, $\Lambda=5$.}
\end{figure}
\begin{figure}[tbp]
\centering
\includegraphics[width=7cm]{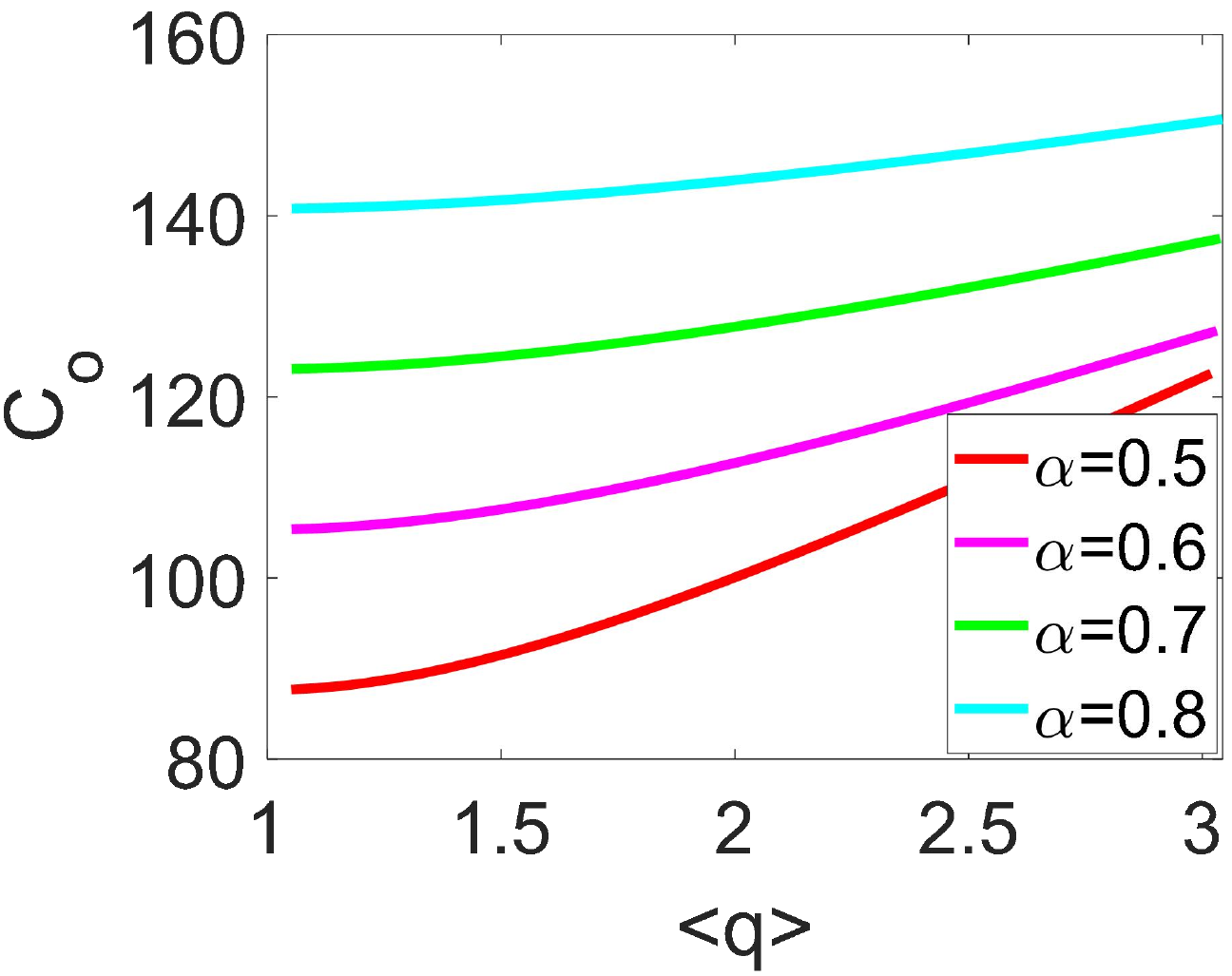}
\caption{\label{fig:5c} Variations in the coherence versus the average value of the scale factor $\langle q\rangle$. The parameters are taken as: $V_{0}=40$, $Q_{0}=1$, $\Lambda=5$.}
\end{figure}
\begin{figure}[tbp]
\centering
\includegraphics[width=7cm]{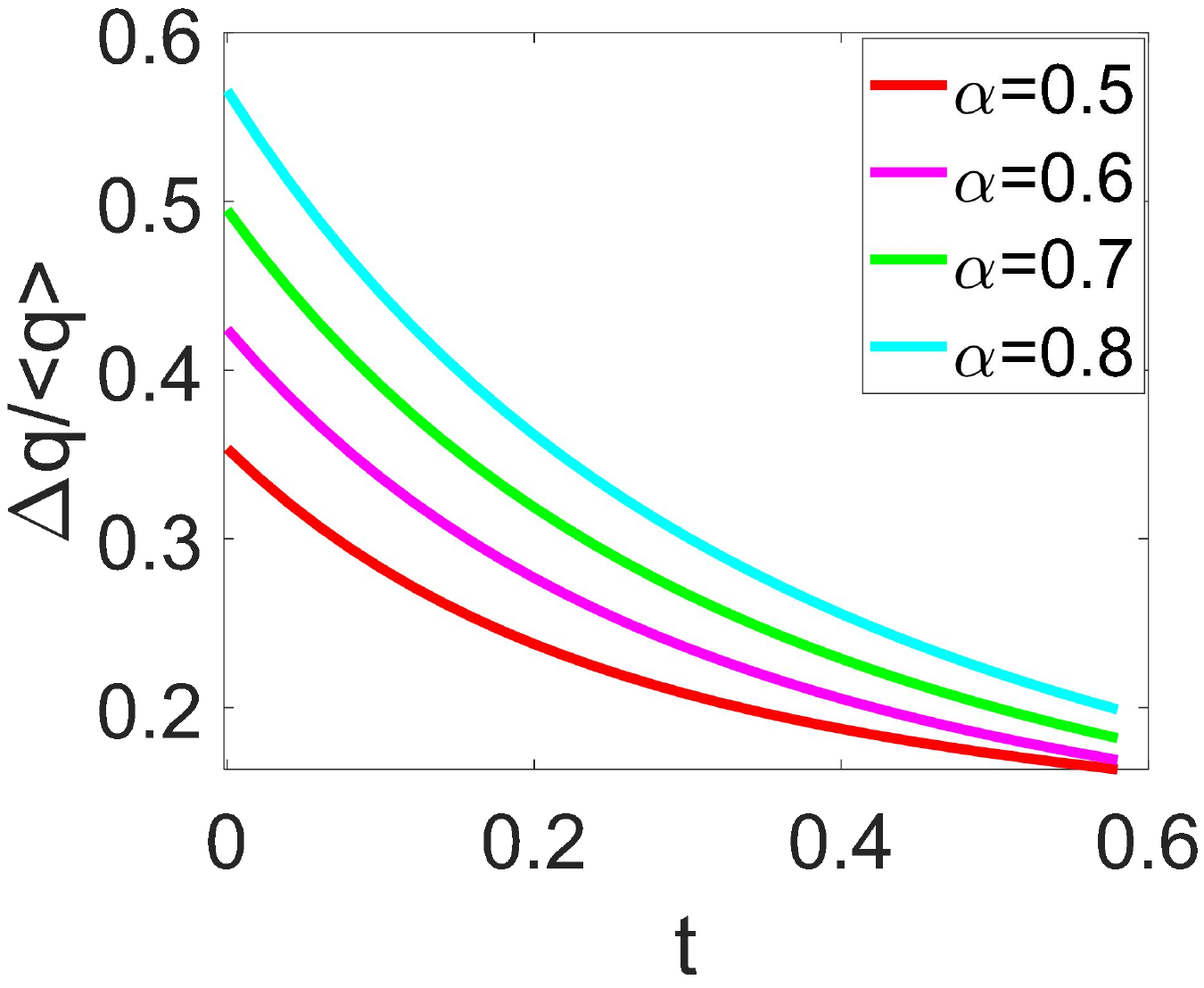}
\caption{\label{fig:5d} Variations in the relative quantum fluctuation. The horizontal and vertical axes represent the time variable and relative fluctuation, respectively. The parameters are taken as: $V_{0}=40$, $Q_{0}=1$, $\Lambda=5$.}
\end{figure}
\begin{figure}[tbp]
\centering
\includegraphics[width=7cm]{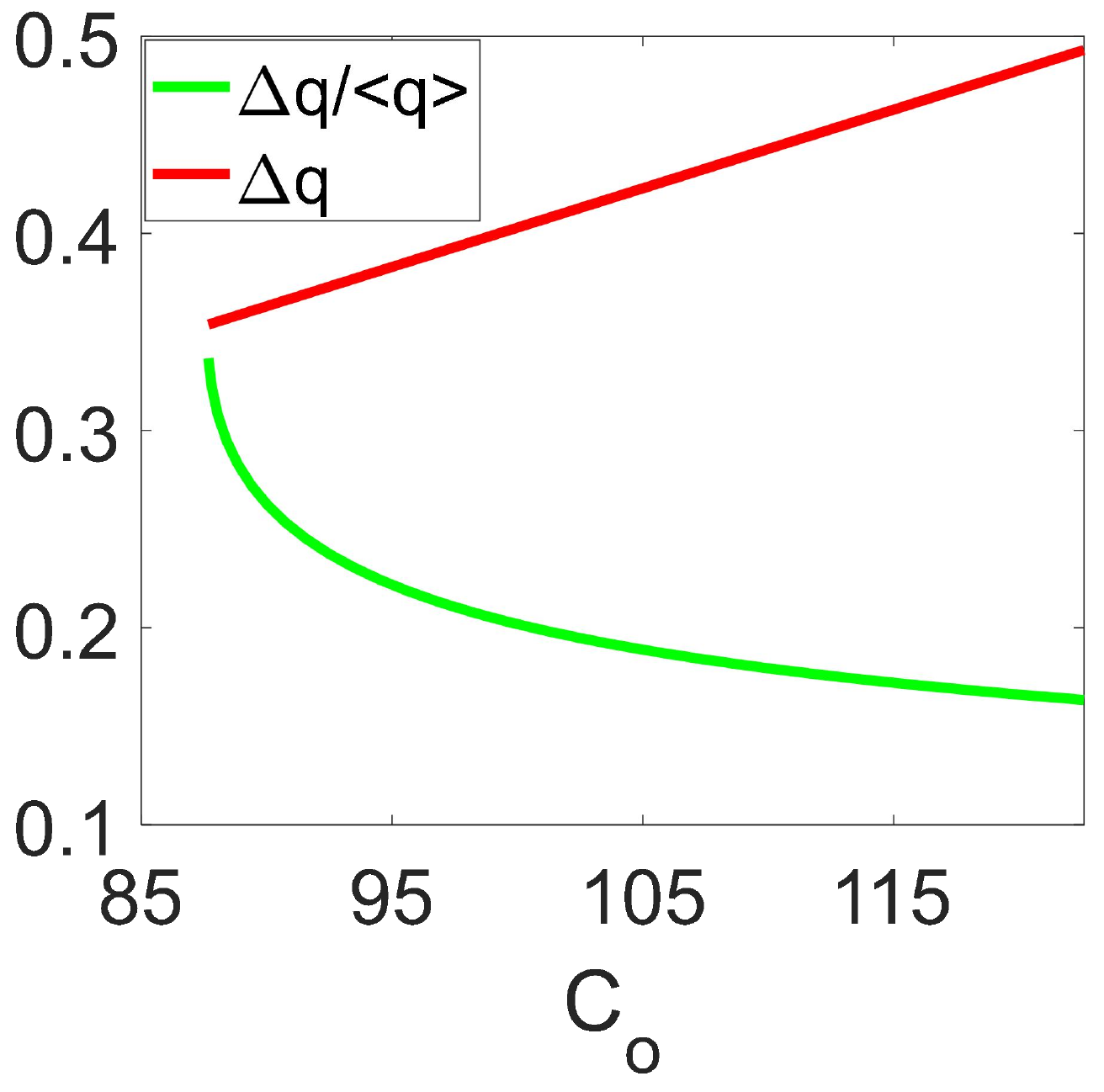}
\caption{\label{fig:5f} Variations in the quantum fluctuation versus the coherence. The horizontal axis represents the coherence.  The parameters are taken as: $\alpha=0.5$, $V_{0}=40$, $Q_{0}=1$, $\Lambda=5$.}
\end{figure}

The evolution of the probability distribution is shown in Fig.~\ref{fig:10a}. In this figure, different colors represent different values of the probabilities. The black curve represents the classical evolution trajectory of the universe driven by the dark energy. Figure~\ref{fig:10a} shows that the classical trajectory of the universe is consistent with the evolution of the wave packet. We point out that although the range of the horizontal axis of Fig.~\ref{fig:10a} is from -2 to 8, we can see from Fig.~\ref{fig:10a} that $\rho(q,t)=0$ when $q<0$. Thus Fig.~\ref{fig:10a} is consistent with the condition $q\geq0$. Figure~\ref{fig:11} shows the variation of the coherence and the Gibbs entropy. From this figure, we can see that the coherence monotonically increases with time. Figure~\ref{fig:5c} shows that the bigger the universe is, the bigger the coherence is. The absolute quantum fluctuation still is $\triangle a=(\alpha^{2}/2+8\pi^{2}t^{2}_{N}/9V^{2}_{0}\alpha^{2})^{1/2}$. It increases with time. Thus the variation of the Gibbs entropy and the absolute quantum fluctuation has the similar trend with the variation of the coherence. Figure~\ref{fig:5d} shows that the relative quantum fluctuation decreases with time. Figure~\ref{fig:5f} shows that the larger absolute quantum fluctuations are associated with the higher coherence. 

In~\cite{HM}, Maeda used the Schr$\mathrm{\ddot{o}}$dinger equation \eqref{eq:1.2} to study the evolution of the quantum universe as driven by a cosmological constant. Maeda shown that the classical trajectory of the universe is consistent with the evolution of the wave packet.  Thus a small quantum universe can evolve to a bigger one. This is consistent with our results. Both results show that the cosmological constant does not give rise to the decoherence of the universe.  We point out that if the matter field has infinity degrees of freedom, it is difficult to solve the Schr$\mathrm{\ddot{o}}$dinger equation \eqref{eq:1.2}. Thus the method in~\cite{HM} is not suitable to study the quantum evolution of the heat radiation dominated universe. However, the method presented in this work can deal with more complicated scenarios.

\section{Quantum transition of the flat FRW universe}

\label{sec:7}
Many studies have considered quantum tunneling in a compact ($k=1$) universe~\cite{JF1,JF2,JF3,JF4,JH,AV1,AV2,AV3,AV4,AV5}; yet, these results can be controversial. Different methods can give various results. For example, the results from the Euclidean path integral can differ from those from the Lorentzian path integral. The Lorentzian path integral quantum cosmology gives various results under differing conditions~\cite{JF1,JDJ} (for example, selecting different integral domains related to the lapse function).

Recently, researchers have realized that the Euclidean path integral may miss some important information~\cite{GN,AM,JF1}. Thus, we use the closed real-time path integral to study quantum transitions of the universe. We fix the lapse function using the gauge condition (such as $\bm{N}=1$). Therefore, the lapse function as a Lagrangian multiplier is not an integral variable (ADM quantization~\cite{WJ}). Such a treatment is often used in quantum cosmology~\cite{JH,AV1,AV2,FT,HM,AOBA,MA}.

One can study the transition probability of the quantum universe by setting the initial state as a wave packet. For a heat radiation dominated universe, assuming the initial state is described by Eq.~\eqref{eq:4.m5}, the transition probability of the universe from the initial state to the final state $|a,t_{N}=1\rangle$  (without loss of generality, we fixed $t_{N}=1$) is
\begin{eqnarray}\begin{split}
\label{eq:7.13}
\rho(a,\alpha_{0})=&\mathscr{N}\frac{|a|}{\big(\frac{1}{10}V_{0}^{2}\alpha_{0}^{4}-2p_{a}^{2}+\frac{9V_{0}^{2}a^{2}}{\frac{32}{5}\pi^{2}V_{0}^{2}\alpha_{0}^{4}-128\pi^{2}p_{a}^{2}}\big)^{\frac{1}{2}}}
\\&\times \mathrm{exp}\Big\{\frac{-\big(p_{a}-\frac{\pi^{2}V_{0}\alpha_{0}^{4}}{15a^{2}}-\frac{3V_{0}a^{2}}{16\pi}\big)^{2}}{\frac{1}{10}V_{0}^{2}\alpha_{0}^{4}-2p_{a}^{2}+\frac{9V_{0}^{2}a^{2}}{\frac{32}{5}\pi^{2}V_{0}^{2}\alpha_{0}^{4}-128\pi^{2}p_{a}^{2}}}\Big\}.
\end{split}
\end{eqnarray}
Substituting Eqs.~\eqref{eq:4.m8} and ~\eqref{eq:4.m11} into Eq.~\eqref{eq:4.m9} and setting $t_{N}=1$ in \eqref{eq:4.m9} transforms Eq.~\eqref{eq:4.m9} into Eq.~\eqref{eq:7.13}.

\begin{figure}[tbp]
\centering
\includegraphics[width=8cm]{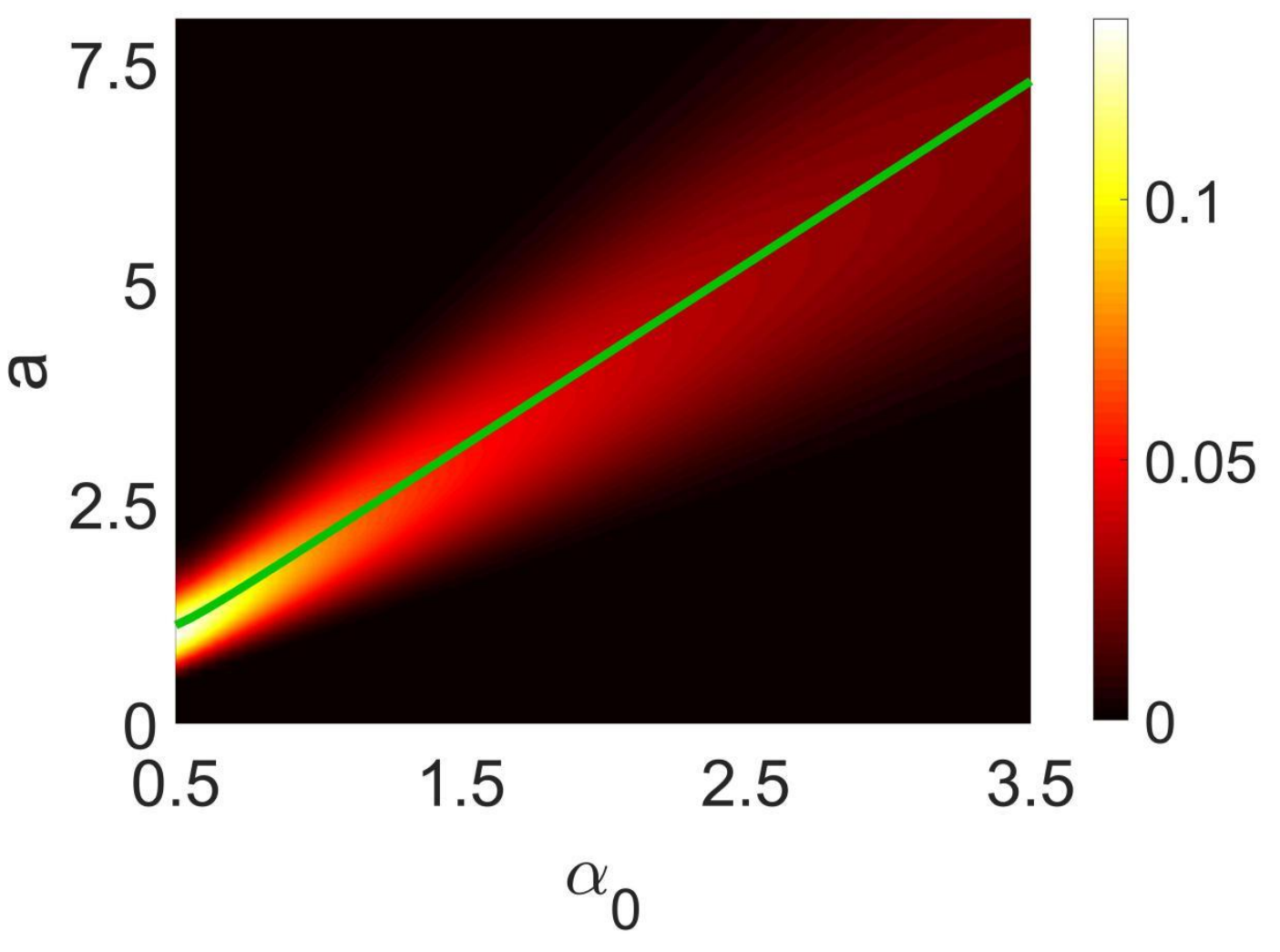}
\caption{\label{fig:9} Variations in the transition probability. The horizontal axis represents the parameter $\alpha_{0}$, and the vertical axis represents the scale factor $a$. The different colors represent various transition probabilities, and the green curve represents the average of the scale factor. The parameters are fixed as $p_{a}=1$ and $V_{0}=50$.}
\end{figure}

Equations~\eqref{eq:7.13} can be diagrammatically presented by Fig.~\ref{fig:9}. In this figure, the different colors represent various transition probabilities, and the green curve represents the average of the scale factor $a$ in time $t_{N}=1$. Our discussions in Sec.~\ref{sec:4}, indicate that $\alpha_{0}\sim T_{0}$ (initial temperature of heat bath). Thus, Fig.~\ref{fig:9} shows the influence of the temperature on the transition probability. As $\alpha_{0}$ increases, the average value of the scale factor (denoted by $\langle a\rangle_{t_{N}=1}$) monotonically increases. This indicates that under higher temperatures, a small quantum universe has a higher chance of a transition to a bigger universe. Figure~\ref{fig:9} shows that $\langle a\rangle_{t_{N}=1}\propto \alpha_{0}$. Recalling that in the classical FRW ($k=0$ and $\mathbf{\emph{N}}=1$) universe, the relation between the scale factor and $\alpha_{0}$ is $a(t)=(32\pi^{3}/90)^{1/4}\alpha_{0}\sqrt{t}$, thus $a(t=1)\propto\alpha_{0}$. Therefore, the conclusion obtained from Fig.~\ref{fig:9} is consistent with the classical result. 

\begin{figure}[tbp]
\centering
\includegraphics[width=14cm]{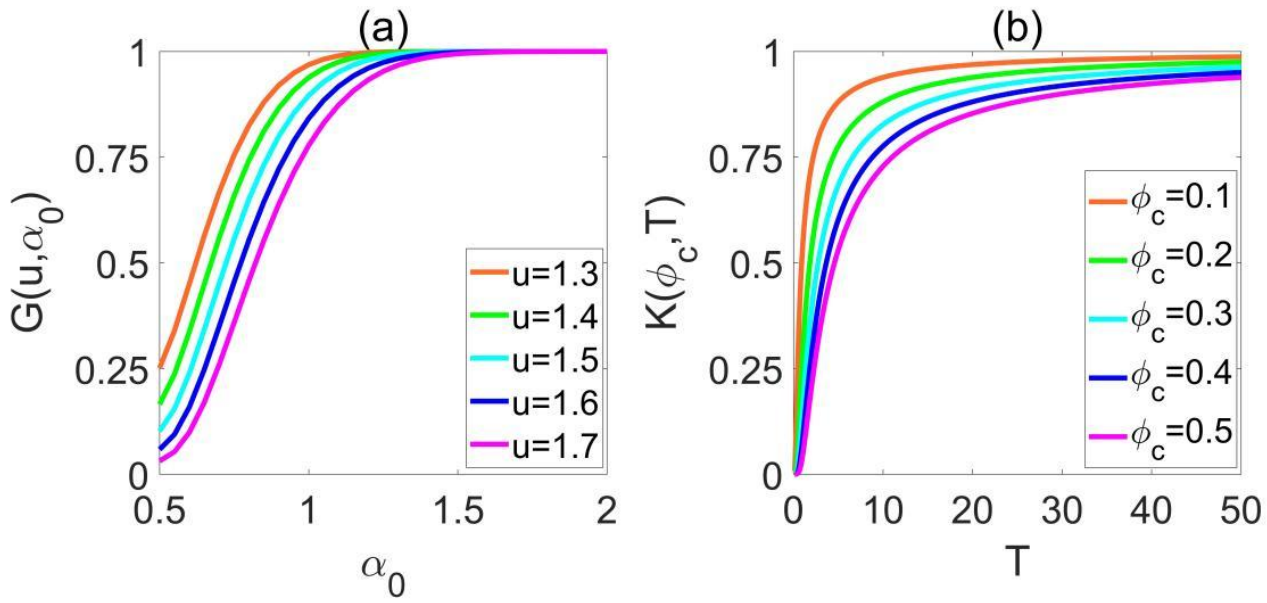}
\caption{\label{fig:10}The left figure represents the total probability of the universe with a scale factor $a\geq u$, where the horizontal axis is the parameter $\alpha_{0}$ and the vertical axis represents the total probability of the universe with $a\geq u$. The parameters are fixed as $p_{a}=1$ and $V_{0}=50$.} The right figure represents the total probability to create a universe with $\phi\geq\phi_{c}$, where the horizontal axis is the temperature and the vertical axis is the function $K(\phi_{c},T)$. The coefficient $\lambda$ is fixed as $\lambda=1$.
\end{figure}

We introduce the function $G(u,\alpha_{0})\equiv \int_{u}^{\infty}\rho(a,\alpha_{0})da$ based on Eq.~\eqref{eq:7.13}. The $G(u,\alpha_{0})$ represents the total transition probability of the universe to a big universe from the initial state of Eq.~\eqref{eq:4.m5}. It is difficult to obtain an analytic form of $G(u,\alpha_{0})$. Thus, the numerical calculation results of $G(u,\alpha_{0})$ are shown in Fig.~\ref{fig:10}(a), which shows that $G(u, \alpha_{0})\rightarrow1$ as $\alpha_{0}$ increases. Therefore, as the initial temperature of the radiation increases, the transition probability of the universe from the initial state of Eq.~\eqref{eq:4.m5} to a bigger universe increases. When $\alpha_{0}$ is enough high, $G(u, \alpha_{0})\rightarrow1$.

In~\cite{LD}, Linde studied the thermal scalar field induced tunneling of the quantum universe. He showed that the tunneling probability is $P(\phi)\approx\mathscr{N}\cdot\mathrm{exp}\{-6.34\phi/(T\sqrt{\lambda})\}$~\cite{LD}. The $\phi$ and $T$ represent the scalar field and temperature, respectively, and $\lambda$ is the coefficient of the $\phi^4$ term in the Lagrangian of the scalar field. Similarly, we define $K(\phi_{c},T)\equiv\int_{\phi_{c}}^{\infty}P(\phi)d\phi$. The meaning of the function $K(\phi_{c},T)$ is the total probability to create a universe with $\phi\geq\phi_{c}$. The $\phi_{c}$ is a real positive number and can be arbitrarily fixed without loss of generality. One can easily prove that $K(\phi_{c},T)=\mathscr{N}\cdot \mathrm{exp}\{-6.34\phi_{c}/(T\sqrt{\lambda})\}$. As shown in Fig.~\ref{fig:10}(b), $K(\phi_{c},T)$ increases (up to the upper limit 1) with temperature $T$. This indicates that a greater temperature increases the chances of creating an universe. Our results are consistent with those obtained by Linde to a certain extent.

In this section we showed that as the radiation temperature increases, it is easier for a small universe to have a transition to a bigger one. We note that the probability of Eq.~\eqref{eq:7.13} may not be simply viewed as the chance of the universe tunneling from ``nothing." This is also different from the meaning of the Hartle--Hawking wave function. Equation~\eqref{eq:7.13} corresponds to the evolution of the initial wave packet in Eq.~\eqref{eq:4.m5}. This represents a spacetime structure with the fluctuations of the scale factor give by $\alpha$. That is, the current results give the transition probability from a small and highly fluctuating spacetime to a larger universe with a definite scale factor.

\section{Conclusions and discussions}

In this work, we non-perturbatively studied the evolution of the quantum universe through the closed real-time path integral. The basic equation of this work is the Liouville--von Neumann expression in Eq.~\eqref{eq:1.3}.
The time variable $t$ in Eqs.~\eqref{eq:1.3} and \eqref{eq:2.11} can be interpreted as the dust field. By introducing the dust field as the clock, Brown and Kucha$\check{\mathrm{r}}$ transformed the Wheeler--DeWitt equation into the form of Schr\"{o}dinger's equation~\cite{JD,VT}. This model may help resolve the problem of time in quantum gravity and quantum cosmology. The average value of the Hamiltonian operator for the dust field is usually not equal to zero. This means that the clock can influence the evolution of the universe. However, in any formal physical theory, the clock does have not any influence on the evolution of the related system. We introduce the condition of Eq.~\eqref{eq:2.14} to eliminate the influence of the clock. The physical meaning of the condition in Eq.~\eqref{eq:2.14} is the average value of the Hamiltonian operator of the clock is set to zero. In this work, the impact of this condition on the evolution of the quantum universe is not strong. There are several parameters in our model, and the effect of the condition in Eq.~\eqref{eq:2.14} is to eliminate one of the independent parameters.

We derived the Hamiltonian operator for the massless real scalar field in the flat FRW spacetime. Our method is similar to that in~\cite{SSF}. However, the derivation in~\cite{SSF} is only correct in the de-Sitter spacetime. Our derivation is exact for more general situations. We proved the form of the Hamiltonian operator of Eq.~\eqref{eq:3.26} in some special FRW spacetimes. We used some rationales to infer that the Hamiltonian operator of Eq.~\eqref{eq:3.26} should be reasonable in any kind of FRW ($k=0$) spacetime. Thus, Eq.~\eqref{eq:3.26} can be used to study the quantum universe. In the Hamiltonian operator of Eq.~\eqref{eq:3.26}, we neglected the cosmological particle production and the vacuum energy. For some extremal situations where the cosmological particle production is important, the operator of Eq.~ \eqref{eq:3.26} is not valid.

For the flat FRW universe, there is a divergence problem as space is not compact. This divergence can be attributed to the coordinate volume $V_{0}$. We illustrate that the coordinate volume can be viewed as a global conformal factor that can be eliminated from all equations. Thus, the value of the coordinate volume $V_{0}$ does not influence the results, and the divergence problem related to the coordinate volume can be solved. However, to clearly show that the total system has a global scaling symmetry, we still retain $V_{0}$ in our formulas.

Based on Eq.~\eqref{eq:1.3}, by setting the initial state of spacetime as a Gaussian wave packet, we studied the evolution of the universe in the cases where the environments are heat radiation, non-relativistic matter and dark energy, respectively. Regardless of the matter form, we find that the average value $\langle a\rangle$ of the scale factor increases with time in both the proper time coordinate and the conformal time coordinate. The evolution for the quantum universe described by the wave packet is always consistent with the classical evolution trajectory of the universe described by the scale factor. This indicates that the universe can grow from the initial state. We also find that the coherence, the absolute quantum fluctuation and the Gibbs entropy all increase with time. However, the relative quantum fluctuation decreases with time. Smaller relative quantum fluctuations show the characteristics of the evolution of the Gaussian-like quantum state. For Gaussian distribution, the relative fluctuations are small since the distribution decays exponentially fast. This leads to the smaller relative fluctuations. Therefore, one can use the average or expectation values and the second order fluctuations to characterize the whole system. On the other hand, if the evolution follows other distributions such as streched exponential or power law, the tail part of the distribution can be fatty and the rare fluctuation events can become important. In such cases, the relative fluctuations are not small but significant. The average or expectation values can no longer be representative. The whole statistical distribution is necessary to characterize the system dynamics.

We show that the larger absolute quantum fluctuations are associated with the higher coherence.  The variations in the absolute quantum  fluctuation and the Gibbs entropy are consistent with the variations in the coherence. The coherence increases with time which indicates that the bigger universe  still maintains important quantum natures in our model. For the radiation dominated universe, the higher initial temperature corresponds to more rapid variations of these quantities. We show that for a given size of the radiation dominated universe, the lower temperature corresponds to a more quantum universe. In the proper time coordinate, in order to obtain these results, we introduced certain approximations. However, in the conformal time coordinate, without any approximation, we obtained the same results. Therefore, our results are reasonably reliable. If we think that the spacetime of the observed universe today is classical and quantum in the very early times, then our work shows that more complicated models maybe needed to describe the decoherence process.

We also studied the transition of the quantum universe in the case where the initial state of the spacetime is a Gaussian wave packet. We find that as the temperature of the bath increases, the transition probability of the universe to a bigger one increases.

In addition, for any massless field (no self-interaction) minimally coupled to the flat FRW spacetime, it appears reasonable to think that the Hamiltonian operator of the field has a similar structure  to Eq.~\eqref{eq:3.26}. That is, the Hamiltonian should be equal to the red shift factor times the sum of all quanta energies (neglecting the vacuum energy and cosmological particle production. For the scalar field, the summation is carried out to the momentum. And for other fields, the summation should also include all the internal degrees of freedom. ). Then, using the same method with that in Sec.~\ref{sec:4}, one can show that in the proper time coordinate any massless field minimally coupled to the flat FRW spacetime can generally lead to non-Markovian dynamics.

\section*{Acknowledgements}
Hong Wang was supported by National Natural Science Foundation of China under Grant 21721003 and the Ministry of Science and Technology of China under Grant 2016YFA0203200. Hong Wang thanks the help from Professor Erkang Wang.

\appendix

\section{Derivation for Hamiltonian operator in Eq.~\eqref{eq:3.23}}
\label{sec:B}
According to the definition of the function $g_{k}(t)$ (Eq.~\eqref{eq:3.12}), one easily obtains
\begin{eqnarray}\begin{split}
\label{eq:B5}
\dot{g}_{k}^{*}&(t)g_{k}(t)-\dot{g}_{k}(t)g_{k}^{*}(t)\\&=(2\pi)^{-3}\cdot\frac{\pi}{4}C^{-1}(t)t\Big\{\mathrm{H}^{(2)}_{\nu}(kt)\partial_{t}\mathrm{H}^{(2)*}_{\nu}(kt)\\&-
\mathrm{H}^{(2)*}_{\nu}(kt)\partial_{t}\mathrm{H}^{(2)}_{\nu}(kt)\Big\},
\end{split}
\end{eqnarray}
where $\nu$ is a real number. Substituting the Wronskian determinant of
\begin{equation}
\label{eq:B6}
z\Big\{\mathrm{H}^{(2)}_{\nu}(z)\partial_{z}\mathrm{H}^{(2)*}_{\nu}(z)-\mathrm{H}^{(2)*}_{\nu}(z)\partial_{z}\mathrm{H}^{(2)}_{\nu}(z)\Big\}=\frac{4i}{\pi}
\end{equation}
into Eq.~\eqref{eq:B5} gives
\begin{equation}
\label{eq:B1}
\dot{g}_{k}^{*}(t)g_{k}(t)-\dot{g}_{k}(t)g_{k}^{*}(t)=i (2\pi)^{-3}C^{-1}(t).
\end{equation}
Substituting Eqs.~\eqref{eq:3.8} and \eqref{eq:3.12} into Eq.~\eqref{eq:3.6} provides
\begin{eqnarray}\begin{split}
\label{eq:B7}
\mathscr{H}_{\phi}=&\frac{1}{2}C(t)\sum_{\vec{k}}\sum_{\vec{k'}}\Big\{a_{\vec{k}}a_{\vec{k'}}\dot{g}_{k}(t)\dot{g}_{k'}(t)e^{i\vec{k}\cdot\vec{x}}e^{i\vec{k'}\cdot\vec{x}}
\\&+a^{\dag}_{\vec{k}}a^{\dag}_{\vec{k'}}\dot{g}^{*}_{k}(t)\dot{g}^{*}_{k'}(t)e^{-i\vec{k}\cdot\vec{x}}e^{-i\vec{k'}\cdot\vec{x}}\\&
+a_{\vec{k}}a^{\dag}_{\vec{k'}}\dot{g}_{k}(t)\dot{g}^{*}_{k'}(t)e^{i\vec{k}\cdot\vec{x}}e^{-i\vec{k'}\cdot\vec{x}}
\\& + a^{\dag}_{\vec{k}}a_{\vec{k'}}\dot{g}^{*}_{k}(t)\dot{g}_{k'}(t)e^{-i\vec{k}\cdot\vec{x}}e^{i\vec{k'}\cdot\vec{x}}\\&
-a_{\vec{k}}a_{\vec{k'}}(\vec{k}\cdot\vec{k'})g_{k}(t)g_{k'}(t)e^{i\vec{k}\cdot\vec{x}}e^{i\vec{k'}\cdot\vec{x}}
\\&-a^{\dag}_{\vec{k}}a^{\dag}_{\vec{k'}}(\vec{k}\cdot\vec{k'})g^{*}_{k}(t)g^{*}_{k'}(t)e^{-i\vec{k}\cdot\vec{x}}e^{-i\vec{k'}\cdot\vec{x}}\\&
+a_{\vec{k}}a^{\dag}_{\vec{k'}}(\vec{k}\cdot\vec{k'})g_{k}(t)g^{*}_{k'}(t)e^{i\vec{k}\cdot\vec{x}}e^{-i\vec{k'}\cdot\vec{x}}
\\&+a^{\dag}_{\vec{k}}a_{\vec{k'}}(\vec{k}\cdot\vec{k'})g^{*}_{k}(t)g_{k'}(t)e^{-i\vec{k}\cdot\vec{x}}e^{i\vec{k'}\cdot\vec{x}}
\Big\}.
\end{split}
\end{eqnarray}
Thus, the Hamiltonian operator is
\begin{eqnarray}\begin{split}
\label{eq:B8}
H_{\phi}(t)&=\int dx^{3}\mathscr{H}_{\phi}\\&=\frac{1}{2}(2\pi)^{3}C(t)\sum_{\vec{k}}\Big\{
a_{\vec{k}}a_{-\vec{k}}\big[\big(\dot{g}_{k}(t)\big)^{2}+\vec{k}^{2}\big(g_{k}(t)\big)^{2}\big]\\&\quad
+a^{\dag}_{\vec{k}}a^{\dag}_{-\vec{k}}\big[\big(\dot{g}^{*}_{k}(t)\big)^{2}+\vec{k}^{2}\big(g_{k}^{*}(t)\big)^{2}\big]
\\&\quad+a^{\dag}_{\vec{k}}a_{\vec{k}}\big[\dot{g}_{k}(t)\dot{g}^{*}_{k}(t)+\vec{k}^{2}g_{k}(t)g^{*}_{k}(t)\big]\\&\quad
+a_{\vec{k}}a^{\dag}_{\vec{k}}\big[\dot{g}_{k}(t)\dot{g}^{*}_{k}(t)+\vec{k}^{2}g_{k}(t)g_{k}^{*}(t)\big]\Big\}
\end{split}
\end{eqnarray}

Introducing the definitions for $\varepsilon_{k}(t)$, $\triangle_{k}(t)$, and $\omega_{k}(t)$ (see Eqs.~\eqref{eq:3.16}, \eqref{eq:3.17}, and \eqref{eq:3.18}) gives the Hamiltonian operator of Eq.~\eqref{eq:B8} as
\begin{eqnarray}\begin{split}
\label{eq:B9}
H_{\phi}(t)=&\frac{1}{2}\sum_{\vec{k}}\big\{\varepsilon_{k}(t)(a^{\dag}_{\vec{k}}a_{\vec{k}}+a_{\vec{k}}a^{\dag}_{\vec{k}})+
\triangle_{k}(t)a_{\vec{k}}a_{-\vec{k}}\\&+\triangle^{*}_{k}(t)a^{\dag}_{-\vec{k}}a^{\dag}_{\vec{k}}\big\}.
\end{split}
\end{eqnarray}
Starting from the definition of Eq.~\eqref{eq:3.18}, we have
\begin{eqnarray}\begin{split}
\label{eq:B10}
\omega_{k}^{2}(t)&=\varepsilon_{k}^{2}(t)-\triangle_{k}^{*}(t)\triangle_{k}(t)\\&
=-(2\pi)^{6}C^{2}(t)\vec{k}^{2}\big(\dot{g}_{k}^{*}(t)g_{k}(t)-\dot{g}_{k}(t)g_{k}^{*}(t)\big)^{2}\\&=\vec{k}^{2}.
\end{split}
\end{eqnarray}

Using these definitions $u_{k}(t)$, $v_{k}(t)$, $A_{\vec{k}}(t)$, and $A_{\vec{k}}^{\dag}(t)$ (see Eqs.~\eqref{eq:3.19}, \eqref{eq:3.20}, \eqref{eq:3.21}, and \eqref{eq:3.22}), and combined with Eq.~\eqref{eq:B10} gives the Hamiltonian operator of Eq.~\eqref{eq:B9} in a more compact form~\cite{SSF}
\begin{equation}
\label{eq:B11}
H_{\phi}(t)=\frac{1}{2}\sum_{\vec{k}}|\vec{k}|\big\{A_{\vec{k}}^{\dag}(t)A_{\vec{k}}(t)+A_{\vec{k}}(t)A_{\vec{k}}^{\dag}(t)\big\}.
\end{equation}
In addition, one can prove that~\cite{SSF}
\begin{eqnarray}\begin{split}
\label{eq:B12}
\big[A_{\vec{k_{1}}}(t_{1}),A_{\vec{k_{2}}}^{\dag}(t_{2})\big]=&\big(u_{k_{1}}(t_{1})u^{*}_{k_{1}}(t_{2})-v_{k_{1}}(t_{1})v^{*}_{k_{1}}(t_{2})\big)\\&\times\delta^{3}(\vec{k_{1}}-\vec{k_{2}}).
\end{split}
\end{eqnarray}
Substituting Eqs.~\eqref{eq:B12} into \eqref{eq:B11} provides the Hamiltonian operator of Eq.~\eqref{eq:3.23}.

\section{Derivation for the influence functional $\mathbf{I_{rN}}$}
\label{sec:C}
The influence functional of Eq.~ \eqref{eq:4.10} can be written as
\begin{eqnarray}\begin{split}
\label{eq:C1}
\mathbf{I_{rN}}=&\mathrm{Tr}_{\phi}\Big\{\mathrm{exp}\big\{i\delta t\sum_{n=1}^{N}\big[\hat{H}_{\phi}(a_{n}^{-})\\&-\hat{H}_{\phi}(a_{n}^{+})\big]\big\}\rho_{\phi}(0)\Big\}.
\end{split}
\end{eqnarray}
Substituting Eqs.~\eqref{eq:4.10} and \eqref{eq:4.13} into Eq.~\eqref{eq:C1} allows writting the influence functional of Eq.~\eqref{eq:C1} as
\begin{eqnarray}\begin{split}
\label{eq:C2}
\mathbf{I_{rN}}&=\mathrm{Tr}_{\phi}\Big\{\mathrm{exp}\big[i\delta t\sum_{n=1}^{N}\sum_{\vec{k}}|\vec{k}|A^{\dag}_{\vec{k}}A_{\vec{k}}(\frac{1}{a_{n}^{-}}-\frac{1}{a_{n}^{+}})\big]\\&\times
\prod_{j}\mathrm{exp}(\frac{-k_{j}}{\alpha_{0}}A^{\dag}_{k_{j}}A_{k_{j}})\big(1-\mathrm{exp}(\frac{-k_{j}}{\alpha_{0}})\big)\Big\}.
\end{split}
\end{eqnarray}
Thus,
\begin{eqnarray}\begin{split}
\label{eq:C3}
\mathbf{I_{rN}}=&\mathrm{Tr}_{\phi}\Big\{\mathscr{N}_{\phi}\prod_{\vec{k}}\mathrm{exp}\big\{\big[-\frac{1}{\alpha_{0}}\\&-i\delta t\sum_{n=1}^{N}(\frac{1}{a_{n}^{+}}-\frac{1}{a_{n}^{-}})\big]|\vec{k}|A^{\dag}_{\vec{k}}A_{\vec{k}}\big\}\Big\}.
\end{split}
\end{eqnarray}
Here,
\begin{equation}
\label{eq:C4}
\mathscr{N}_{\phi}\equiv \prod_{\vec{k}}(1-e^{-\frac{\vec{k}}{\alpha_{0}}})
\end{equation}
is a parameter that is independent with the variables $(a_{0}^{\pm},a_{1}^{\pm},\cdot\cdot\cdot,a_{N}^{\pm})$. therefore, it does not impact the properties of the reduced density matrix of Eq.~\eqref{eq:4.9} and can be absorbed into the normalization constant $\mathscr{N}$.

In the particle number representation, the influence functional can be written as
\begin{eqnarray}\begin{split}
\label{eq:C6}
\mathbf{I_{rN}}&=\sum_{n_{1},...,n_{k},...}\prod_{\vec{k}}\mathrm{exp}\Big\{\big[-\frac{1}{\alpha_{0}}\\&\quad-i\delta t\sum_{n=1}^{N}(\frac{1}{a_{n}^{+}}-\frac{1}{a_{n}^{-}})\big]|\vec{k}|n_{k}\Big\}\\&=
\prod_{\vec{k}}\frac{1}{1-\mathrm{exp}\Big\{\big[-\frac{1}{\alpha_{0}}-i\delta t\sum_{n=1}^{N}(\frac{1}{a_{n}^{+}}-\frac{1}{a_{n}^{-}})\big]|\vec{k}|\Big\}}.
\end{split}
\end{eqnarray}
Here, $n_{k}$ represents the particle number with momentum $\vec{k}$. Equation~\eqref{eq:C6} gives
\begin{eqnarray}\begin{split}
\label{eq:C7}
\mathrm{ln}\;\mathbf{I_{rN}}=&-\sum_{\vec{k}}\mathrm{ln}\Big\{1-\mathrm{exp}\big\{\big[-\frac{1}{\alpha_{0}}\\&-i\delta t\sum_{n=1}^{N}(\frac{1}{a_{n}^{+}}-\frac{1}{a_{n}^{-}})\big]|\vec{k}|\big\}\Big\}.
\end{split}
\end{eqnarray}
Taking the continuous limit and integrating over the momentum $\vec{k}$ indicates
\begin{eqnarray}\begin{split}
\label{eq:C8}
\mathrm{ln}\;\mathbf{I_{rN}}&=-\frac{V_{0}}{(2\pi)^{3}}\cdot\int d\vec{k}^{3}\;\mathrm{ln}\Big\{1-\mathrm{exp}\big\{\big[-\frac{1}{\alpha_{0}}\\&\quad-i\delta t\sum_{n=1}^{N}(\frac{1}{a_{n}^{+}}-\frac{1}{a_{n}^{-}})\big]|\vec{k}|\big\}\Big\}\\&=
\frac{V_{0}}{90}\pi^{2}\alpha_{0}^{3}\cdot\frac{1}{\big(1+i\alpha_{0}\delta t\sum_{n=1}^{N}(\frac{1}{a_{n}^{+}}-\frac{1}{a_{n}^{-}})\big)^{3}}.
\end{split}
\end{eqnarray}
Thus, one can easily obtain the influence functional in Eq.~\eqref{eq:4.15}.

\end{document}